\documentclass[aps,prl,twocolumn,superscriptaddress,showpacs,floatfix,longbibliography]{revtex4-1}  
\usepackage{amsmath,amssymb,wasysym,graphicx}
\usepackage{times}
\usepackage[varg]{txfonts}
\usepackage{textcomp}
\usepackage{subfigure}
\usepackage{tabu}
\usepackage{color}
\usepackage{xcolor}
\usepackage[colorlinks=true,citecolor=blue,urlcolor=blue,linkcolor=blue,hyperindex]{hyperref}
\usepackage{braket}
\usepackage{dsfont}
\usepackage{overpic}
\usepackage{clrscode3e}

\usepackage[normalem]{ulem}
\usepackage{verbatim}
\allowdisplaybreaks

\usepackage{tikz}
\usetikzlibrary{shapes}
\usetikzlibrary{tikzmark,calc}
\newcommand{\tetrahedron}{\vcenter{\hbox{ \begin{tikzpicture}[scale=0.25] \coordinate (A) at (-0.56,-0.56); \coordinate (B) at (-0.7,0.0); \coordinate (C) at (0.,-0.7); \coordinate (D) at (0.14,0.14); \draw (A) -- (B); \draw (A) -- (C); \draw (A) -- (D); \draw (B) -- (C); \draw (B) -- (D); \draw (C) -- (D); \end{tikzpicture} }}}

\begin{document}
	\title{Quantum Fisher Information as a Thermal Probe in Frustrated Magnets through Insights from Quantum Spin Ice}

\author{Chengkang Zhou}
\thanks{These authors contributed equally to this work.}
\affiliation{Department of Physics and HK Institute of Quantum Science \& Technology, The University of Hong Kong, Pokfulam Road, Hong Kong}

\author{Zhengbang Zhou}
\thanks{These authors contributed equally to this work.}
\affiliation{Department of Physics, University of Toronto, Toronto, Ontario M5S 1A7, Canada}

\author{F\'elix Desrochers}
\affiliation{Department of Physics, University of Toronto, Toronto, Ontario M5S 1A7, Canada}
\affiliation{%
Department of Physics, Harvard University, Cambridge, MA 02138, USA
}%

\author{Yong Baek Kim}
\affiliation{Department of Physics, University of Toronto, Toronto, Ontario M5S 1A7, Canada}

\author{Zi Yang Meng}
\affiliation{Department of Physics and HK Institute of Quantum Science \& Technology, The University of Hong Kong, Pokfulam Road, Hong Kong}

\date{\today}
	
\begin{abstract}
Quantum Fisher information (QFI) is a measure of multipartite entanglement accessible via inelastic neutron scattering. Here we demonstrate that QFI reveals thermal and dynamical properties of quantum spin ice (QSI), a three-dimensional quantum spin liquid with fractionalized excitations. By developing a multi-directed loop update quantum Monte Carlo algorithm, along with exact diagonalization and gauge mean-field theory, we compute the QFI for the pyrochlore lattice. The temperature and momentum dependence of QFI maps the phase diagram, distinguishing the ferromagnetic ordered phase, its critical region, the zero-flux QSI, and the $\pi$-flux QSI. QFI also captures two crossover scales: from trivial paramagnet to classical spin ice, then to QSI. We discuss the $\pi$-flux QSI in light of experiments on cerium-based pyrochlores. Our results suggest that QFI not only detects entanglement but also serves as a sensitive thermal and dynamical probe for frustrated quantum magnets.
\end{abstract}
\maketitle

\section{Introduction}\label{sec1}

Entanglement is arguably the most important concept in quantum physics. 
It is at the heart of our modern understanding of phases of matter and transitions between them. As a notable example, entanglement is the fundamental characteristic of quantum spin liquids (QSLs)~\cite{wen2002quantum, levin2005string, gu2009tensor,zhaoMeasuring2022} 
--- paramagnetic phases of frustrated spin systems that fail to magnetically order down to zero temperature, and host fractional excitations and emergent gauge fields~\cite{anderson1987resonating, savary2016quantumspinliquids, zhou2017quantum, knolle2019field, broholm2020quantum}. QSLs are characterized by long-range entanglement, which may be measured for gapped QSL by a non-vanishing topological entanglement entropy~\cite{levin2006detecting, kitaev2006topological,isakovTopological2011,zhaoMeasuring2022,chenTopological2022}. However, despite its role as a theoretical cornerstone and its usefulness in numerical studies~\cite{grover2011entanglement, jiang2012identifying,isakovTopological2011,zhaoScaling2022,chenTopological2022,Extracting2024Liao}, topological entanglement entropy is not experimentally accessible in solid-state platforms, where only local correlations are typically measured.

One still has experimental access to other measures of entanglement. A particularly useful one that has recently been at the forefront of the experimental search for QSLs is the quantum Fisher information (QFI)~\cite{Determining2016Shitara,lambert2019estimates, Witnessing2021Scheie, laurell2021quantifying, menon2023multipartite, laurell2025witnessing, Can2025shimokawa}. Initially employed to define the maximal achievable precision in parameter estimation for a given quantum state in the quantum metrology community~\cite{braunstein1994statistical, petz1996geometries, petz2002covariance, paris2009quantum, escher2011general, tino2014atom}, the QFI is directly related to the dynamical susceptibility~\cite{hauke2016measuring}, which is routinely measured in inelastic neutron scattering experiments~\cite{lovesey1984theory, boothroyd2020principles, bramwell2014neutron}. The QFI density $f_{Q}$ provides a lower bound on the multipartite entanglement in the system~\cite{toth2012multipartite,hyllus2012fisher}.
This bound can be used to experimentally differentiate between QSLs and other, more trivial states, such as random singlet states driven by strong disorder~\cite{Witnessing2025Sabharwal, Can2025shimokawa}. These states may otherwise be challenging to distinguish experimentally, as they can both lead to similar signatures, such as continua of excitations in inelastic neutron scattering or a lack of experimentally observable finite-temperature phase transitions~\cite{lee1996spin, paddison2017continuous,kimchi2018valence,gao2023disorder,ross2016static,sarkar2017spin,greenblatt2009rounding}. Nevertheless, it should be emphasized that a large value of the QFI, althought promising, does not provide evidence for the realization of a QSL in and of itself. Indeed, trivial ordered states sufficiently close to a quantum critical point may have an arbitrarily large value of $f_{Q}$~\cite{hauke2016measuring}. Be as it may, if precise theoretical predictions for the momentum and temperature dependence of the QFI for prospective QSLs exist, they offer stringent quantitative predictions which, if measured, may provide significantly more convincing evidence than qualitative features, like the presence of broad continua of excitations. 

It is worth noting that the QFI is not meant to be a universal entanglement classifier; rather, it provides an operator-resolved and experimentally aligned witness for multipartite entanglement depth. Compared with other witnesses, QFI is distinguished in collective spin systems because it is simultaneously (i) well-defined for mixed thermal states, (ii) scalable to large many-body settings without requiring reduced density matrices, and (iii) directly linked to the excitation spectrum through a frequency integral of the dynamical response. By contrast, pairwise measures such as concurrence \cite{Witnessing2025Sabharwal, Tutorial2025Scheie} probe only two-spin entanglement, they neither certify multipartite entanglement depth nor remain informative when entanglement is predominantly multipartite and delocalized, as is typical in collective correlated phases.

In this letter, we make such detailed predictions for the QFI on one of the most paradigmatic QSL: quantum spin ice (QSI)~\cite{Pyrochlore2004Hermele, castro2006ice, benton2012seeing,castelnovo2012spin, gingras2014quantum, udagawa2021spin}. QSI is a three-dimensional QSL that is the ground state of an XXZ model with dominant Ising and subleading transverse couplings on the pyrochlore lattice. It realizes the deconfined (Coulomb) phase of compact $U(1)$ gauge theory and, as such, hosts emergent photon excitations, spin-$1/2$ spinons that act as emergent electric charges, and magnetic monopoles~\cite{Dynamics2018Huang}. QSI is the ideal platform for making specific predictions for the QFI, as it is one of the few known examples of an experimentally relevant model that stabilizes a well-understood QSL that is numerically accessible over a parameter regime (i.e., ferromagnetic transverse couplings) with sign-problem-free quantum Monte Carlo (QMC)~\cite{Unusual2008Banerjee,shannon2012quantum,Coulomb2015Lv,Numerical2015Kato,Dynamics2018Huang,huang2020extended}. Furthermore, several compounds have historically been considered as possible experimental realization of QSI such as Tb$_2$Ti$_2$O$_7$~\cite{kao2003understanding, guitteny2013anisotropic, petit2012spin, takatsu2016quadrupole, taniguchi2013long}, Pr$_2$(Sn,Zr,Hf)$_2$O$_7$~\cite{zhou2008dynamic, onoda2010quantum, matsuhira2004low, kimura2013quantum, petit2016antiferroquadrupolar, sibille2018experimental}, and Yb$_2$Ti$_2$O$_7$~\cite{ross2011quantum, gingras2014quantum, kermarrec2017ground, robert2015spin, thompson2017quasiparticle, arpino2017impact, petit2020way}. The most recently considered candidate materials are the Cerium-based pyrochlore compounds Ce$_2$(Zr,Hf,Sn)$_2$O$_7$~\cite{huang2014quantum, sibille2015candidate, gaudet2019quantum, smith2022case, Beare2923MuSr, Neutron2025Gao, smith2025single, yahne2022dipolar, poree2025dipolar, smith2025two, kermarrec2025magnetization}. For this last family of compounds, no magnetic order has been reported in Ce$_2$(Zr,Hf)$_2$O$_7$. Experimental determinations of their microscopic couplings indicate that Ce$_2$Zr$_2$O$_7$ and Ce$_2$Hf$_2$O$_7$ likely fall in a region of parameter space that stabilizes, so-called, $\pi$-flux QSI~\cite{smith2022case,bhardwaj2022sleuthing,smith2023quantum, Poiree2022, poree2025dipolar, smith2025two, kermarrec2025magnetization} where a constant flux of the emergent gauge field is threading hexagonal plaquettes such that translation acts projectively on spinons excitations~\cite{lee2012generic, savary2021quantum, li2017symmetry, chen2017spectral, yao2020pyrochlore, desrochers2022symmetry, desrochers2023spectroscopic}. Energy integrated and inelastic neutron scattering measurements consistent with theoretical predictions have also been reported~\cite{smith2022case, gao2019experimental, Neutron2025Gao, smith2023quantum, smith2025two, desrochers2023spectroscopic}, and a cubic scaling of the low-temperature heat capacity has even been recently measured in Ce$_2$Zr$_2$O$_7$~\cite{Neutron2025Gao}. Precise theoretical predictions for the QFI should thus be experimentally measurable and provide rigorous tests for these and future candidate materials.

We employ large-scale QMC simulations~\cite{Generalized2005Wessel,Computational2010Sandvik,Stochastic2003Sandvik,QMC2002Sandvik} to evaluate the QFI of the XXZ model with ferromagnetic transverse coupling. To this end, we develop a multi-directed loop (MDL) update algorithm to efficiently sample the highly frustrated (3+1)d configurational space; the details are provided in the Supplemental Material (SM)~\cite{suppl}. Exact diagonalization (ED) is also used to compute the QFI~\cite{thermal2012sugiura, canonical2013sugiura} for both anti-ferromagnetic and ferromagnetic transverse couplings. We show that ED and QMC results are consistent for ferromagnetic $XY$ exchange, which justifies extending our ED result to the anti-ferromagnetic exchange case, where $\pi$-flux QSI is the ground state. These numerical results are further compared with predictions from the gauge mean-field theory (GMFT) parton construction~\cite{savary2012coulombic, lee2012generic, savary2013spin, savary2021quantum, desrochers2022symmetry, desrochers2023spectroscopic, zhou2024magnetic, zhou2025towards}, which provides a mapping from the spin Hamiltonian to a lattice U(1) gauge theory coupled to bosonic matter fields in the low-temperature limit.

We find that the QFI at finite temperature is sensitive to thermal phase transitions and crossovers, as well as regions of large critical fluctuations, depending on the momentum position. For example, the QFI associated with the transverse spin components $S^{\pm}$ at the $\Gamma=(0,0,0)$ point clearly maps out the thermal phase diagram of the pyrochlore XXZ model, revealing the two crossover temperature scales between the high-temperature paramagnet and classical spin ice as well as between classical spin ice and QSI~\cite{Numerical2015Kato, Dynamics2018Huang}. Its evolution also delineates the ferromagnetic (FM) thermal phase transition for large ferromagnetic transverse couplings~\cite{Unusual2008Banerjee, Coulomb2015Lv}. In contrast, the QFI at the $\Gamma^\prime=(4\pi,4\pi,0)$ point appears to be sensitive to critical fluctuations near the phase transition between QSI and the ferromagnetic ordered phases. Altogether, our work provides the first large-scale, unbiased computation of experimentally accessible entanglement properties of QSI, offering exhaustive numerical results that can be quantitatively compared with future experimental results on current candidate materials.

\section{Results}\label{sec2-3}

\begin{figure*}[htp!]
	\centering
	\includegraphics[width=\textwidth]{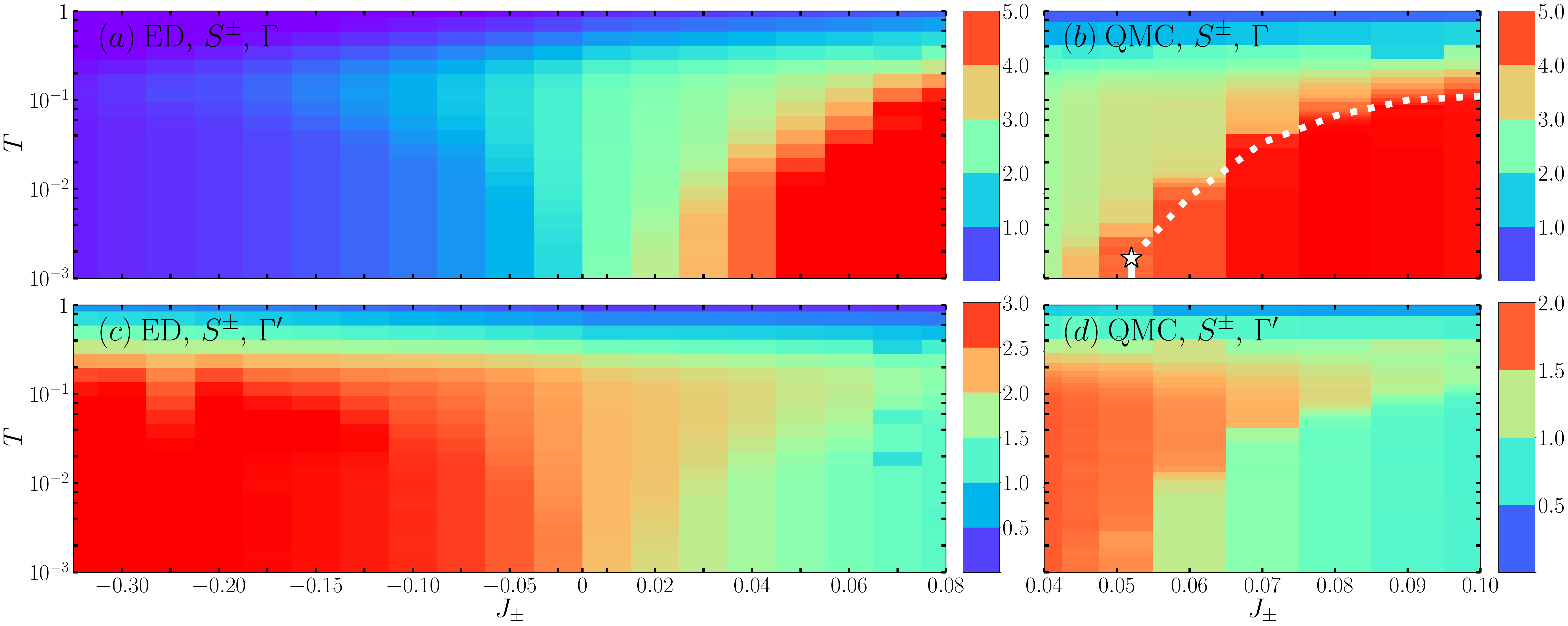}
	\caption{\textbf{Heat maps of the QFI as functions of temperature $T$ and $J_{\pm}$.} 
	Panels (a) and (b) show the QFI density $f_{Q}(S^{\pm}_\mathbf{q},T)$ in the $S^{\pm}$ channel at $\Gamma=(0,0,0)$, while panels (c,d) show it at $\Gamma^\prime=(4\pi,4\pi,0)$. Panel (a) and (c) are obtained from the ED calculation of 16-site cluster with $J_{\pm}$ ranging from $-0.045$ to $0.08$, and panels (b) and (d) are from the QMC simulation of $4\times L^3$ sizes ($L=4$) with $J_{\pm}$ ranging from $0.04$ to $0.10$. In panels (a) and (b), $f_{Q}(S^{\pm}_\Gamma,T)$ maps out the thermodynamic phase boundaries between the ferromagnetic phase (FM) and QSI$_0$. The temperature dependence of $f_{Q}(S^{\pm}_\Gamma,T)$ from QMC further discerns the crossover temperature scales between the high-temperature paramagnetic regime to classical spin ice and eventually to the QSI$_0$ regime (see also Fig.~\ref{fig:QFI_line} (a)). In panels (c) and (d), $f_{Q}(S^{\pm}_{\Gamma^\prime},T)$ reflects the strength of the fluctuations in the thermal and quantum phase diagram, with strong QFI at the classical critical region above the FM phase and stronger QFI in the classical spin ice and QSI$_0$ region. Moreover, the strongest QFI signal, represented by the red region in panel (c) for $J_{\pm}<0$, reflects the QSI$_\pi$ regime (see also Fig.~\ref{fig:QFI_line} (b)). 
	In panel (b), the white dashed line denotes the phase boundary between the paramagnetic and FM phases, whereas the white solid line ($J_{\pm,c}=0.052$) represents the transition between the FM and QSI$_0$ phases. The white star marks the crossover temperature scale from the CSI regime to the QSI$_0$ regime ($T=12\times0.052^3\approx 0.00169$) at $J_{\pm,c}$.}
	\label{fig:QFI_phase}
\end{figure*}

\subsection{Model}\label{sec2}

The pyrochlore lattice is a face-centered cubic lattice with four sublattices per unit cell, forming a network of corner-sharing tetrahedra. We consider spin-$1/2$ on the pyrochlore lattice that are interacting through an XXZ model of the form
\begin{equation}
\label{eq:XXZ_Ham}
H = J_{\mathrm{z}}\sum_{\langle i, j\rangle} S_i^{\mathrm{z}} S_j^{\mathrm{z}}
- J_{\pm}\sum_{\langle i, j\rangle}\left(S_i^{+} S_j^{-}+S_i^{-} S_j^{+}\right),
\end{equation}
where the summation $\langle i,j \rangle$ runs over all the nearest-neighbor pairs. We focus on the regime with a dominant antiferromagnetic Ising coupling (i.e., $J_{\mathrm{z}}>|J_\pm|>0$), which provides geometric frustration within each tetrahedron and energetically favors ``classical spin ice configurations'' where $\sum_{i\in \hspace{-0.5mm}\tetrahedron}\hspace{-0.5mm} S^{\mathrm{z}}_i=0$ with the sum over all spins in a given tetrahedron. In the perturbative regime $|J_{\pm}|\ll J_{\mathrm{z}}$, the transverse coupling allows for tunneling between the classically degenerate configurations that respect this local energetic constraint~\cite{Pyrochlore2004Hermele}. This tunnelling stabilizes 0-flux QSI ($\mathrm{QSI}_0$) for $0<J_{\pm}/J_{\mathrm{z}}<0.052$ at low temperatures ($T\lesssim 12|J_{\pm}^3|/J_{\mathrm{z}}^2$) as shown in QMC~\cite{Unusual2008Banerjee,shannon2012quantum,Coulomb2015Lv,Numerical2015Kato,Dynamics2018Huang}. When $-1<J_{\pm}/J_{\mathrm{z}}<0$, $\pi$-flux QSI ($\mathrm{QSI}_\pi$) is instead stabilized, where the flux refers to the emergent static magnetic flux threading the hexagonal plaquettes of the pyrochlore lattice in the ground state~\cite{Pyrochlore2004Hermele,lee2012generic,chen2017spectral,patri2020distinguishing,benton2020ground,desrochers2022symmetry, desrochers2023spectroscopic, chern2023pseudofermion}. For $J_{\pm}/J_{\mathrm{z}}>0.052$, the system undergoes a fluctuation-induced first-order transition~\cite{halperin1974first, makhfudz2014fluctuation} into an XY ferromagnetic (FM) ordered phase below a critical temperature $T_c$ \cite{Unusual2008Banerjee,Coulomb2015Lv,Numerical2015Kato}. Hereafter, we set $J_{\mathrm{z}}=1$ as the unit of energy. For reference, the dominant exchange is usually slightly smaller than 1K in Cerium-based dipolar-octupolar candidates~\cite{smith2022case,bhardwaj2022sleuthing,smith2023quantum, Poiree2022, poree2025dipolar, smith2025two, kermarrec2025magnetization}.

To quantify the entanglement properties of a given state, the QFI provides a lower bound on multipartite entanglement, also known as entanglement depth~\cite{Fisher2012Hyllus,Tutorial2025Scheie,Witnessing2025Sabharwal}.
For a collective generator $\mathcal{O}=\sum_{i=1}^N \mathcal{O}_i$ with local eigenvalue range
$\Delta\lambda=\lambda_{\max}-\lambda_{\min}$, any $m$-producible state $\rho$ obeys~\cite{hyllus2012fisher, toth2012multipartite, Witnessing2021Scheie, Tutorial2025Scheie, liu2016quantum}
\begin{equation}
\frac{F_Q[\rho,\mathcal{O}]}{N} \le m(\Delta\lambda)^2.\label{eq:QFI_Bound}
\end{equation}Equivalently, with $f_Q(\mathcal{O}):=F_Q[\rho,\mathcal{O}]/N$, we can define the normalized QFI (nQFI) as the lower bound of entanglement depth, such that 
\begin{equation}
\mathrm{nQFI}(\mathcal{O}):=\frac{f_Q(\mathcal{O})}{(\Delta\lambda)^2} > m\Longrightarrow
\text{entanglement depth } \ge m+1,\label{eq:nQFI}
\end{equation}
For the operator $\mathcal{O}=S^\alpha_\mathbf{q}:=\sum_{i}S_{\mathbf{R}_i}^\alpha e^{i\mathbf{q}\cdot\mathbf{R}_i}$, the QFI density is related to the dynamical structure factor $A^\alpha(\mathbf{q}, \omega):=\frac{1}{2\pi N}\int dt\langle S^{\alpha\dagger}_\mathbf{q}(t)S^\alpha_{\mathbf{q}}(0)\rangle e^{i\omega t}$ at momentum $\mathbf{q}$ by~\cite{hauke2016measuring}
\begin{equation}
	f_{Q}(S^\alpha_\mathbf{q}, T)=4 \int_{0}^{\infty} d \omega \tanh \left(\frac{\omega}{2T}\right)\left(1-e^{- \omega/T}\right) A^\alpha(\mathbf{q}, \omega).
	\label{eq:QFI}
\end{equation}
Here, $T$ is the temperature and $\alpha$ labels different pseudospin components, such as $\alpha \in\{\mathrm{x},\mathrm{y},\mathrm{z}\}$. We see that the QFI density $f_{Q}(S^\alpha_\mathbf{q}, T)$ can be obtained by integrating $A^\alpha(\mathbf{q}, \omega)$ over all frequencies at a fixed temperature and momentum. Experimental measurements of the QFI have already been reported for quasi-1D $\mathrm{KCuF}_2$ 
~\cite{Witnessing2021Scheie} and for 2D frustrated triangular lattice material $\mathrm{KYbSe}_2$~\cite{Proximate2024scheie}.


To simulate the QSI$_0$ phase, we developed a MDL-QMC update that enables efficient simulations of the QSI$_0$ phase on the pyrochlore lattice. This task is difficult to achieve with the conventional directed loop update algorithm as in the QSI$_0$ phase, the simplest quantum fluctuation that connects different classical spin-ice states involves six spin flips around a hexagon, which is hard to realize by a simple directed loop update. Our MDL algorithm overcomes this limitation by allowing the insertion of multiple operator pairs, thereby naturally generating higher-order processes during the Monte Carlo update and making the QSI$_0$ phase accessible (see SM~\cite{suppl}). We then measure the imaginary time correlation function $G^{\pm}=\frac{1}{2N}\sum_{\gamma,\nu}\langle S^{+}_{-\mathbf{q},\gamma}(\tau) S^{-}_{\mathbf{q},\nu}(0)+S^{-}_{-\mathbf{q},\gamma}(\tau) S^{+}_{\mathbf{q},\nu}(0)\rangle$ and $G^{\mathrm{z}}=\frac{1}{N}\sum_{\gamma,\nu}\langle S^{\mathrm{z}}_{-\mathbf{q},\gamma}(\tau) S^{\mathrm{z}}_{\mathbf{q},\nu}(0)\rangle$, where $\gamma$, $\nu$ label the four pyrochlore sublattices, the momentum-transfer $\mathbf{q}$ is measured in the 3D pyrochlore Brillouin zone (BZ), and  $\tau\in[0,\beta]$ denotes the imaginary time with $\beta=1/T$ the inverse temperature. 
We utilize the stochastic analytic continuation (SAC) scheme~\cite{Stochastic1998sandvik,Identifying2004beach,Using2008syliuaasen,Constrained2016sandvik,Progress2023shao} to convert the imaginary time correlation function to a real-frequency dynamic structure factor. This QMC+SAC scheme has been successfully applied to a variety of lattice models, producing reliable spectral properties ranging from magnon and amplitude modes in a magnetically order state~\cite{Nearly2017shao,Amplitude2021zhou} to fractionalized excitations in QSL and QSI models~\cite{Dynamical2018Sun,Dynamics2018Huang,Fractionalized2021wang,Emergent2025chen}. 

We further compute the QFI density using ED and GMFT.
The ED calculations are performed over a 16-site periodic cubic conventional cluster (see SM~\cite{suppl}), providing access to momentum positions $\Gamma$, $\Gamma^\prime$, and $X$. We first compare the ED and GMFT results with our QMC data before extending these calculations into the $\pi$-flux $J_\pm < 0$ regime, where most QSI candidate materials reside and QMC encounters a sign problem.

\subsection{Numerical Results}\label{sec3}

Our main results are summarized in Fig.~\ref{fig:QFI_phase}. The white dashed line marks the phase boundary between the paramagnetic and FM phases, whereas the white solid line ($J_{\pm,c}=0.052$) indicates the transition from the FM to the QSI$_0$ phases in panel (b). Let us first define the total spinon QFI $f_Q(S^\pm_{\mathbf{q}},T) = f_Q(S^\mathrm{x}_{\mathbf{q}}+S^\mathrm{y}_{\mathbf{q}},T) = f_Q(S^\mathrm{x}_{\mathbf{q}},T) + f_Q(S^\mathrm{y}_{\mathbf{q}},T)$ under the XXZ model. Panels (a, b) and (c, d) show $f_{Q}(S^{\pm}_\Gamma,T)$ and $f_{Q}(S^{\pm}_{\Gamma'},T)$ respectively at different values of $J_{\pm}$. Among which, (a, c) are obtained from ED and (b, d) are from QMC on a $L=4$ lattice ($4\times L^3$ sites). We additionally computed the QFI obtained from GMFT and compared it with QMC simulations at $J_{\pm}=0.045$ for system sizes $L=3$ and $L=4$. We find that, while QMC and ED results are consistent with each other, GMFT results are only off by a factor of approximately $6/7$ compared to QMC and ED at low temperatures (see SM~\cite{suppl}). This is impressive given the mean-field nature of the GMFT. As such, we argue that the ED and GMFT results can be readily extended into the $J_\pm < 0$ regime.

We would like to first highlight that QFI shows non-trivial entanglement depth across various phases. For spin-$1/2$ systems, we find $\text{nQFI}(S^\pm_{\mathbf{q}})=f_Q(S^\pm_{\mathbf{q}},T)/2$ (see SM~\cite{suppl}). As shown in Fig.~\ref{fig:QFI_phase}(a,b), $f_{Q}\left(S^{\pm}_\Gamma,T\right)\gtrsim 3$ in QSI$_0$ and $\gtrsim 5$ in the FM phase, implying at least 2- and 3-partite entanglement, respectively, in sharp contrast to the high-temperature paramagnet (PM), where $f_{Q}\left(S^{\pm}_\Gamma,T\right)=0$. On the other hand, we see that $f_{Q}(S^{\pm}_\Gamma,T)$ of QSI$_\pi$ is $\sim 0$. We hereby stress that QFI is a mere lower bound of entanglement depth --- such results do not reflect that QSI$_\pi$ is trivially entangled nor that the FM phase is more entangled than that of the QSI$_\pi$ phase. In fact, QFI is heavily dependent on the momentum positions. In principle, to aptly categorize the lower bound of the entanglement depth of a certain phase, one should rigorously search through the momentum space to find the largest QFI. One guiding principle for determining said momentum position is to examine the equal-time spin structure factor (ETSF), since the QFI density at $T\rightarrow 0$ becomes $\langle S^{+}_{-\mathbf{q}}S^{-}_{\mathbf{q}}+S^{-}_{-\mathbf{q}}S^{+}_{\mathbf{q}}\rangle/2.
$ We demonstrate this point by presenting $f_{Q}(S^{\pm}_{\Gamma'},T)$ in Fig.~\ref{fig:QFI_phase} (c, d), where $\Gamma'=(4\pi,4\pi,0)$, at different values of $J_{\pm}$. This momentum position corresponds to a local maximum of the ETSF for QSI$_\pi$~\cite{desrochers2023spectroscopic, desrochers2022symmetry} and, consequently, a local maximum for QFI at low temperatures\cite{Tutorial2025Scheie,Witnessing2021Scheie}. Indeed, here $f_{Q}(S^{\pm}_{\Gamma^\prime},T)$ of QSI$_\pi$ is $\gtrsim 3$, whereas for QSI$_0$ and FM they are $\gtrsim 2$ and $\gtrsim 1$ respectively. From this, we can conclude that QSI$_\pi$ is at least 2-partite entangled.

\begin{figure}
	\centering
	\includegraphics[width=\columnwidth]{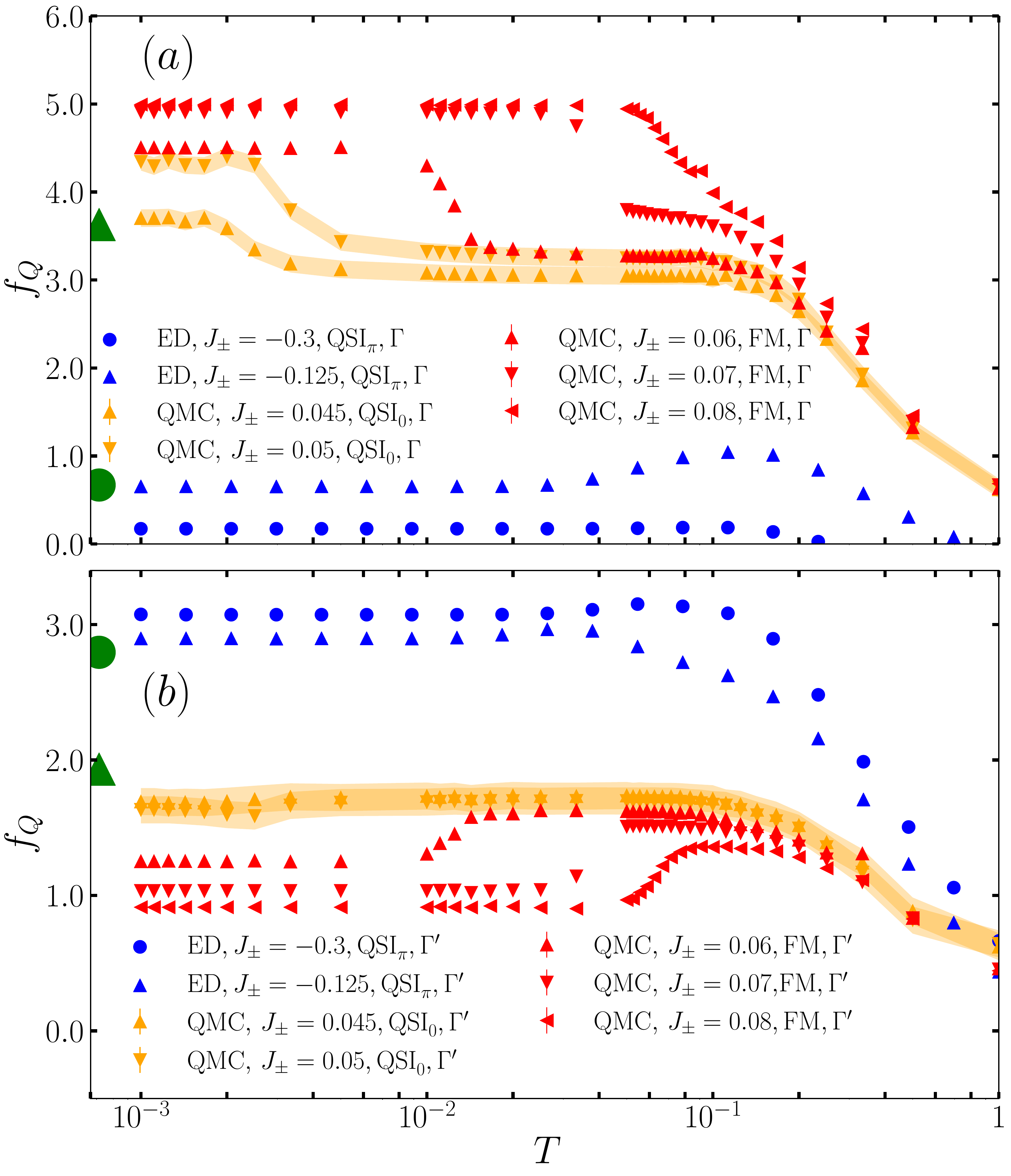}	\caption{\textbf{Temperature evolution of the QFI in different phases obtained with various computational methods.} The QFI $f_{Q}(S^{\pm}_\mathbf{q},T)$ are shown for (a) the $S^{\pm}$ channel at the $\Gamma$ point, (b) the $S^{\pm}$ channel at the $\Gamma^{\prime}$ point. The blue disk points represent results from ED calculations with $J_{\pm}=-0.3$ and $J_{\pm}=-0.125$ in the $\mathrm{QSI}_{\pi}$ regime, while the organge triangle correspond to QMC simulations with $J_{\pm}=0.045$ and $J_{\pm}=0.05$ in the $\mathrm{QSI}_{0}$ regime. The orange shaded areas accompanying the data highlight the crossover from paramagnetic regime to classical spin ice regime at $T\sim 1$ and that from the classical spin ice regime to QSI$_0$ at $T\sim |J^3_{\pm}|$. The red triangles indicate QMC results with $J_{\pm}$ ranging from $0.06$ to $0.08$ in the FM regime. 
	The green circles near the y-axis indicate the GMFT calculation results at $J_{\pm}=-0.3$ (QSI$_\pi$) at zero temperature, while the green triangle corresponds to the result at $J_{\pm}=0.045$ (QSI$_0$).
	The QMC, ED, and GMFT results are consistent (see SM~\cite{suppl} for details).
    }
	\label{fig:QFI_line}
\end{figure}

Beyond providing a lower bound on entanglement depth, the QFI is a channel-resolved measure of quantum coherence, and therefore depends explicitly on the choice of probe operator. For thermal states, the temperature kernel in Eq.~\ref{eq:QFI} effectively suppresses quasi-static contributions with $\omega\ll T$, which are strongly thermally populated and typically correspond to slow, relaxational fluctuations that dominate classical noise in equal-time correlators and static susceptibilities~\cite{hauke2016measuring, laurell2025witnessing,frerot2016quantum}. In contrast, the QFI preferentially weights coherent dynamical fluctuations at $\omega\gtrsim T$ in the specific channel accessed by $S^\alpha_{\mathbf q}$, so changing $\alpha$ and $\mathbf q$ directly tunes which excitations contribute most strongly~\cite{hauke2016measuring,scheie2021witnessing,laurell2025witnessing}. This built-in quantum selectivity and operator tunability make $f_Q$ a sharp diagnostic of the thermal crossovers and phase boundaries. In particular, at finite temperature, $f_{Q}(S^{\pm}_\Gamma,T)$ faithfully tracks the two crossover temperatures of QSI$_0$ regime as shown in Fig.~\ref{fig:QFI_line}(a) for $J_\pm=0.045$ and $0.05$. From previous QMC calculations of the specific heat~\cite{Dynamics2018Huang}, it is known that as one cools down from the high-temperature paramagnetic (PM) phase, the system first crosses over into the classical spin ice (CSI) at $T\sim 1$ before finally entering the QSI$_0$ regime at $T\sim 
|J_\pm^3|$. These crossover temperatures are marked by the two peaks in the specific heat. But these features are equally well represented in the QFI density. In Fig.~\ref{fig:QFI_line} (a), $f_{Q}(S^{\pm}_\Gamma, T)$ first starts increasing from 0 in the PM phase at $T\sim 1$ when we cross over into CSI. After which, $f_{Q}(S^{\pm}_\Gamma, T)$ plateaus until the second temperature scales $T\sim |J^3_{\pm}|$ where $f_{Q}(S^{\pm}_\Gamma, T)$ rises again as we cross over into the QSI$_0$ phase. Finally, $f_{Q}(S^{\pm}_\Gamma, T)$ reaches another plateau, which is consistent with the GMFT calculation (the green triangle in Fig.\ref{fig:QFI_line}(a)) up to a factor of approximately $\sim 6/7$. Similarly, the phase boundary between the FM phase and the high-temperature PM is clearly delineated by the increase in $f_{Q}(S^{\pm}_\Gamma,T)$, as shown in Fig.~\ref{fig:QFI_line}(a). As such, the QFI also appears to be sensitive to this thermal phase transition~\cite{Unusual2008Banerjee}. 

On the other hand, the two temperature crossover scales are also clearly demarcated by QFI for QSI$_\pi$ phase, as shown in Fig.~\ref{fig:QFI_phase}(a,c): both $f_Q(S^\pm_\Gamma,T)$ and $f_Q(S^\pm_{\Gamma'},T)$ increase near $T\sim1$; at $T\sim|J_\pm|^{3}$, $f_Q(S^\pm_\Gamma,T)$ decreases while $f_Q(S^\pm_{\Gamma'},T)$ increases for $J_\pm\lesssim -0.3$. At $J_\pm\sim-0.3$, the two temperature scales coalesce, as shown by the essentially flat QFI below the first temperature scale $1$K in Fig.~\ref{fig:QFI_line}(a,b). The behavior is consistent with Ce$_2$Zr$_2$O$_7$ (best-fit $J_{\pm}\approx -0.3$~\cite{smith2022case, smith2023quantum, smith2025single}), where only one peak is observed in specific heat measurement~\cite{smith2023quantum}. In contrasts, specific heat measurements on Ce$_2$Hf$_2$O$_7$ ($J_{\pm}\approx -0.125$ in the QSI scenario)~\cite{smith2025two, poree2025dipolar, bhardwaj2025thermodynamics, kermarrec2025magnetization} show two well-separated peaks~\cite{smith2025two} mirrored by the predicted two inflection points of QFI, most pronounced in $f_Q(S^\pm_\Gamma,T)$ as shown in Fig.~\ref{fig:QFI_line}(a).


The evolution of QFI also strongly depends on the momentum position. The two aforementioned temperature scales manifest themselves in a strikingly different way when looking at $f_{Q}(S^{\pm}_{\Gamma'},T)$ in Fig.~\ref{fig:QFI_line}(b) as opposed to $f_{Q}(S^{\pm}_{\Gamma},T)$ in Fig.~\ref{fig:QFI_line}(a). As we cool down to the CSI regime from the high-temperature PM at the temperature scale $J_\mathrm{z}$, $f_{Q}(S^{\pm}_{\Gamma'},T)$ rapidly rises and eventually plateaus. But after the second temperature scale is met, $f_{Q}(S^{\pm}_{\Gamma'},T)$ now dips instead of increasing, giving rise to this characteristic hump as shown in Fig.~\ref{fig:QFI_line}(b), which precisely marks the region of large critical fluctuations. This speaks to the versatility of QFI as a potentially powerful experimental probe --- by adjusting the momentum positions, we can design probes that are sensitive to specific features of interest.

\begin{figure}
	\centering
	\includegraphics[width=\columnwidth]{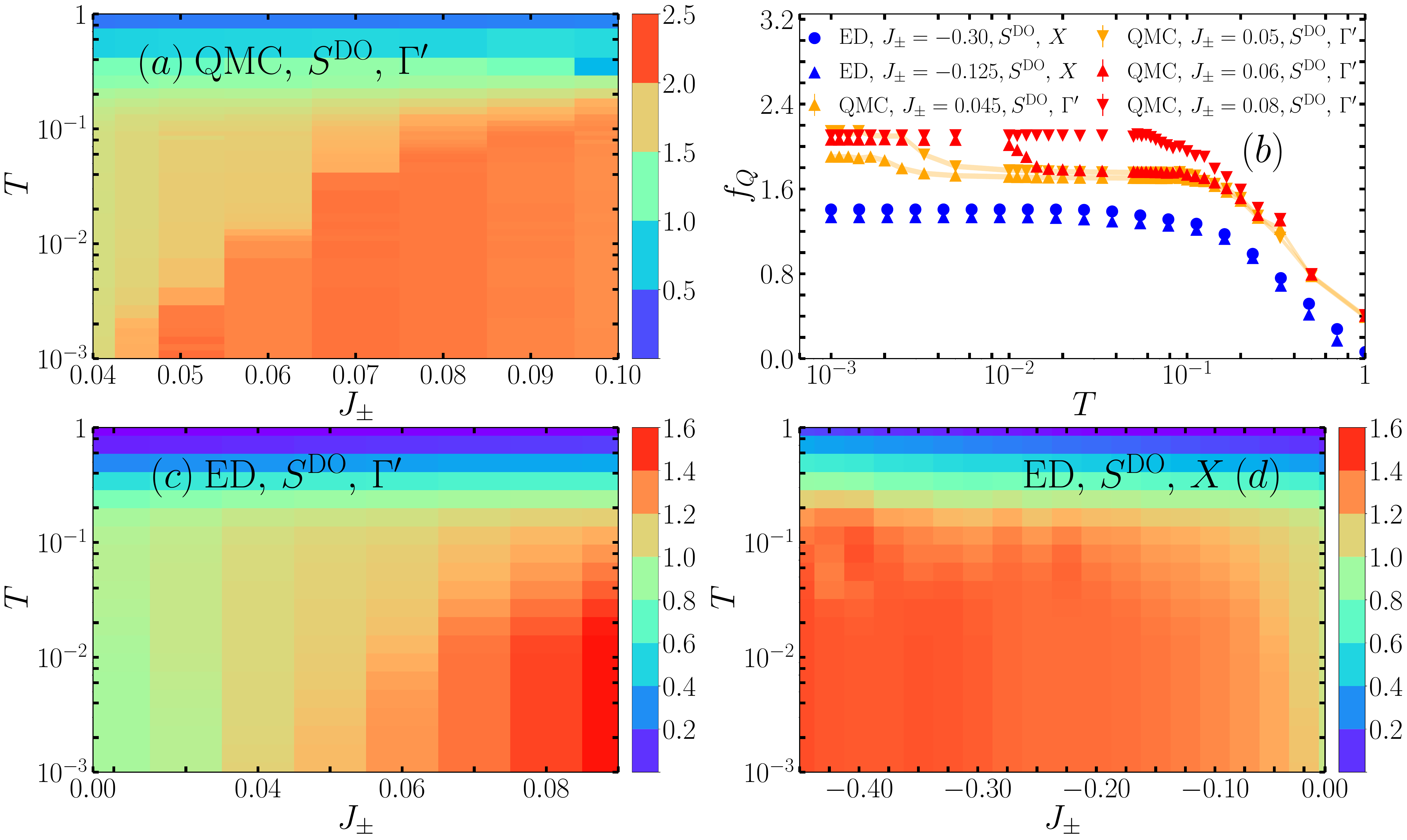}
	\caption{\textbf{QFI in experimental coordinates for Cerium-based pyrochlore compounds.} 
    Panel (a) presents a heat map of the QMC computed QFI $f_{Q}(S^{\text{DO}}_{\Gamma'}, T)$, with $J_{\pm}$ ranging from $0.04$ to $0.10$. Panel (b) is the line cut of panel (a) from $J_{\pm}=0.045$ to $0.08$. The crossover temperature scales from the paramagnetic phase to CSI at $T\sim 1$ and from CSI to QSI$_0$ at $T\sim |J^3_{\pm}|$ are clearly manifest in the shaded data, highlighted in orange. Panel (c) shows ED results of $f_{Q}(S^{\text{DO}}_{\Gamma'}, T)$ for $J_{\pm}$ between $0.01$ and $0.08$. Panel (d) focuses $f_{Q}(S^{\text{DO}}_X, T)$ with $X=(0,0,2\pi)$ in the range from $-0.425$ to $0.00$.}
	\label{fig:QFI_experiment}
\end{figure}

\section{Discussion}\label{sec4}

So far, we have considered the QFI of an abstract XXZ model. The same framework connects directly to real neutron scattering experiments once the microscopic nature of the local moments and their coupling to the neutron’s magnetic moment are specified (see detailed derivation in SM.~\cite{suppl}).
In particular, we present predictions in Fig.~\ref{fig:QFI_experiment} for the Ce-based dipolar–octupolar pyrochlore Ce$_2$Zr$_2$O$_7$\cite{smith2023quantum,smith2022case,smith2024experimental}—a leading QSI candidate. While Ce ions carry both dipolar and octupolar moments, neutron scattering primarily couples to the dipolar moments, which correspond to transverse spin components $S^\pm$ in the current model \cite{huang2014quantum, rau2019frustrated}. Fig.~\ref{fig:QFI_experiment} shows the corresponding QFI $f(S^{\text{DO}}_\mathbf{q}, T)$ that takes into account the transverse momentum projector in the inelastic neutron scattering.

In Fig.~\ref{fig:QFI_experiment}, panels (a,b) are obtained from QMC simulation for $L=4$ and panels (c,d) are the ED results. One sees that in the global frame, the QFI again vividly captures the thermodynamic phase structure. The crossover temperature scale from paramagnetic region at high temperature $(T>J_\mathrm{z})$ to the CSI and its plateau at intermediate temperature $(|J_{\pm}^3|<T<J_\mathrm{z})$ and eventually the QSI$_0$ at low temperature $(T<|J^3_{\pm}|)$ manifest in the both the ED and QMC data. The evolution of QFI closely resembles that of the thermal entropy measurements~\cite{Dynamics2018Huang}.

One important point is that, in principle, the bounds on $f_{Q}(S^{\mathrm{DO}}_{\mathbf{q}},T)$ for an $m$-producible state are momentum dependent because of the transverse projector. But we can establish a $\mathbf q$-independent bound as derived in SM~\cite{suppl}, giving us the following relationship: $\text{nQFI}(S^{\text{DO}}_{\mathbf{q}}) = \frac{3}{2}f_Q(S^\text{DO}_{\mathbf{q}})$. As a result, values extracted directly from our computed data in a neutron scattering setting already certify nontrivial entanglement depth: $\text{nQFI}(S^{\mathrm{DO}}_{\Gamma'},T)\gtrsim2.2$ for 0-flux QSI and $\text{nQFI}(S^{\mathrm{DO}}_{\Gamma'},T)\gtrsim3.7$ for all-in-all-out order (FM in the local spin frame) respectively. 
At the $X=(0,0,2\pi)$ point, $\text{nQFI}(S^{\mathrm{DO}}_{X},T)\gtrsim2.3$ certifies that the $\pi$-flux QSI is at least 3-partite entangled. Moreover, in principle, polarized neutron measurements with the non-spin-flip (NSF) and spin-flip (SF) channels are the optimal probes to find the best entanglement depth bound. As shown in SM~\cite{suppl}, $\text{nQFI}(S^{\mathrm{(N)SF}}_{\mathbf{q}}) = 3 f_Q(S^{\mathrm{(N)SF}}_{\mathbf{q}})$, therefore, even a modest $f_Q(S^{\mathrm{(N)SF}}_{\mathbf{q}})$ can signal deep multipartite entanglement.

Based on the results and analysis above, we believe the QFI can actually offer richer information than initially anticipated. Besides providing insightful information about the entanglement depth, it can be used as both a thermal and dynamical probe in the investigation of quantum magnets, with quantum spin ice as an archetypal manifestation. Based on spin spectra in both local and global frames, the temperature and momentum dependence of QFI can faithfully trace out the ground state and finite-temperature phase boundaries in the QSI system, consistent with previous theoretical and numerical knowledge. These properties of QFI can bridge the fundamental entanglement concept to practical model computations and applications to experiments.
In this way, other exotic quantum critical points, QSL phases, and their associated theory, computation, and experimental verification, could be expected to unite from the entanglement perspective.

\noindent{\bf Methods}

The numerical results for the pyrochlore lattice XXZ model are obtained from three methods: quantum Monte Carlo (QMC) simulations with a multi-directed loop (MDL) update scheme, exact diagonalization (ED), and gauge mean-field theory (GMFT) calculations. In addition, we transfer our observations to the neutron scattering section that connects directly to the real experiments. Here, we briefly introduce all three numerical methods and this transformation. The details of these methods are provided in the Supplemental Material~\cite{suppl}.

{\bf Quantum Monte Carlo simulations with the multi-directed loop update}
The QMC simulations are performed based on the stochastic series expansion (SSE) framework~\cite{QMC2002Sandvik,Generalized2005Wessel,Computational2010Sandvik}, where the configuration space is constructed by expanding the partition function, $Z=\text{Tr}(e^{-\beta H})$, as a Taylor series into an operator string. We apply the diagonal update, directed loop (DL) update, and multi-directed loop (MDL) update schemes in our simulation. Our simulations are performed on system sizes $L=3$ and $4$. The MDL update is specifically designed to efficiently sample the highly frustrated (3+1)d configurational space of the pyrochlore lattice XXZ model, where the simplest quantum fluctuation that connects different classical spin-ice states involves six spin flips around a hexagon. This nature limits the sampling efficiency of the traditional DL update. To overcome this limitation, the MDL update allows the insertion of multiple operator pairs, thereby achieving efficient sampling and making the QSI$_0$ phase accessible (see SM~\cite{suppl} for details). In this update method, we first randomly select a hexagon in the configurational space and insert three pairs of operators, $S_i^+$ and $S_i^-$, on three alternate sites of the hexagon. Then, for each pair of operators, we randomly choose one operator to be the head of the loop, while the other operator becomes the tail. These three heads then propagate through the operator string until all of these heads meet their tails again at the same propagation step, at which point all loops are closed and the update process is completed.

In addition, we measure the imaginary time correlation function by tracing the evolution of the configuration space in the QMC simulation~\cite{Amplitude2021zhou}. We then utilize the stochastic analytic continuation (SAC) scheme~\cite{Stochastic1998sandvik,Identifying2004beach,Using2008syliuaasen,Constrained2016sandvik,Progress2023shao} to convert the imaginary time correlation function to a real-frequency dynamic structure factor.

{\bf Exact diagonalization}
The exact diagonalization scheme follows that of the microcanonical thermal pure quantum method (mTPQ)~\cite{thermal2012sugiura, canonical2013sugiura}. 
The spectral function is obtained from an mTPQ representative state $|\psi_k\rangle$ using the broadened resolvent of a chosen operator  $\mathcal{O}$~\cite{jaklic1994ftlm, prelovsek2013lanczoschapter},
\begin{equation}
    G_{\mathcal{O}}(z;\beta_k)
    =
    \frac{\langle \psi_k|\mathcal{O}^\dagger\,(z-\widetilde{H})^{-1}\mathcal{O}|\psi_k\rangle}
         {\langle \psi_k|\psi_k\rangle},
    \label{eq:resolvent_def}
\end{equation}
with $\widetilde{H}\equiv H-E$ and $z=\omega+i\eta$. Here $\eta>0$ controls the Lorentzian broadening, and the shift $E$ fixes the absolute frequency reference. We choose $E$ as the ground-state energy, obtained independently from a separate Lanczos run, so that the spectrum is reported with the correct energy offset. The associated spectral density follows as
\begin{equation}
    I_{\mathcal{O}}(\omega;\beta_k) = -\frac{1}{\pi}\,\mathrm{Im}\,G_{\mathcal{O}}(\omega+i\eta;\beta_k).
    \label{eq:spectral_def}
\end{equation}
Specifically, we perform the ED calculations on a 16-site periodic lattice, where four tetrahedra construct a large tetrahedron. Such a geometry respects the full symmetry of the pyrochlore lattice.

{\bf Gauge mean-field theory calculations}
The gauge mean-field theory (GMFT) calculations are performed to calculate the dynamical structure factor and QFI by canonically mapping the spin Hamiltonian to that of a lattice $U(1)$ gauge theory coupled to bosonic matter fields. We note that the results of GMFT align well with the QMC results with a factor of approximately $6/7$ (see SM~\cite{suppl} for details). 

{\bf Neutron scattering cross section}
The neutron scattering cross-section is proportional to
\begin{equation}
    \begin{aligned}
        A^{\text{DO}}(\mathbf{q},\omega) =& \sum_{\mu,\nu} \left(\hat{\mathbf{z}}_\mu \cdot \hat{\mathbf{z}}_\nu-\frac{\left(\hat{\mathbf{z}}_\mu \cdot \mathbf{q}\right)\left( \hat{\mathbf{z}}_\nu \cdot \mathbf{q}\right)}{|\mathbf{q}|^2}\right) A^{\pm}_{\mu,\nu}(\mathbf{q},\omega), \label{eq:method_ADO}
    \end{aligned}
\end{equation}
where $\mu$ and $\nu$ are the four sublattice indices of the pyrochlore lattice, and $A^{\pm}_{\mu,\nu}(\mathbf{Q},\omega)$ is the spectral function obtained in the local frame. $\hat{\mathbf{z}}_\mu$ is the local $z$-axis direction of sublattice $\mu$,
which are given by
\begin{equation}
\begin{aligned}
    \mathbf{z}_0 &= \frac{1}{\sqrt{3}}(1,1,1) &
    \mathbf{z}_1 &= \frac{1}{\sqrt{3}}(1,-1,-1) \\
    \mathbf{z}_2 &= \frac{1}{\sqrt{3}}(-1,1,-1) &
    \mathbf{z}_3 &= \frac{1}{\sqrt{3}}(-1,-1,1)
\end{aligned}
\end{equation}

{\bf Quantum Fisher information}
A meaningful QFI bound is defined with respect to a particular collective observable $\mathcal{O}=\sum_i\mathcal{O}_i$, such that $f_Q(\mathcal{O}) \leq m(\Delta\lambda)^2$ when evaluated over an $m$-producible state and $\Delta\lambda$ is the spectral width of the individual observables $\mathcal{O}_i$. However, in the main text, we use $S_\mathbf{q}^\alpha=\sum_i S_i^\alpha e^{i\mathbf{q}\cdot\mathbf{R}_i}$, which is not always Hermitian. Nevertheless, with the definition in Eq.~\eqref{eq:QFI}, we now show how to extract the essential properties of the QFI $f_Q$, such as the entanglement lower bound. A detailed derivation is provided in the Supplemental Material~\cite{suppl}.


For a given operator $\mathcal{O}_\mathbf{q}=\sum_i\mathcal{O}_ie^{i\mathbf{q}\mathbf{R}_i}$, we use the Hermitian $\mathcal{O}_{c,\mathbf{q}} =(\mathcal{O}_\mathbf{q}+\mathcal{O}_\mathbf{q}^\dagger)/2$ and the antihermitian $\,\mathcal{O}_{s,\mathbf{q}} =(\mathcal{O}_\mathbf{q}-\mathcal{O}_\mathbf{q}^\dagger)/2i$ parts to construct a QFI matrix (QFIM) $F$~\cite{liu2019quantum,fiderer2021general,dominik2018simple}, whose elements are specified by:
\begin{equation}
    F_{ab} (T):=4 \int d \omega \tanh \left(\frac{\omega}{2T}\right)\left(1-e^{- \omega/T}\right) \mathcal{A}_{ab}(\mathbf{q}, \omega),
    \label{eq:QFIM}
\end{equation}
where
\begin{equation}
    \mathcal{A}_{ab}(\mathbf{q}, \omega) = \frac{1}{2\pi N}\int dt e^{i\omega t}\langle \mathcal{O}_{a,\mathbf{q}}(t) \mathcal{O}_{b,\mathbf{q}}(0)\rangle,
\end{equation}
with $a,b\in\{c,s\}$. $f_Q(\mathcal{O}_\mathbf{q})$ defined in Eq.~\eqref{eq:QFI} is equivalent to $\operatorname{Tr}(F)$ (see SM~\cite{suppl} for more details). Therefore, since both diagonal elements are bounded by Eq.~\eqref{eq:QFI_Bound}, for an $m$-producible state, $f_Q(\mathcal{O}_\mathbf{q})=f_Q(\mathcal{O}_{c,\mathbf{q}})+f_Q(\mathcal{O}_{s,\mathbf{q}})$ is simply bounded by the sum of the individual QFI bounds. 
After detailed derivation, we find that:
\begin{equation}
f_Q(\mathcal{O}_\mathbf{q},T)\leq m \left(\Delta(\mathcal{O}_{i})\right)^2, \label{eq:method_generalized_bound}
\end{equation}
where $\Delta(\mathcal{O}_{i})$ is the spectral width of the operator $\mathcal{O}_{i}$. When $\mathcal{O}_\mathbf{q}=S^\alpha_\mathbf{q}$ ($\alpha \in\{\mathrm{x},\mathrm{y},\mathrm{z}\}$) and $S^\pm_\mathbf{q}$, the corresponding entanglement depth lower bound condition is that when
\begin{equation}
f_Q(S^\alpha_\mathbf{q},T) > m, \qquad f_Q(S^\pm_\mathbf{q},T) > 2m,
\end{equation}
the state is at least $(m+1)$-partite entangled.

Furthermore, we show how to extract essential properties of the QFI for the neutron scattering cross section via Eq.~\eqref{eq:method_ADO}. This is noteworthy because there does not exist any collective operator, $S^{\text{DO}}$, such that its two-point correlation function, $\langle S^{\text{DO}\dagger}S^{\text{DO}}\rangle$, is equivalent to $A^{\text{DO}}$. Nevertheless, we can rewrite it as a sum of the QFI induced by the non-spin-flip (NSF) channel with underlying operator $S_\mathbf{q}^{\text{NSF}} = \sum_{\mathbf{R}_\mu} \left(\hat{\mathbf{p}}\cdot\hat{\mathbf{z}}_\mu \right)\tau^z_{\mathbf{R}_\mu}e^{i\mathbf{q}\cdot\mathbf{R}_\mu}$ and the spin-flip (SF) channel with operator $ S_\mathbf{q}^{\text{SF}} = \sum_{\mathbf{R}_\mu} \left(\hat{\mathbf{v}}\cdot\hat{\mathbf{z}}_\mu \right)\tau^z_{\mathbf{R}_\mu}e^{i\mathbf{q}\cdot\mathbf{R}_\mu}$, where $\mathbf{p}\perp\mathbf{q}$ defines a neutron polarization vector and $\mathbf{v}=\mathbf{p}\times\mathbf{q}$ defines the other transverse direction. 
Therefore, $f_Q(S^{\text{DO}}_\mathbf{q}) = f_Q(S^{\text{NSF}}_\mathbf{q}) +f_Q(S^{\text{SF}}_\mathbf{q})$, since $A^{\text{DO}}(\mathbf{q})=\langle S^{\text{NSF}^\dagger}_\mathbf{q} S^{\text{NSF}}_\mathbf{q}\rangle + \langle S^{\text{SF}^\dagger}_\mathbf{q} S^{\text{SF}}_\mathbf{q}\rangle$ by construction. This defines $f_Q(S^{\text{DO}})$ as another sum witness whose individual components are well-bounded via Eq.~\eqref{eq:method_generalized_bound}. 
Thus, $f_Q(S^{\text{DO}}_\mathbf{q})$ is bounded by the sum of the underlying NSF and SF channels. Applying Eq.~\eqref{eq:method_generalized_bound} to $S^{\text{(N)SF}}_\mathbf{q}$, we arrive at 
\begin{equation}
f_Q(S^{\text{(N)SF}}_\mathbf{q}, T) > m/3, \qquad f_Q(S^{\text{DO}}_\mathbf{q},T) > 2m/3,
\end{equation}
which implies that the state is at least $(m+1)$-partite entangled. See SM~\cite{suppl} for a detailed derivation.

\noindent\textbf{Data availability}
The data that support the findings of this study are provided at \href{https://doi.org/10.25442/hku.32404548}{https://doi.org/10.25442/hku.32404548.}

\noindent\textbf{Code availability}
The main codes in this paper are available at \href{https://doi.org/10.25442/hku.32412273}{https://doi.org/10.25442/hku.32412273.}

\noindent\textbf{Acknowledgements}
We acknowledge inspiring discussion with Bruce Gaulin. CKZ and ZYM thank the discussions with Menghan Song and Ting-Tung Wang on entanglement-related topics And we thank HPC2021 system under the Information Technology Services at the University of Hong Kong, as well as the Beijing Paratera Tech Corp., Ltd~\cite{paratera} for providing HPC resources that have contributed to the research results reported within this paper. 

\noindent\textbf{Funding}
Z.Y.M. acknowledge the support from the Research Grants Council (RGC) of
Hong Kong (Project Nos. 17309822, C7037-22GF, 17302223, 17301924), the ANR/RGC Joint Research Scheme sponsored by RGC of Hong Kong and French National Research Agency (Project No. A\_HKU703/22). Z.Z., F.D., and Y.B.K. were supported by the Natural Sciences and Engineering Research Council of Canada (NSERC) Grant No. RGPIN-2023-03296 and the Centre of Quantum Materials at the University of Toronto. Computations at the University of Toronto were performed on the Cedar and Fir clusters, which are hosted by the Digital Research Alliance of Canada. F.D. is further supported by the Vanier Canada Graduate Scholarship (CGV-186886). 

\noindent\textbf{Authors contributions}
Z.Y.M. and Y.B.K. supervised the project. C.K.Z. developed the multi-directed loop update scheme and performed the QMC simulations. Z.Z. and F.D. performed the ED calculations and the GMFT calculations. All authors contributed to writing the manuscript.

\noindent\textbf{Competing interests}
The authors declare no competing interests.

\bibliography{ref}

@article{Can2025shimokawa,
  title={Can experimentally-accessible measures of entanglement distinguish quantum spin liquids from disorder-driven ``random singlet" phases?},
  author={Shimokawa, Tokuro and Sabharwal, Snigdh and Shannon, Nic},
  journal={arXiv:2505.11874},
  year={2025},
  url={https://arxiv.org/abs/2505.11874v1}
}

@article{Neutron2025Gao,
author = {Gao, Bin and Desrochers, F{\'e}lix and Tam, David W. and Kirschbaum, Diana M. and Steffens, Paul and Hiess, Arno and Nguyen, Duy Ha and Su, Yixi and Cheong, Sang-Wook and Paschen, Silke and Kim, Yong Baek and Dai, Pengcheng},
year = {2025},
title = {Neutron scattering and thermodynamic evidence for emergent photons and fractionalization in a pyrochlore spin ice},
journal = {Nature Physics},
pages = {1203 - 1210},
volume = {21},
url = {https://doi.org/10.1038/s41567-025-02922-9},
doi = {10.1038/s41567-025-02922-9}
}

@article{Witnessing2021Scheie,
  title = {Witnessing entanglement in quantum magnets using neutron scattering},
  author = {Scheie, A. and Laurell, Pontus and Samarakoon, A. M. and Lake, B. and Nagler, S. E. and Granroth, G. E. and Okamoto, S. and Alvarez, G. and Tennant, D. A.},
  journal = {Phys. Rev. B},
  volume = {103},
  issue = {22},
  pages = {224434},
  numpages = {16},
  year = {2021},
  month = {Jun},
  publisher = {American Physical Society},
  doi = {10.1103/PhysRevB.103.224434},
  url = {https://link.aps.org/doi/10.1103/PhysRevB.103.224434}
}

@article{Proximate2024scheie,
  title={Proximate spin liquid and fractionalization in the triangular antiferromagnet KYbSe2},
  author={Scheie, AO and Ghioldi, EA and Xing, J and Paddison, JAM and Sherman, NE and Dupont, M and Sanjeewa, LD and Lee, Sangyun and Woods, AJ and Abernathy, D and others},
  journal={Nat. Phys.},
  volume={20},
  number={1},
  pages={74--81},
  year={2024},
  publisher={Nature Publishing Group UK London},
  url={https://www.nature.com/articles/s41567-023-02259-1}
}

@article{zhaoMeasuring2022,
author = {Zhao, Jiarui and Chen, Bin-Bin and Wang, Yan-Cheng and Yan, Zheng and Cheng, Meng and Meng, Zi Yang},
year = {2022},
title = {Measuring R{\'e}nyi entanglement entropy with high efficiency and precision in quantum Monte Carlo simulations},
journal = {npj Quantum Materials},
pages = {69},
volume = {7},
url = {https://doi.org/10.1038/s41535-022-00476-0},
doi = {10.1038/s41535-022-00476-0}
}

@article{chenTopological2022,
  title = {Topological disorder parameter: A many-body invariant to characterize gapped quantum phases},
  author = {Chen, Bin-Bin and Tu, Hong-Hao and Meng, Zi Yang and Cheng, Meng},
  journal = {Phys. Rev. B},
  volume = {106},
  issue = {9},
  pages = {094415},
  numpages = {18},
  year = {2022},
  month = {Sep},
  publisher = {American Physical Society},
  doi = {10.1103/PhysRevB.106.094415},
  url = {https://link.aps.org/doi/10.1103/PhysRevB.106.094415}
}

@article{Extracting2024Liao,
  title = {Extracting universal corner entanglement entropy during the quantum Monte Carlo simulation},
  author = {Da Liao, Yuan and Song, Menghan and Zhao, Jiarui and Meng, Zi Yang},
  journal = {Phys. Rev. B},
  volume = {110},
  issue = {23},
  pages = {235111},
  numpages = {9},
  year = {2024},
  month = {Dec},
  publisher = {American Physical Society},
  doi = {10.1103/PhysRevB.110.235111},
  url = {https://link.aps.org/doi/10.1103/PhysRevB.110.235111}
}

@article{QSL2018Wang,
  title = {Quantum Spin Liquid with Even Ising Gauge Field Structure on Kagome Lattice},
  author = {Wang, Yan-Cheng and Zhang, Xue-Feng and Pollmann, Frank and Cheng, Meng and Meng, Zi Yang},
  journal = {Phys. Rev. Lett.},
  volume = {121},
  issue = {5},
  pages = {057202},
  numpages = {6},
  year = {2018},
  month = {Aug},
  publisher = {American Physical Society},
  doi = {10.1103/PhysRevLett.121.057202},
  url = {https://link.aps.org/doi/10.1103/PhysRevLett.121.057202}
}

@article{Dynamical2018Sun,
  title = {Dynamical Signature of Symmetry Fractionalization in Frustrated Magnets},
  author = {Sun, Guang-Yu and Wang, Yan-Cheng and Fang, Chen and Qi, Yang and Cheng, Meng and Meng, Zi Yang},
  journal = {Phys. Rev. Lett.},
  volume = {121},
  issue = {7},
  pages = {077201},
  numpages = {6},
  year = {2018},
  month = {Aug},
  publisher = {American Physical Society},
  doi = {10.1103/PhysRevLett.121.077201},
  url = {https://link.aps.org/doi/10.1103/PhysRevLett.121.077201}
}

@article{Markov2010Suwa,
  title = {Markov Chain Monte Carlo Method without Detailed Balance},
  author = {Suwa, Hidemaro and Todo, Synge},
  journal = {Phys. Rev. Lett.},
  volume = {105},
  issue = {12},
  pages = {120603},
  numpages = {4},
  year = {2010},
  month = {Sep},
  publisher = {American Physical Society},
  doi = {10.1103/PhysRevLett.105.120603},
  url = {https://link.aps.org/doi/10.1103/PhysRevLett.105.120603}
}

@INPROCEEDINGS{Computational2010Sandvik,
  author = {Sandvik, Anders W},
  title = {Computational studies of quantum spin systems},
  booktitle = {AIP Conference Proceedings},
  year = {2010},
  volume = {1297},
  number = {1},
  pages = {135--338},
  organization = {AIP},
  url = {https://aip.scitation.org/doi/abs/10.1063/1.3518900}
}

@article{QMC2002Sandvik,
  title = {Quantum Monte Carlo with directed loops},
  author = {Sylju\aa{}sen, Olav F. and Sandvik, Anders W.},
  journal = {Phys. Rev. E},
  volume = {66},
  issue = {4},
  pages = {046701},
  numpages = {28},
  year = {2002},
  month = {Oct},
  publisher = {American Physical Society},
  doi = {10.1103/PhysRevE.66.046701},
  url = {https://link.aps.org/doi/10.1103/PhysRevE.66.046701}
}

@article{Generalized2005Wessel,
  title = {Generalized directed loop method for quantum Monte Carlo simulations},
  author = {Alet, Fabien and Wessel, Stefan and Troyer, Matthias},
  journal = {Phys. Rev. E},
  volume = {71},
  issue = {3},
  pages = {036706},
  numpages = {16},
  year = {2005},
  month = {Mar},
  publisher = {American Physical Society},
  doi = {10.1103/PhysRevE.71.036706},
  url = {https://link.aps.org/doi/10.1103/PhysRevE.71.036706}
}

@article{Pyrochlore2004Hermele,
  title = {Pyrochlore photons: The $U(1)$ spin liquid in a $S=\frac{1}{2}$ three-dimensional frustrated magnet},
  author = {Hermele, Michael and Fisher, Matthew P. A. and Balents, Leon},
  journal = {Phys. Rev. B},
  volume = {69},
  issue = {6},
  pages = {064404},
  numpages = {21},
  year = {2004},
  month = {Feb},
  publisher = {American Physical Society},
  doi = {10.1103/PhysRevB.69.064404},
  url = {https://link.aps.org/doi/10.1103/PhysRevB.69.064404}
}

@article{Dynamics2018Huang,
  title = {Dynamics of Topological Excitations in a Model Quantum Spin Ice},
  author = {Huang, Chun-Jiong and Deng, Youjin and Wan, Yuan and Meng, Zi Yang},
  journal = {Phys. Rev. Lett.},
  volume = {120},
  issue = {16},
  pages = {167202},
  numpages = {6},
  year = {2018},
  month = {Apr},
  publisher = {American Physical Society},
  doi = {10.1103/PhysRevLett.120.167202},
  url = {https://link.aps.org/doi/10.1103/PhysRevLett.120.167202}
}

@article{Coulomb2015Lv,
  title = {Coulomb Liquid Phases of Bosonic Cluster Mott Insulators on a Pyrochlore Lattice},
  author = {Lv, Jian-Ping and Chen, Gang and Deng, Youjin and Meng, Zi Yang},
  journal = {Phys. Rev. Lett.},
  volume = {115},
  issue = {3},
  pages = {037202},
  numpages = {5},
  year = {2015},
  month = {Jul},
  publisher = {American Physical Society},
  doi = {10.1103/PhysRevLett.115.037202},
  url = {https://link.aps.org/doi/10.1103/PhysRevLett.115.037202}
}

@article{Witnessing2025Sabharwal,
  title = {Witnessing disorder in quantum magnets},
  author = {Sabharwal, Snigdh and Shimokawa, Tokuro and Shannon, Nic},
  journal = {Phys. Rev. Res.},
  volume = {7},
  issue = {2},
  pages = {023271},
  numpages = {16},
  year = {2025},
  month = {Jun},
  publisher = {American Physical Society},
  doi = {10.1103/95fl-rxl3},
  url = {https://link.aps.org/doi/10.1103/95fl-rxl3}
}

@article{Fisher2012Hyllus,
  title = {Fisher information and multiparticle entanglement},
  author = {Hyllus, Philipp and Laskowski, Wies\l{}aw and Krischek, Roland and Schwemmer, Christian and Wieczorek, Witlef and Weinfurter, Harald and Pezz\'e, Luca and Smerzi, Augusto},
  journal = {Phys. Rev. A},
  volume = {85},
  issue = {2},
  pages = {022321},
  numpages = {10},
  year = {2012},
  month = {Feb},
  publisher = {American Physical Society},
  doi = {10.1103/PhysRevA.85.022321},
  url = {https://link.aps.org/doi/10.1103/PhysRevA.85.022321}
}

@article{Tutorial2025Scheie,
   title={Tutorial: Extracting entanglement signatures from neutron spectroscopy},
   volume={5},
   ISSN={2950-2578},
   url={http://dx.doi.org/10.1016/j.mtquan.2024.100020},
   DOI={10.1016/j.mtquan.2024.100020},
   journal={Materials Today Quantum},
   publisher={Elsevier BV},
   author={Scheie, Allen and Laurell, Pontus and Simeth, Wolfgang and Dagotto, Elbio and Tennant, D. Alan},
   year={2025},
   month=mar, pages={100020} }

@article{Unusual2008Banerjee,
  title = {Unusual Liquid State of Hard-Core Bosons on the Pyrochlore Lattice},
  author = {Banerjee, Argha and Isakov, Sergei V. and Damle, Kedar and Kim, Yong Baek},
  journal = {Phys. Rev. Lett.},
  volume = {100},
  issue = {4},
  pages = {047208},
  numpages = {4},
  year = {2008},
  month = {Jan},
  publisher = {American Physical Society},
  doi = {10.1103/PhysRevLett.100.047208},
  url = {https://link.aps.org/doi/10.1103/PhysRevLett.100.047208}
}

@article{Numerical2015Kato,
  title = {Numerical Evidence of Quantum Melting of Spin Ice: Quantum-to-Classical Crossover},
  author = {Kato, Yasuyuki and Onoda, Shigeki},
  journal = {Phys. Rev. Lett.},
  volume = {115},
  issue = {7},
  pages = {077202},
  numpages = {5},
  year = {2015},
  month = {Aug},
  publisher = {American Physical Society},
  doi = {10.1103/PhysRevLett.115.077202},
  url = {https://link.aps.org/doi/10.1103/PhysRevLett.115.077202}
}

@article{Stochastic2003Sandvik,
  title = {Stochastic series expansion method for quantum Ising models with arbitrary interactions},
  author = {Sandvik, Anders W.},
  journal = {Phys. Rev. E},
  volume = {68},
  issue = {5},
  pages = {056701},
  numpages = {9},
  year = {2003},
  month = {Nov},
  publisher = {American Physical Society},
  doi = {10.1103/PhysRevE.68.056701},
  url = {https://link.aps.org/doi/10.1103/PhysRevE.68.056701}
}

@article{canonical2013sugiura,
  title = {Canonical Thermal Pure Quantum State},
  author = {Sugiura, Sho and Shimizu, Akira},
  journal = {Phys. Rev. Lett.},
  volume = {111},
  issue = {1},
  pages = {010401},
  numpages = {5},
  year = {2013},
  month = {Jul},
  publisher = {American Physical Society},
  doi = {10.1103/PhysRevLett.111.010401},
  url = {https://link.aps.org/doi/10.1103/PhysRevLett.111.010401}
  }

@article{thermal2012sugiura,
  title = {Thermal Pure Quantum States at Finite Temperature},
  author = {Sugiura, Sho and Shimizu, Akira},
  journal = {Phys. Rev. Lett.},
  volume = {108},
  issue = {24},
  pages = {240401},
  numpages = {4},
  year = {2012},
  month = {Jun},
  publisher = {American Physical Society},
  doi = {10.1103/PhysRevLett.108.240401},
  url = {https://link.aps.org/doi/10.1103/PhysRevLett.108.240401}
}

@misc{suppl,
  note = {The Supplemental Material details our multi-directed-loop QMC algorithm for efficient sampling of the QSI model, the exact diagonalization and the gauge field theory calculation, and the computation of neutron scattering cross section in experimental setting.}
}

@ARTICLE{paratera,
journal={Beijing PARATERA
Tech CO.,Ltd},
url = {https://cloud.paratera.com}
}

@article{Stochastic1998sandvik,
  title = {Stochastic method for analytic continuation of quantum Monte Carlo data},
  author = {Sandvik, Anders W.},
  journal = {Phys. Rev. B},
  volume = {57},
  issue = {17},
  pages = {10287--10290},
  numpages = {0},
  year = {1998},
  month = {May},
  publisher = {American Physical Society},
  doi = {10.1103/PhysRevB.57.10287},
  url = {https://link.aps.org/doi/10.1103/PhysRevB.57.10287}
}

@ARTICLE{Identifying2004beach,
       author = {{Beach}, K.~S.~D.},
        title = "{Identifying the maximum entropy method as a special limit of stochastic analytic continuation}",
      journal = {arXiv e-prints},
         year = 2004,
        month = mar,
          eid = {cond-mat/0403055},
        pages = {cond-mat/0403055},
          doi = {10.48550/arXiv.cond-mat/0403055},
archivePrefix = {arXiv},
       eprint = {cond-mat/0403055},
 primaryClass = {cond-mat.str-el},
       adsurl = {https://ui.adsabs.harvard.edu/abs/2004cond.mat..3055B}
}

@article{Constrained2016sandvik,
  title = {Constrained sampling method for analytic continuation},
  author = {Sandvik, Anders W.},
  journal = {Phys. Rev. E},
  volume = {94},
  issue = {6},
  pages = {063308},
  numpages = {5},
  year = {2016},
  month = {Dec},
  publisher = {American Physical Society},
  doi = {10.1103/PhysRevE.94.063308},
  url = {https://link.aps.org/doi/10.1103/PhysRevE.94.063308}
}

@article{Using2008syliuaasen,
  title = {Using the average spectrum method to extract dynamics from quantum Monte Carlo simulations},
  author = {Sylju\aa{}sen, Olav F.},
  journal = {Phys. Rev. B},
  volume = {78},
  issue = {17},
  pages = {174429},
  numpages = {10},
  year = {2008},
  month = {Nov},
  publisher = {American Physical Society},
  doi = {10.1103/PhysRevB.78.174429},
  url = {https://link.aps.org/doi/10.1103/PhysRevB.78.174429}
}

@ARTICLE{Emergent2025chen,
       author = {{Chen}, Chuang and {Seifert}, Urban F.~P. and {Feng}, Kexin and {Starykh}, Oleg A. and {Balents}, Leon and {Meng}, Zi Yang},
        title = "{Emergent gauge flux in {QED}$_3$ with flavor chemical potential: application to magnetized U(1) Dirac spin liquids}",
      journal = {arXiv e-prints},
         year = 2025,
        month = aug,
          eid = {arXiv:2508.08528},
        pages = {arXiv:2508.08528},
          doi = {10.48550/arXiv.2508.08528},
archivePrefix = {arXiv},
       eprint = {2508.08528},
 primaryClass = {cond-mat.str-el},
       adsurl = {https://ui.adsabs.harvard.edu/abs/2025arXiv250808528C}
}

@article{Nearly2017shao,
  title = {Nearly Deconfined Spinon Excitations in the Square-Lattice Spin-$1/2$ Heisenberg Antiferromagnet},
  author = {Shao, Hui and Qin, Yan Qi and Capponi, Sylvain and Chesi, Stefano and Meng, Zi Yang and Sandvik, Anders W.},
  journal = {Phys. Rev. X},
  volume = {7},
  issue = {4},
  pages = {041072},
  numpages = {26},
  year = {2017},
  month = {Dec},
  publisher = {American Physical Society},
  doi = {10.1103/PhysRevX.7.041072},
  url = {https://link.aps.org/doi/10.1103/PhysRevX.7.041072}
}

@article{Amplitude2021zhou,
  title = {Amplitude Mode in Quantum Magnets via Dimensional Crossover},
  author = {Zhou, Chengkang and Yan, Zheng and Wu, Han-Qing and Sun, Kai and Starykh, Oleg A. and Meng, Zi Yang},
  journal = {Phys. Rev. Lett.},
  volume = {126},
  issue = {22},
  pages = {227201},
  numpages = {6},
  year = {2021},
  month = {Jun},
  publisher = {American Physical Society},
  doi = {10.1103/PhysRevLett.126.227201},
  url = {https://link.aps.org/doi/10.1103/PhysRevLett.126.227201}
}

@article{Dynamical2018ma,
  title = {Dynamical signature of fractionalization at a deconfined quantum critical point},
  author = {Ma, Nvsen and Sun, Guang-Yu and You, Yi-Zhuang and Xu, Cenke and Vishwanath, Ashvin and Sandvik, Anders W. and Meng, Zi Yang},
  journal = {Phys. Rev. B},
  volume = {98},
  issue = {17},
  pages = {174421},
  numpages = {12},
  year = {2018},
  month = {Nov},
  publisher = {American Physical Society},
  doi = {10.1103/PhysRevB.98.174421},
  url = {https://link.aps.org/doi/10.1103/PhysRevB.98.174421}
}

@article{Fractionalized2021wang,
author = {Wang, Yan-Cheng and Cheng, Meng and Witczak-Krempa, William and Meng, Zi Yang},
year = {2021},
title = {Fractionalized conductivity and emergent self-duality near topological phase transitions},
journal = {Nature Communications},
pages = {5347},
volume = {12},
url = {https://doi.org/10.1038/s41467-021-25707-z},
doi = {10.1038/s41467-021-25707-z}
}

@article{Progress2023shao,
  title={Progress on stochastic analytic continuation of quantum Monte Carlo data},
  author={Shao, Hui and Sandvik, Anders W},
  journal={	Phys. Rep.},
  volume={1003},
  pages={1--88},
  year={2023},
  publisher={Elsevier},
  url = {https://doi.org/10.1016/j.physrep.2022.11.002}
}

@article{zhaoScaling2022,
  title = {Scaling of Entanglement Entropy at Deconfined Quantum Criticality},
  author = {Zhao, Jiarui and Wang, Yan-Cheng and Yan, Zheng and Cheng, Meng and Meng, Zi Yang},
  journal = {Phys. Rev. Lett.},
  volume = {128},
  issue = {1},
  pages = {010601},
  numpages = {6},
  year = {2022},
  month = {Jan},
  publisher = {American Physical Society},
  doi = {10.1103/PhysRevLett.128.010601},
  url = {https://link.aps.org/doi/10.1103/PhysRevLett.128.010601}
}

@article{isakovTopological2011,
author = {Isakov, Sergei V. and Hastings, Matthew B. and Melko, Roger G.},
year = {2011},
title = {Topological entanglement entropy of a Bose--Hubbard spin liquid},
journal = {Nature Physics},
pages = {772 - 775},
volume = {7},
url = {https://doi.org/10.1038/nphys2036},
doi = {10.1038/nphys2036}
}

@article{gagliano1987,
  title = {Dynamical Properties of Quantum Many-Body Systems at Zero Temperature},
  author = {Gagliano, E. R. and Balseiro, C. A.},
  journal = {Phys. Rev. Lett.},
  volume = {59},
  issue = {26},
  pages = {2999--3002},
  numpages = {0},
  year = {1987},
  month = {Dec},
  publisher = {American Physical Society},
  doi = {10.1103/PhysRevLett.59.2999},
  url = {https://link.aps.org/doi/10.1103/PhysRevLett.59.2999}
}

@article{hallberg1995,
  title = {Density-matrix algorithm for the calculation of dynamical properties of low-dimensional systems},
  author = {Hallberg, Karen A.},
  journal = {Phys. Rev. B},
  volume = {52},
  issue = {14},
  pages = {R9827--R9830},
  numpages = {0},
  year = {1995},
  month = {Oct},
  publisher = {American Physical Society},
  doi = {10.1103/PhysRevB.52.R9827},
  url = {https://link.aps.org/doi/10.1103/PhysRevB.52.R9827}
}

@incollection{prelovsek2013lanczoschapter,
  author    = {Prelov{\v{s}}ek, Peter and Bon{\v{c}}a, Janez},
  title     = {Ground State and Finite Temperature Lanczos Methods},
  booktitle = {Strongly Correlated Systems: Numerical Methods},
  editor    = {Avella, Adolfo and Mancini, Ferdinando},
  series    = {Springer Series in Solid-State Sciences},
  volume    = {176},
  pages     = {1--30},
  publisher = {Springer},
  address   = {Berlin, Heidelberg},
  year      = {2013},
  doi       = {10.1007/978-3-642-35106-8\_1}
}

@article{jaklic1994ftlm,
  author  = {Jakli{\v{c}}, Janez and Prelov{\v{s}}ek, Peter},
  title   = {Lanczos method for the calculation of finite-temperature quantities in correlated systems},
  journal = {Phys. Rev. B},
  volume  = {49},
  pages   = {5065--5068},
  year    = {1994},
  doi     = {10.1103/PhysRevB.49.5065}
}

@article{Determining2016Shitara,
  title = {Determining the continuous family of quantum Fisher information from linear-response theory},
  author = {Shitara, Tomohiro and Ueda, Masahito},
  journal = {Phys. Rev. A},
  volume = {94},
  issue = {6},
  pages = {062316},
  numpages = {7},
  year = {2016},
  month = {Dec},
  publisher = {American Physical Society},
  doi = {10.1103/PhysRevA.94.062316},
  url = {https://link.aps.org/doi/10.1103/PhysRevA.94.062316}
}

@article{braunstein1994statistical,
  title={Statistical distance and the geometry of quantum states},
  author={Braunstein, Samuel L and Caves, Carlton M},
  journal={Physical Review Letters},
  volume={72},
  number={22},
  pages={3439},
  year={1994},
  publisher={APS},
  url={https://journals.aps.org/prl/abstract/10.1103/PhysRevLett.72.3439}
}

@article{escher2011general,
  title={General framework for estimating the ultimate precision limit in noisy quantum-enhanced metrology},
  author={Escher, BM and de Matos Filho, Ruynet Lima and Davidovich, Luiz},
  journal={Nature Physics},
  volume={7},
  number={5},
  pages={406--411},
  year={2011},
  publisher={Nature Publishing Group UK London},
  url={https://www.nature.com/articles/nphys1958}
}

@book{tino2014atom,
  title={Atom interferometry},
  author={Tino, Guglielmo M and Kasevich, Mark A},
  volume={188},
  year={2014},
  publisher={IOS Press},
  address = {Amsterdam},
}

@article{petz1996geometries,
  title={Geometries of quantum states},
  author={Petz, D{\'e}nes and Sud{\'a}r, Csaba},
  journal={Journal of Mathematical Physics},
  volume={37},
  number={6},
  pages={2662--2673},
  year={1996},
  publisher={American Institute of Physics},
  url={https://pubs.aip.org/aip/jmp/article-abstract/37/6/2662/447689/Geometries-of-quantum-states}
}

@article{petz2002covariance,
  title={Covariance and Fisher information in quantum mechanics},
  author={Petz, D{\'e}nes},
  journal={Journal of Physics A: Mathematical and General},
  volume={35},
  number={4},
  pages={929},
  year={2002},
  publisher={IOP Publishing},
  url={https://iopscience.iop.org/article/10.1088/0305-4470/35/4/305/meta?casa_token=z2aq_bPQo2cAAAAA:AMvSNnhjr_R209cXn3kkari6Ygv_GJXOlvrjmtXsMKNWl8vu3XOJOxmL2hZD23JM_eSGQPjN2dtztp9dRA3wnDH8NQ}
}

@article{paris2009quantum,
  title={Quantum estimation for quantum technology},
  author={Paris, Matteo GA},
  journal={International Journal of Quantum Information},
  volume={7},
  number={supp01},
  pages={125--137},
  year={2009},
  publisher={World Scientific},
  url={https://www.worldscientific.com/doi/abs/10.1142/S0219749909004839}
}

@article{toth2012multipartite,
  title={Multipartite entanglement and high-precision metrology},
  author={T{\'o}th, G{\'e}za},
  journal={Physical Review A---Atomic, Molecular, and Optical Physics},
  volume={85},
  number={2},
  pages={022322},
  year={2012},
  publisher={APS},
  url={https://journals.aps.org/pra/abstract/10.1103/PhysRevA.85.022322}
}

@article{hyllus2012fisher,
  title={Fisher information and multiparticle entanglement},
  author={Hyllus, Philipp and Laskowski, Wies{\l}aw and Krischek, Roland and Schwemmer, Christian and Wieczorek, Witlef and Weinfurter, Harald and Pezz{\'e}, Luca and Smerzi, Augusto},
  journal={Physical Review A---Atomic, Molecular, and Optical Physics},
  volume={85},
  number={2},
  pages={022321},
  year={2012},
  publisher={APS},
  url={https://journals.aps.org/pra/abstract/10.1103/PhysRevA.85.022321}
}

@article{hauke2016measuring,
  title={Measuring multipartite entanglement through dynamic susceptibilities},
  author={Hauke, Philipp and Heyl, Markus and Tagliacozzo, Luca and Zoller, Peter},
  journal={Nature Physics},
  volume={12},
  number={8},
  pages={778--782},
  year={2016},
  publisher={Nature Publishing Group UK London},
  url={https://www.nature.com/articles/nphys3700}
}

@article{lambert2019estimates,
  title={Estimates of the quantum Fisher information in the {S}=1 antiferromagnetic Heisenberg spin chain with uniaxial anisotropy},
  author={Lambert, James and S{\o}rensen, Erik S},
  journal={Physical Review B},
  volume={99},
  number={4},
  pages={045117},
  year={2019},
  publisher={APS},
  url={https://journals.aps.org/prb/abstract/10.1103/PhysRevB.99.045117}
}

@article{laurell2021quantifying,
  title={Quantifying and controlling entanglement in the quantum magnet {Cs$_2$CoCl$_4$}},
  author={Laurell, Pontus and Scheie, Allen and Mukherjee, Chiron J and Koza, Michael M and Enderle, Mechtild and Tylczynski, Zbigniew and Okamoto, Satoshi and Coldea, Radu and Tennant, D Alan and Alvarez, Gonzalo},
  journal={Physical Review Letters},
  volume={127},
  number={3},
  pages={037201},
  year={2021},
  publisher={APS},
  url={https://journals.aps.org/prl/abstract/10.1103/PhysRevLett.127.037201}
}

@article{laurell2025witnessing,
  title={Witnessing entanglement and quantum correlations in condensed matter: {A} review},
  author={Laurell, Pontus and Scheie, Allen and Dagotto, Elbio and Tennant, D Alan},
  journal={Advanced Quantum Technologies},
  volume={8},
  number={3},
  pages={2400196},
  year={2025},
  publisher={Wiley Online Library},
  url={https://advanced.onlinelibrary.wiley.com/doi/full/10.1002/qute.202400196}
}

@article{menon2023multipartite,
  title={Multipartite entanglement in the one-dimensional spin-1 2 Heisenberg antiferromagnet},
  author={Menon, Varun and Sherman, Nicholas E and Dupont, Maxime and Scheie, Allen O and Tennant, D Alan and Moore, Joel E},
  journal={Physical Review B},
  volume={107},
  number={5},
  pages={054422},
  year={2023},
  publisher={APS},
  url={https://journals.aps.org/prb/abstract/10.1103/PhysRevB.107.054422}
}

@article{levin2005string,
  title = {String-net condensation: A physical mechanism for topological phases},
  author = {Levin, Michael A. and Wen, Xiao-Gang},
  journal = {Phys. Rev. B},
  volume = {71},
  issue = {4},
  pages = {045110},
  numpages = {21},
  year = {2005},
  month = {Jan},
  publisher = {American Physical Society},
  doi = {10.1103/PhysRevB.71.045110},
  url = {https://link.aps.org/doi/10.1103/PhysRevB.71.045110}
}

@article{wen2002quantum,
  title = {Quantum orders and symmetric spin liquids},
  author = {Wen, Xiao-Gang},
  journal = {Phys. Rev. B},
  volume = {65},
  issue = {16},
  pages = {165113},
  numpages = {37},
  year = {2002},
  month = {Apr},
  publisher = {American Physical Society},
  doi = {10.1103/PhysRevB.65.165113},
  url = {https://link.aps.org/doi/10.1103/PhysRevB.65.165113}
}

@article{gu2009tensor,
  title = {Tensor-product representations for string-net condensed states},
  author = {Gu, Zheng-Cheng and Levin, Michael and Swingle, Brian and Wen, Xiao-Gang},
  journal = {Phys. Rev. B},
  volume = {79},
  issue = {8},
  pages = {085118},
  numpages = {10},
  year = {2009},
  month = {Feb},
  publisher = {American Physical Society},
  doi = {10.1103/PhysRevB.79.085118},
  url = {https://link.aps.org/doi/10.1103/PhysRevB.79.085118}
}

@article{levin2006detecting,
  title = {Detecting Topological Order in a Ground State Wave Function},
  author = {Levin, Michael and Wen, Xiao-Gang},
  journal = {Phys. Rev. Lett.},
  volume = {96},
  issue = {11},
  pages = {110405},
  numpages = {4},
  year = {2006},
  month = {Mar},
  publisher = {American Physical Society},
  doi = {10.1103/PhysRevLett.96.110405},
  url = {https://link.aps.org/doi/10.1103/PhysRevLett.96.110405}
}

@article{kitaev2006topological,
  title = {Topological Entanglement Entropy},
  author = {Kitaev, Alexei and Preskill, John},
  journal = {Phys. Rev. Lett.},
  volume = {96},
  issue = {11},
  pages = {110404},
  numpages = {4},
  year = {2006},
  month = {Mar},
  publisher = {American Physical Society},
  doi = {10.1103/PhysRevLett.96.110404},
  url = {https://link.aps.org/doi/10.1103/PhysRevLett.96.110404}
}

@article{jiang2012identifying,
  title={Identifying topological order by entanglement entropy},
  author={Jiang, Hong-Chen and Wang, Zhenghan and Balents, Leon},
  journal={Nature Physics},
  volume={8},
  number={12},
  pages={902--905},
  year={2012},
  publisher={Nature Publishing Group},
  url={https://www.nature.com/articles/nphys2465}
}

@article{grover2011entanglement,
  title={Entanglement entropy of gapped phases and topological order in three dimensions},
  author={Grover, Tarun and Turner, Ari M and Vishwanath, Ashvin},
  journal={Physical Review B---Condensed Matter and Materials Physics},
  volume={84},
  number={19},
  pages={195120},
  year={2011},
  publisher={APS},
  url={https://journals.aps.org/prb/abstract/10.1103/PhysRevB.84.195120}
}

@article{anderson1987resonating,
  title={The resonating valence bond state in {La$_2$CuO$_4$} and superconductivity},
  author={Anderson, Philip W},
  journal={science},
  volume={235},
  number={4793},
  pages={1196--1198},
  year={1987},
  publisher={American Association for the Advancement of Science},
  url={https://www.science.org/doi/abs/10.1126/science.235.4793.1196}
}

@article{savary2016quantumspinliquids,
  title={Quantum spin liquids: a Review},
  author={Savary, Lucile and Balents, Leon},
  journal={Reports on Progress in Physics},
  volume={80},
  number={1},
  pages={016502},
  year={2016},
  publisher={IOP Publishing},
  url={https://iopscience.iop.org/article/10.1088/0034-4885/80/1/016502/meta?casa_token=cX7yevlBRLUAAAAA:9vpWype835AmPn9X_5wQinoKv27N4H4IDlvvcrjzhIQkkSz45TXyT5k6F1_zGn_3CruCI9LIFFKHg2A9XFedxXwBWrNI5A}
}

@article{broholm2020quantum,
  title={Quantum spin liquids},
  author={Broholm, C and Cava, RJ and Kivelson, SA and Nocera, DG and Norman, MR and Senthil, T},
  journal={Science},
  volume={367},
  number={6475},
  pages={eaay0668},
  year={2020},
  publisher={American Association for the Advancement of Science},
  url={https://www.science.org/doi/full/10.1126/science.aay0668}
}

@article{zhou2017quantum,
  title = {Quantum spin liquid states},
  author = {Zhou, Yi and Kanoda, Kazushi and Ng, Tai-Kai},
  journal = {Rev. Mod. Phys.},
  volume = {89},
  issue = {2},
  pages = {025003},
  numpages = {50},
  year = {2017},
  month = {Apr},
  publisher = {American Physical Society},
  doi = {10.1103/RevModPhys.89.025003},
  url = {https://link.aps.org/doi/10.1103/RevModPhys.89.025003}
}

@article{knolle2019field,
  title={A field guide to spin liquids},
  author={Knolle, Johannes and Moessner, Roderich},
  journal={Annual Review of Condensed Matter Physics},
  volume={10},
  pages={451--472},
  year={2019},
  publisher={Annual Reviews},
  url={https://www.annualreviews.org/content/journals/10.1146/annurev-conmatphys-031218-013401}
}

@article{gingras2014quantum,
  title={Quantum spin ice: a search for gapless quantum spin liquids in pyrochlore magnets},
  author={Gingras, Michel JP and McClarty, Paul A},
  journal={Reports on Progress in Physics},
  volume={77},
  number={5},
  pages={056501},
  year={2014},
  publisher={IOP Publishing},
  url={https://iopscience.iop.org/article/10.1088/0034-4885/77/5/056501/meta}
}

@article{castro2006ice,
  title={Ice: a strongly correlated proton system},
  author={Castro Neto, AH and Pujol, Pierre and Fradkin, Eduardo},
  journal={Physical Review B---Condensed Matter and Materials Physics},
  volume={74},
  number={2},
  pages={024302},
  year={2006},
  publisher={APS},
  url={https://journals.aps.org/prb/abstract/10.1103/PhysRevB.74.024302}
}

@article{castelnovo2012spin,
  title={Spin ice, fractionalization, and topological order},
  author={Castelnovo, C and Moessner, R and Sondhi, Shivaji Lal},
  journal={Annu. Rev. Condens. Matter Phys.},
  volume={3},
  number={1},
  pages={35--55},
  year={2012},
  publisher={Annual Reviews},
  url={https://www.annualreviews.org/content/journals/10.1146/annurev-conmatphys-020911-125058}
}

@book{udagawa2021spin,
  title={Spin Ice},
  author={Udagawa, Masafumi and Jaubert, Ludovic},
  year={2021},
  publisher={Springer},
  address = {Cham},
}

@article{benton2012seeing,
  title = {Seeing the light: experimental signatures of emergent electromagnetism in a quantum spin ice},
  author = {Benton, Owen and Sikora, Olga and Shannon, Nic},
  journal = {Phys. Rev. B},
  volume = {86},
  issue = {7},
  pages = {075154},
  numpages = {25},
  year = {2012},
  month = {Aug},
  publisher = {American Physical Society},
  doi = {10.1103/PhysRevB.86.075154},
  url = {https://link.aps.org/doi/10.1103/PhysRevB.86.075154}
}

@article{huang2020extended,
  title = {Extended Coulomb liquid of paired hardcore boson model on a pyrochlore lattice},
  author = {Huang, Chun-Jiong and Liu, Changle and Meng, Ziyang and Yu, Yue and Deng, Youjin and Chen, Gang},
  journal = {Phys. Rev. Research},
  volume = {2},
  issue = {4},
  pages = {042022},
  numpages = {6},
  year = {2020},
  month = {Oct},
  publisher = {American Physical Society},
  doi = {10.1103/PhysRevResearch.2.042022},
  url = {https://link.aps.org/doi/10.1103/PhysRevResearch.2.042022}
}

@article{shannon2012quantum,
  title = {Quantum Ice: A Quantum Monte Carlo Study},
  author = {Shannon, Nic and Sikora, Olga and Pollmann, Frank and Penc, Karlo and Fulde, Peter},
  journal = {Phys. Rev. Lett.},
  volume = {108},
  issue = {6},
  pages = {067204},
  numpages = {5},
  year = {2012},
  month = {Feb},
  publisher = {American Physical Society},
  doi = {10.1103/PhysRevLett.108.067204},
  url = {https://link.aps.org/doi/10.1103/PhysRevLett.108.067204}
}

@article{smith2022case,
  title = {Case for a {U}(1)$_{\ensuremath{\pi}}$ Quantum Spin Liquid Ground State in the Dipole-Octupole Pyrochlore {Ce$_2$Zr$_2$O$_7$}},
  author = {Smith, E. M. and Benton, O. and Yahne, D. R. and Placke, B. and Sch\"afer, R. and Gaudet, J. and Dudemaine, J. and Fitterman, A. and Beare, J. and Wildes, A. R. and Bhattacharya, S. and DeLazzer, T. and Buhariwalla, C. R. C. and Butch, N. P. and Movshovich, R. and Garrett, J. D. and Marjerrison, C. A. and Clancy, J. P. and Kermarrec, E. and Luke, G. M. and Bianchi, A. D. and Ross, K. A. and Gaulin, B. D.},
  journal = {Phys. Rev. X},
  volume = {12},
  issue = {2},
  pages = {021015},
  numpages = {19},
  year = {2022},
  month = {Apr},
  publisher = {American Physical Society},
  doi = {10.1103/PhysRevX.12.021015},
  url = {https://link.aps.org/doi/10.1103/PhysRevX.12.021015}
}

@article{sibille2015candidate,
  title = {Candidate Quantum Spin Liquid in the Ce$^{3+}$ Pyrochlore Stannate {Ce$_2$Sn$_2$O$_7$}},
  author = {Sibille, Romain and Lhotel, Elsa and Pomjakushin, Vladimir and Baines, Chris and Fennell, Tom and Kenzelmann, Michel},
  journal = {Phys. Rev. Lett.},
  volume = {115},
  issue = {9},
  pages = {097202},
  numpages = {5},
  year = {2015},
  month = {Aug},
  publisher = {American Physical Society},
  doi = {10.1103/PhysRevLett.115.097202},
  url = {https://link.aps.org/doi/10.1103/PhysRevLett.115.097202}
}

@article{li2017symmetry,
  title = {Symmetry enriched U(1) topological orders for dipole-octupole doublets on a pyrochlore lattice},
  author = {Li, Yao-Dong and Chen, Gang},
  journal = {Phys. Rev. B},
  volume = {95},
  issue = {4},
  pages = {041106},
  numpages = {6},
  year = {2017},
  month = {Jan},
  publisher = {American Physical Society},
  doi = {10.1103/PhysRevB.95.041106},
  url = {https://link.aps.org/doi/10.1103/PhysRevB.95.041106}
}

@article{gao2019experimental,
  title={Experimental signatures of a three-dimensional quantum spin liquid in effective spin-1/2 {Ce$_2$Zr$_2$O$_7$} pyrochlore},
  author={Gao, Bin and Chen, Tong and Tam, David W and Huang, Chien-Lung and Sasmal, Kalyan and Adroja, Devashibhai T and Ye, Feng and Cao, Huibo and Sala, Gabriele and Stone, Matthew B and others},
  journal={Nature Physics},
  volume={15},
  number={10},
  pages={1052--1057},
  year={2019},
  publisher={Nature Publishing Group},
  url={https://www.nature.com/articles/s41567-019-0577-6}
}

@article{smith2023quantum,
  title={Quantum Spin Ice Response to a Magnetic Field in the Dipole-Octupole Pyrochlore ${\mathrm{Ce}}_{2}{\mathrm{Zr}}_{2}{\mathrm{O}}_{7}$},
  author = {Smith, E. M. and Dudemaine, J. and Placke, B. and Sch\"afer, R. and Yahne, D. R. and DeLazzer, T. and Fitterman, A. and Beare, J. and Gaudet, J. and Buhariwalla, C. R. C. and Podlesnyak, A. and Xu, Guangyong and Clancy, J. P. and Movshovich, R. and Luke, G. M. and Ross, K. A. and Moessner, R. and Benton, O. and Bianchi, A. D. and Gaulin, B. D.},
  journal = {Phys. Rev. B},
  volume = {108},
  issue = {5},
  pages = {054438},
  numpages = {33},
  year = {2023},
  month = {Aug},
  publisher = {American Physical Society},
  doi = {10.1103/PhysRevB.108.054438},
  url = {https://link.aps.org/doi/10.1103/PhysRevB.108.054438}
}

@article{gaudet2019quantum,
  title = {Quantum Spin Ice Dynamics in the Dipole-Octupole Pyrochlore Magnet ${\mathrm{Ce}}_{2}{\mathrm{Zr}}_{2}{\mathrm{O}}_{7}$},
  author = {Gaudet, J. and Smith, E. M. and Dudemaine, J. and Beare, J. and Buhariwalla, C. R. C. and Butch, N. P. and Stone, M. B. and Kolesnikov, A. I. and Xu, Guangyong and Yahne, D. R. and Ross, K. A. and Marjerrison, C. A. and Garrett, J. D. and Luke, G. M. and Bianchi, A. D. and Gaulin, B. D.},
  journal = {Phys. Rev. Lett.},
  volume = {122},
  issue = {18},
  pages = {187201},
  numpages = {6},
  year = {2019},
  month = {May},
  publisher = {American Physical Society},
  doi = {10.1103/PhysRevLett.122.187201},
  url = {https://link.aps.org/doi/10.1103/PhysRevLett.122.187201}
}

@article{yahne2022dipolar,
  title = {Dipolar Spin Ice Regime Proximate to an All-In-All-Out N\'eel Ground State in the Dipolar-Octupolar Pyrochlore {Ce$_{2}$Sn$_{2}$O$_{7}$}},
  author = {Yahne, D. R. and Placke, B. and Sch\"afer, R. and Benton, O. and Moessner, R. and Powell, M. and Kolis, J. W. and Pasco, C. M. and May, A. F. and Frontzek, M. D. and Smith, E. M. and Gaulin, B. D. and Calder, S. and Ross, K. A.},
  journal = {Phys. Rev. X},
  volume = {14},
  issue = {1},
  pages = {011005},
  numpages = {16},
  year = {2024},
  month = {Jan},
  publisher = {American Physical Society},
  doi = {10.1103/PhysRevX.14.011005},
  url = {https://link.aps.org/doi/10.1103/PhysRevX.14.011005}
}

@article{smith2025single,
  title={Single-Crystal Diffuse Neutron Scattering Study of the Dipole-Octupole Quantum Spin-Ice Candidate {Ce$_2$Zr$_2$O$_7$}: No Apparent Octupolar Correlations Above {T}= 0.05 {K}},
  author={Smith, EM and Sch{\"a}fer, R and Dudemaine, J and Placke, Benedikt and Yuan, B and Morgan, Z and Ye, F and Moessner, Roderich and Benton, Owen and Bianchi, AD and others},
  journal={Physical Review X},
  volume={15},
  number={2},
  pages={021033},
  year={2025},
  publisher={APS},
  url={https://journals.aps.org/prx/abstract/10.1103/PhysRevX.15.021033}
}

@article{Beare2923MuSr,
  title = {$\ensuremath{\mu}\mathrm{SR}$ study of the dipole-octupole quantum spin ice candidate ${\mathrm{Ce}}_{2}{\mathrm{Zr}}_{2}{\mathrm{O}}_{7}$},
  author = {Beare, J. and Smith, E. M. and Dudemaine, J. and Sch\"afer, R. and Rutherford, M. R. and Sharma, S. and Fitterman, A. and Marjerrison, C. A. and Williams, T. J. and Aczel, A. A. and Dunsiger, S. R. and Bianchi, A. D. and Gaulin, B. D. and Luke, G. M.},
  journal = {Phys. Rev. B},
  volume = {108},
  issue = {17},
  pages = {174411},
  numpages = {14},
  year = {2023},
  month = {Nov},
  publisher = {American Physical Society},
  doi = {10.1103/PhysRevB.108.174411},
  url = {https://link.aps.org/doi/10.1103/PhysRevB.108.174411}
}

@article{Poiree2022,
  title = {Crystal-field states and defect levels in candidate quantum spin ice {Ce$_2$Hf$_2$O$_7$}},
  author = {Por\'ee, Victor and Lhotel, Elsa and Petit, Sylvain and Krajewska, Aleksandra and Puphal, Pascal and Clark, Adam H. and Pomjakushin, Vladimir and Walker, Helen C. and Gauthier, Nicolas and Gawryluk, Dariusz J. and Sibille, Romain},
  journal = {Phys. Rev. Mater.},
  volume = {6},
  issue = {4},
  pages = {044406},
  numpages = {11},
  year = {2022},
  month = {Apr},
  publisher = {American Physical Society},
  doi = {10.1103/PhysRevMaterials.6.044406},
  url = {https://link.aps.org/doi/10.1103/PhysRevMaterials.6.044406}
}

@article{poree2025dipolar,
   title={Dipolar-octupolar correlations and hierarchy of exchange interactions in {Ce$_2$Hf$_2$O$_7$}},
   volume={112},
   ISSN={2469-9969},
   url={http://dx.doi.org/10.1103/j451-ztvr},
   DOI={10.1103/j451-ztvr},
   number={18},
   journal={Physical Review B},
   publisher={American Physical Society (APS)},
   author={Porée, Victor and Bhardwaj, Anish and Lhotel, Elsa and Petit, Sylvain and Gauthier, Nicolas and Yan, Han and Pomjakushin, Vladimir and Ollivier, Jacques and Quilliam, Jeffrey A. and Nevidomskyy, Andriy H. and Changlani, Hitesh J. and Sibille, Romain},
   year={2025},
   month=nov }

@article{smith2025two,
  title={Two-peak heat capacity accounts for Rln(2) entropy and ground state access in the dipole-octupole pyrochlore {Ce$_2$Hf$_2$O$_7$}},
  author={Smith, EM and Fitterman, A and Sch{\"a}fer, R and Placke, B and Woods, A and Lee, S and Huang, SH-Y and Beare, J and Sharma, S and Chatterjee, D and others},
  journal={Physical Review Letters},
  volume={135},
  number={8},
  pages={086702},
  year={2025},
  publisher={APS},
  url={https://journals.aps.org/prl/abstract/10.1103/4qxy-l8pg}
}

@article{smith2024experimental,
  title={Experimental Insights into Quantum Spin Ice Physics in Dipole--Octupole Pyrochlore Magnets},
  author={Smith, Evan M and Lhotel, Elsa and Petit, Sylvain and Gaulin, Bruce D},
  journal={Annual Review of Condensed Matter Physics},
  volume={16},
  year={2024},
  publisher={Annual Reviews},
  url={https://www.annualreviews.org/content/journals/10.1146/annurev-conmatphys-041124-015101},
}

@article{kermarrec2025magnetization,
  title={Magnetization and magnetostriction measurements of the dipole-octupole quantum spin ice candidate {Ce$_2$Hf$_2$O$_7$}},
  author={Kermarrec, Edwin and Chen, Guanyue and Okamoto, Hiromu and Huang, Chun-Jiong and Yan, Han and Yan, Jian and Takeda, Hikaru and Shimizu, Yusei and Smith, Evan M and Fitterman, Avner and others},
  journal={arXiv preprint arXiv:2509.09189},
  year={2025},
  url={https://arxiv.org/abs/2509.09189}
}

@article{huang2014quantum,
  title = {Quantum Spin Ices and Topological Phases from Dipolar-Octupolar Doublets on the Pyrochlore Lattice},
  author = {Huang, Yi-Ping and Chen, Gang and Hermele, Michael},
  journal = {Phys. Rev. Lett.},
  volume = {112},
  issue = {16},
  pages = {167203},
  numpages = {5},
  year = {2014},
  month = {Apr},
  publisher = {American Physical Society},
  doi = {10.1103/PhysRevLett.112.167203},
  url = {https://link.aps.org/doi/10.1103/PhysRevLett.112.167203}
}

@article{bhardwaj2022sleuthing,
  title={Sleuthing out exotic quantum spin liquidity in the pyrochlore magnet {Ce$_2$Zr$_2$O$_7$}},
  author={Bhardwaj, Anish and Zhang, Shu and Yan, Han and Moessner, Roderich and Nevidomskyy, Andriy H and Changlani, Hitesh J},
  journal={npj Quantum Materials},
  volume={7},
  number={1},
  pages={1--8},
  year={2022},
  publisher={Nature Publishing Group},
  url={https://www.nature.com/articles/s41535-022-00458-2}
}

@book{lovesey1984theory,
  title={Theory of neutron scattering from condensed matter},
  author={Lovesey, Stephen W},
  year={1984},
  publisher={Oxford University Press},
  address = {Oxford},
}

@book{boothroyd2020principles,
  title={Principles of neutron scattering from condensed matter},
  author={Boothroyd, Andrew T},
  year={2020},
  publisher={Oxford University Press},
  address = {Oxford},
}

@article{bramwell2014neutron,
  title={Neutron scattering from quantum condensed matter},
  author={Bramwell, Steven T and Keimer, Bernhard},
  journal={Nature Materials},
  volume={13},
  number={8},
  pages={763--767},
  year={2014},
  publisher={Nature Publishing Group UK London},
  url={https://www.nature.com/articles/nmat4045}
}

@article{lee1996spin,
  title={Spin-glass and non--spin-glass features of a geometrically frustrated magnet},
  author={Lee, S-H and Broholm, C and Aeppli, G and Ramirez, AP and Perring, TG and Carlile, CJ and Adams, M and Jones, TJL and Hessen, B},
  journal={EPL (Europhysics Letters)},
  volume={35},
  number={2},
  pages={127},
  year={1996},
  publisher={IOP Publishing},
  url={https://iopscience.iop.org/article/10.1209/epl/i1996-00543-x/meta?casa_token=Tl3mGCRY6owAAAAA:GDGWOvX7cMgiAqZeHvsv-EEwNRY8x0ghSYGDFbqpEvMFnZbwWKe3Bl6IsHNKd6LK7EcwwMAXGHyVa0PqAnBv6t1QhA}
}

@article{lee2012generic,
  title = {Generic quantum spin ice},
  author = {Lee, SungBin and Onoda, Shigeki and Balents, Leon},
  journal = {Phys. Rev. B},
  volume = {86},
  issue = {10},
  pages = {104412},
  numpages = {11},
  year = {2012},
  month = {Sep},
  publisher = {American Physical Society},
  doi = {10.1103/PhysRevB.86.104412},
  url = {https://link.aps.org/doi/10.1103/PhysRevB.86.104412}
}

@article{savary2013spin,
  title = {Spin liquid regimes at nonzero temperature in quantum spin ice},
  author = {Savary, Lucile and Balents, Leon},
  journal = {Phys. Rev. B},
  volume = {87},
  issue = {20},
  pages = {205130},
  numpages = {14},
  year = {2013},
  month = {May},
  publisher = {American Physical Society},
  doi = {10.1103/PhysRevB.87.205130},
  url = {https://link.aps.org/doi/10.1103/PhysRevB.87.205130}
}

@article{savary2012coulombic,
  title = {Coulombic Quantum Liquids in Spin-$1/2$ Pyrochlores},
  author = {Savary, Lucile and Balents, Leon},
  journal = {Phys. Rev. Lett.},
  volume = {108},
  issue = {3},
  pages = {037202},
  numpages = {5},
  year = {2012},
  month = {Jan},
  publisher = {American Physical Society},
  doi = {10.1103/PhysRevLett.108.037202},
  url = {https://link.aps.org/doi/10.1103/PhysRevLett.108.037202}
}

@incollection{savary2021quantum,
  title={Quantum Coherence: Quantum Spin Ice and Lattice Gauge Theory},
  author={Savary, Lucile and Balents, Leon},
  booktitle={Spin Ice},
  pages={239--271},
  year={2021},
  publisher={Springer},
  address={Cham}
}

@article{desrochers2022symmetry,
  title = {Symmetry fractionalization in the gauge mean-field theory of quantum spin ice},
  author = {Desrochers, F\'elix and Chern, Li Ern and Kim, Yong Baek},
  journal = {Phys. Rev. B},
  volume = {107},
  issue = {6},
  pages = {064404},
  numpages = {20},
  year = {2023},
  month = {Feb},
  publisher = {American Physical Society},
  doi = {10.1103/PhysRevB.107.064404},
  url = {https://link.aps.org/doi/10.1103/PhysRevB.107.064404}
}

@article{desrochers2023spectroscopic,
  title = {Spectroscopic Signatures of Fractionalization in Octupolar Quantum Spin Ice},
  author = {Desrochers, F\'elix and Kim, Yong Baek},
  journal = {Phys. Rev. Lett.},
  volume = {132},
  issue = {6},
  pages = {066502},
  numpages = {7},
  year = {2024},
  month = {Feb},
  publisher = {American Physical Society},
  doi = {10.1103/PhysRevLett.132.066502},
  url = {https://link.aps.org/doi/10.1103/PhysRevLett.132.066502}
}

@article{yao2020pyrochlore,
  title = {Pyrochlore {U}(1) spin liquid of mixed-symmetry enrichments in magnetic fields},
  author = {Yao, Xu-Ping and Li, Yao-Dong and Chen, Gang},
  journal = {Phys. Rev. Res.},
  volume = {2},
  issue = {1},
  pages = {013334},
  numpages = {11},
  year = {2020},
  month = {Mar},
  publisher = {American Physical Society},
  doi = {10.1103/PhysRevResearch.2.013334},
  url = {https://link.aps.org/doi/10.1103/PhysRevResearch.2.013334}
}

@article{chen2017spectral,
  title = {Spectral periodicity of the spinon continuum in quantum spin ice},
  author = {Chen, Gang},
  journal = {Phys. Rev. B},
  volume = {96},
  issue = {8},
  pages = {085136},
  numpages = {6},
  year = {2017},
  month = {Aug},
  publisher = {American Physical Society},
  doi = {10.1103/PhysRevB.96.085136},
  url = {https://link.aps.org/doi/10.1103/PhysRevB.96.085136}
}

@article{ross2011quantum,
  title = {Quantum Excitations in Quantum Spin Ice},
  author = {Ross, Kate A. and Savary, Lucile and Gaulin, Bruce D. and Balents, Leon},
  journal = {Phys. Rev. X},
  volume = {1},
  issue = {2},
  pages = {021002},
  numpages = {10},
  year = {2011},
  month = {Oct},
  publisher = {American Physical Society},
  doi = {10.1103/PhysRevX.1.021002},
  url = {https://link.aps.org/doi/10.1103/PhysRevX.1.021002}
}

@article{onoda2010quantum,
  title = {Numerical Evidence of Quantum Melting of Spin Ice: Quantum-to-Classical Crossover},
  author = {Kato, Yasuyuki and Onoda, Shigeki},
  journal = {Phys. Rev. Lett.},
  volume = {115},
  issue = {7},
  pages = {077202},
  numpages = {5},
  year = {2015},
  month = {Aug},
  publisher = {American Physical Society},
  doi = {10.1103/PhysRevLett.115.077202},
  url = {https://link.aps.org/doi/10.1103/PhysRevLett.115.077202}
}

@article{robert2015spin,
  title={Spin dynamics in the presence of competing ferromagnetic and antiferromagnetic correlations in {Yb$_2$Ti$_2$O$_7$}},
  author = {Robert, J. and Lhotel, E. and Remenyi, G. and Sahling, S. and Mirebeau, I. and Decorse, C. and Canals, B. and Petit, S.},
  journal = {Phys. Rev. B},
  volume = {92},
  issue = {6},
  pages = {064425},
  numpages = {15},
  year = {2015},
  month = {Aug},
  publisher = {American Physical Society},
  doi = {10.1103/PhysRevB.92.064425},
  url = {https://link.aps.org/doi/10.1103/PhysRevB.92.064425}
}

@article{sibille2018experimental,
  title={Experimental signatures of emergent quantum electrodynamics in {Pr$_2$Hf$_2$O$_7$}},
  author={Sibille, Romain and Gauthier, Nicolas and Yan, Han and Ciomaga Hatnean, Monica and Ollivier, Jacques and Winn, Barry and Filges, Uwe and Balakrishnan, Geetha and Kenzelmann, Michel and Shannon, Nic and others},
  journal={Nature Physics},
  volume={14},
  number={7},
  pages={711--715},
  year={2018},
  publisher={Nature Publishing Group UK London},
  url={https://www.nature.com/articles/s41567-018-0116-x}
}

@article{thompson2017quasiparticle,
  title={Quasiparticle breakdown and spin Hamiltonian of the frustrated quantum pyrochlore {Yb$_2$Ti$_2$O$_7$} in a magnetic field},
  author={Thompson, JD and McClarty, Paul A and Prabhakaran, D and Cabrera, I and Guidi, T and Coldea, R},
  journal={Physical review letters},
  volume={119},
  number={5},
  pages={057203},
  year={2017},
  publisher={APS},
  url={https://journals.aps.org/prl/abstract/10.1103/PhysRevLett.119.057203}
}

@article{kao2003understanding,
  title={Understanding paramagnetic spin correlations in the spin-liquid pyrochlore {Tb$_2$Ti$_2$O$_7$}},
  author={Kao, Ying-Jer and Enjalran, Matthew and Del Maestro, Adrian and Molavian, Hamid R and Gingras, Michel JP},
  journal={Physical Review B},
  volume={68},
  number={17},
  pages={172407},
  year={2003},
  publisher={APS},
  url={https://journals.aps.org/prb/abstract/10.1103/PhysRevB.68.172407}
}

@article{guitteny2013anisotropic,
  title={Anisotropic propagating excitations and quadrupolar effects in {Tb$_2$Ti$_2$O$_7$}},
  author={Guitteny, Sol{\`e}ne and Robert, Julien and Bonville, Pierre and Ollivier, Jacques and Decorse, Claudia and Steffens, Paul and Boehm, Martin and Mutka, Hannu and Mirebeau, Isabelle and Petit, Sylvain},
  journal={Physical Review Letters},
  volume={111},
  number={8},
  pages={087201},
  year={2013},
  publisher={APS},
  url={https://journals.aps.org/prl/abstract/10.1103/PhysRevLett.111.087201}
}

@article{petit2012spin,
  title={Spin liquid correlations, anisotropic exchange, and symmetry breaking in {Tb$_2$Ti$_2$O$_7$}},
  author={Petit, Sylvain and Bonville, Pierre and Robert, Julien and Decorse, Claudia and Mirebeau, Isabelle},
  journal={Physical Review B---Condensed Matter and Materials Physics},
  volume={86},
  number={17},
  pages={174403},
  year={2012},
  publisher={APS},
  url={https://journals.aps.org/prb/abstract/10.1103/PhysRevB.86.174403}
}

@article{takatsu2016quadrupole,
  title={Quadrupole order in the frustrated pyrochlore {Tb$_{2+x}$Ti$_{2-x}$O$_{7+y}$}},
  author={Takatsu, H and Onoda, S and Kittaka, S and Kasahara, A and Kono, Y and Sakakibara, T and Kato, Y and F{\aa}k, B and Ollivier, J and Lynn, Jeffrey W and others},
  journal={Physical Review Letters},
  volume={116},
  number={21},
  pages={217201},
  year={2016},
  publisher={APS},
  url={https://journals.aps.org/prl/abstract/10.1103/PhysRevLett.116.217201}
}

@article{taniguchi2013long,
  title={Long-range order and spin-liquid states of polycrystalline {Tb$_{2+x}$Ti$_{2-x}$O$_{7+y}$}},
  author={Taniguchi, T and Kadowaki, Hiroaki and Takatsu, Hiroshi and F{\aa}k, B and Ollivier, J and Yamazaki, Teruo and Sato, TJ and Yoshizawa, H and Shimura, Y and Sakakibara, T and others},
  journal={Physical Review B---Condensed Matter and Materials Physics},
  volume={87},
  number={6},
  pages={060408},
  year={2013},
  publisher={APS},
  url={https://journals.aps.org/prb/abstract/10.1103/PhysRevB.87.060408}
}

@article{kimura2013quantum,
  title={Quantum fluctuations in spin-ice-like {Pr$_2$Zr$_2$O$_7$}},
  author={Kimura, Kenta and Nakatsuji, S and Wen, JJ and Broholm, C and Stone, MB and Nishibori, E and Sawa, H},
  journal={Nature communications},
  volume={4},
  number={1},
  pages={1934},
  year={2013},
  publisher={Nature Publishing Group UK London},
  url={https://www.nature.com/articles/ncomms2914}
}

@article{petit2016antiferroquadrupolar,
  title={Antiferroquadrupolar correlations in the quantum spin ice candidate {Pr$_2$Zr$_2$O$_7$}},
  author={Petit, Sylvain and Lhotel, Elsa and Guitteny, Sol{\`e}ne and Florea, O and Robert, Julien and Bonville, Pierre and Mirebeau, Isabelle and Ollivier, Jacques and Mutka, Hannu and Ressouche, Eric and others},
  journal={Physical Review B},
  volume={94},
  number={16},
  pages={165153},
  year={2016},
  publisher={APS},
  url={https://journals.aps.org/prb/abstract/10.1103/PhysRevB.94.165153}
}

@article{kermarrec2017ground,
  title={Ground state selection under pressure in the quantum pyrochlore magnet {Yb$_2$Ti$_2$O$_7$}},
  author={Kermarrec, E and Gaudet, J and Fritsch, K and Khasanov, R and Guguchia, Z and Ritter, C and Ross, KA and Dabkowska, HA and Gaulin, BD},
  journal={Nature communications},
  volume={8},
  number={1},
  pages={14810},
  year={2017},
  publisher={Nature Publishing Group UK London},
  url={https://www.nature.com/articles/ncomms14810}
}

@article{arpino2017impact,
  title={Impact of stoichiometry of {Yb$_2$Ti$_2$O$_7$} on its physical properties},
  author={Arpino, KE and Trump, BA and Scheie, AO and McQueen, TM and Koohpayeh, SM},
  journal={Physical Review B},
  volume={95},
  number={9},
  pages={094407},
  year={2017},
  publisher={APS},
  url={https://journals.aps.org/prb/abstract/10.1103/PhysRevB.95.094407}
}

@article{zhou2008dynamic,
  title={Dynamic Spin Ice: {Pr$_2$Sn$_2$O$_7$}},
  author={Zhou, HD and Wiebe, CR and Janik, JA and Balicas, L and Yo, YJ and Qiu, Y and Copley, JRD and Gardner, JS},
  journal={Physical review letters},
  volume={101},
  number={22},
  pages={227204},
  year={2008},
  publisher={APS},
  url={https://journals.aps.org/prl/abstract/10.1103/PhysRevLett.101.227204}
}

@article{matsuhira2004low,
  title={Low-temperature magnetic properties of the geometrically frustrated pyrochlore {Pr$_2$Sn$_2$O$_7$}},
  author={Matsuhira, K and Sekine, C and Paulsen, C and Hinatsu, Y},
  journal={Journal of Magnetism and Magnetic Materials},
  volume={272},
  pages={E981--E982},
  year={2004},
  publisher={Elsevier},
  url={https://www.sciencedirect.com/science/article/abs/pii/S0304885303021929}
}

@article{petit2020way,
  title={On the way to understanding {Yb$_2$Ti$_2$O$_7$}},
  author={Petit, Sylvain},
  journal={Proceedings of the National Academy of Sciences},
  volume={117},
  number={47},
  pages={29263--29264},
  year={2020},
  publisher={National Academy of Sciences},
  url={https://www.pnas.org/doi/abs/10.1073/pnas.2020105117}
}

@article{zhou2025towards,
  title={Towards a global phase diagram of Ce-based dipolar-octupolar pyrochlore magnets under magnetic fields},
  author={Zhou, Zhengbang and Kim, Yong Baek},
  journal={Physical Review B},
  volume={112},
  number={6},
  pages={L060407},
  year={2025},
  publisher={APS},
  url={https://journals.aps.org/prb/abstract/10.1103/8mhy-qwsw}
}

@article{zhou2024magnetic,
  title = {Magnetic field response of dipolar-octupolar quantum spin ice},
  author = {Zhou, Zhengbang and Desrochers, F\'elix and Kim, Yong Baek},
  journal = {Phys. Rev. B},
  volume = {110},
  issue = {17},
  pages = {174441},
  numpages = {22},
  year = {2024},
  month = {Nov},
  publisher = {American Physical Society},
  doi = {10.1103/PhysRevB.110.174441},
  url = {https://link.aps.org/doi/10.1103/PhysRevB.110.174441}
}

@article{paddison2017continuous,
  title={Continuous excitations of the triangular-lattice quantum spin liquid {YbMgGaO$_4$}},
  author={Paddison, Joseph AM and Daum, Marcus and Dun, Zhiling and Ehlers, Georg and Liu, Yaohua and Stone, Matthew B and Zhou, Haidong and Mourigal, Martin},
  journal={Nature Physics},
  volume={13},
  number={2},
  pages={117--122},
  year={2017},
  publisher={Nature Publishing Group UK London},
  url={https://www.nature.com/articles/nphys3971}
}

@article{kimchi2018valence,
  title={Valence bonds in random quantum magnets: Theory and application to {YbMgGaO$_4$}},
  author={Kimchi, Itamar and Nahum, Adam and Senthil, T},
  journal={Physical Review X},
  volume={8},
  number={3},
  pages={031028},
  year={2018},
  publisher={APS},
  url={https://journals.aps.org/prx/abstract/10.1103/PhysRevX.8.031028}
}

@article{gao2023disorder,
  title={Disorder-induced excitation continuum in a spin-1 2 cobaltate on a triangular lattice},
  author={Gao, Bin and Chen, Tong and Huang, Chien-Lung and Qiu, Yiming and Xu, Guangyong and Liebman, Jesse and Chen, Lebing and Stone, Matthew B and Feng, Erxi and Cao, Huibo and others},
  journal={Physical Review B},
  volume={108},
  number={2},
  pages={024431},
  year={2023},
  publisher={APS},
  url={https://journals.aps.org/prb/abstract/10.1103/PhysRevB.108.024431}
}

@article{ross2016static,
  title={Static and dynamic XY-like short-range order in a frustrated magnet with exchange disorder},
  author={Ross, Kathryn A and Krizan, Jason W and Rodriguez-Rivera, Jose A and Cava, Robert Joseph and Broholm, Collin L},
  journal={Physical Review B},
  volume={93},
  number={1},
  pages={014433},
  year={2016},
  publisher={APS},
  url={https://journals.aps.org/prb/abstract/10.1103/PhysRevB.93.014433}
}

@article{sarkar2017spin,
  title={Spin freezing in the disordered pyrochlore magnet {NaCaCo$_2$F$_7$}: NMR studies and Monte Carlo simulations},
  author={Sarkar, Rajib and Krizan, Jason W and Br{\"u}ckner, Felix and Andrade, Eric de Castro and Rachel, Stephan and Vojta, Matthias and Cava, Robert Joseph and Klauss, H-H},
  journal={Physical Review B},
  volume={96},
  number={23},
  pages={235117},
  year={2017},
  publisher={APS},
  url={https://journals.aps.org/prb/abstract/10.1103/PhysRevB.96.235117}
}

@article{greenblatt2009rounding,
  title={Rounding of First Order Transitions in Low-Dimensional Quantum Systems with Quenched Disorder},
  author={Greenblatt, Rafael L and Aizenman, Michael and Lebowitz, Joel L},
  journal={Physical review letters},
  volume={103},
  number={19},
  pages={197201},
  year={2009},
  publisher={APS},
  url={https://journals.aps.org/prl/abstract/10.1103/PhysRevLett.103.197201}
}

@article{rau2019frustrated,
  title={Frustrated quantum rare-earth pyrochlores},
  author={Rau, Jeffrey G and Gingras, Michel JP},
  journal={Annual Review of Condensed Matter Physics},
  volume={10},
  number={1},
  pages={357--386},
  year={2019},
  publisher={Annual Reviews},
  url={https://www.annualreviews.org/content/journals/10.1146/annurev-conmatphys-022317-110520}
}

@article{patri2020distinguishing,
  title = {Distinguishing dipolar and octupolar quantum spin ices using contrasting magnetostriction signatures},
  author = {Patri, Adarsh S. and Hosoi, Masashi and Kim, Yong Baek},
  journal = {Phys. Rev. Res.},
  volume = {2},
  issue = {2},
  pages = {023253},
  numpages = {12},
  year = {2020},
  month = {Jun},
  publisher = {American Physical Society},
  doi = {10.1103/PhysRevResearch.2.023253},
  url = {https://link.aps.org/doi/10.1103/PhysRevResearch.2.023253}
}

@article{benton2020ground,
  title = {Ground-state phase diagram of dipolar-octupolar pyrochlores},
  author = {Benton, Owen},
  journal = {Phys. Rev. B},
  volume = {102},
  issue = {10},
  pages = {104408},
  numpages = {11},
  year = {2020},
  month = {Sep},
  publisher = {American Physical Society},
  doi = {10.1103/PhysRevB.102.104408},
  url = {https://link.aps.org/doi/10.1103/PhysRevB.102.104408}
}

@article{chern2023pseudofermion,
  title = {Pseudofermion functional renormalization group study of dipolar-octupolar pyrochlore magnets},
  author = {Chern, Li Ern and Desrochers, F\'elix and Kim, Yong Baek and Castelnovo, Claudio},
  journal = {Phys. Rev. B},
  volume = {109},
  issue = {18},
  pages = {184421},
  numpages = {17},
  year = {2024},
  month = {May},
  publisher = {American Physical Society},
  doi = {10.1103/PhysRevB.109.184421},
  url = {https://link.aps.org/doi/10.1103/PhysRevB.109.184421}
}

@article{halperin1974first,
  title={First-order phase transitions in superconductors and smectic-A liquid crystals},
  author={Halperin, BI and Lubensky, TC and Ma, Shang-keng},
  journal={Physical Review Letters},
  volume={32},
  number={6},
  pages={292},
  year={1974},
  publisher={APS},
  url={https://journals.aps.org/prl/abstract/10.1103/PhysRevLett.32.292}
}

@article{makhfudz2014fluctuation,
  title={Fluctuation-induced first-order quantum phase transition of the U (1) spin liquid in a pyrochlore quantum spin ice},
  author={Makhfudz, Imam},
  journal={Physical Review B},
  volume={89},
  number={2},
  pages={024401},
  year={2014},
  publisher={APS},
  url={https://journals.aps.org/prb/abstract/10.1103/PhysRevB.89.024401}
}

@article{bhardwaj2025thermodynamics,
  title={Thermodynamics of the dipole-octupole pyrochlore magnet {Ce$_2$Hf$_2$O$_7$} in applied magnetic fields},
  author={Bhardwaj, Anish and Por{\'e}e, Victor and Yan, Han and Gauthier, Nicolas and Lhotel, Elsa and Petit, Sylvain and Quilliam, Jeffrey A and Nevidomskyy, Andriy H and Sibille, Romain and Changlani, Hitesh J},
  journal={Physical Review B},
  volume={111},
  number={15},
  pages={155137},
  year={2025},
  publisher={APS},
  url={https://journals.aps.org/prb/abstract/10.1103/PhysRevB.111.155137}
}

@article{liu2019quantum,
   title={Quantum Fisher information matrix and multiparameter estimation},
   volume={53},
   ISSN={1751-8121},
   url={http://dx.doi.org/10.1088/1751-8121/ab5d4d},
   DOI={10.1088/1751-8121/ab5d4d},
   number={2},
   journal={Journal of Physics A: Mathematical and Theoretical},
   publisher={IOP Publishing},
   author={Liu, Jing and Yuan, Haidong and Lu, Xiao-Ming and Wang, Xiaoguang},
   year={2019},
   month=dec, pages={023001} }

@article{liu2016quantum,
   title={Quantum Fisher information and symmetric logarithmic derivative via anti-commutators},
   volume={49},
   ISSN={1751-8121},
   url={http://dx.doi.org/10.1088/1751-8113/49/27/275302},
   DOI={10.1088/1751-8113/49/27/275302},
   number={27},
   journal={Journal of Physics A: Mathematical and Theoretical},
   publisher={IOP Publishing},
   author={Liu, Jing and Chen, Jie and Jing, Xiao-Xing and Wang, Xiaoguang},
   year={2016},
   month=may, pages={275302} }

@article{fiderer2021general,
  title = {General Expressions for the Quantum Fisher Information Matrix with Applications to Discrete Quantum Imaging},
  author = {Fiderer, Lukas J. and Tufarelli, Tommaso and Piano, Samanta and Adesso, Gerardo},
  journal = {PRX Quantum},
  volume = {2},
  issue = {2},
  pages = {020308},
  numpages = {15},
  year = {2021},
  month = {Apr},
  publisher = {American Physical Society},
  doi = {10.1103/PRXQuantum.2.020308},
  url = {https://link.aps.org/doi/10.1103/PRXQuantum.2.020308}
}

@article{dominik2018simple,
  title = {Simple expression for the quantum Fisher information matrix},
  author = {Šafránek, Dominik},
  journal = {Phys. Rev. A},
  volume = {97},
  issue = {4},
  pages = {042322},
  numpages = {6},
  year = {2018},
  month = {Apr},
  publisher = {American Physical Society},
  doi = {10.1103/PhysRevA.97.042322},
  url = {https://link.aps.org/doi/10.1103/PhysRevA.97.042322}
}

@article{frerot2016quantum,
   title={Quantum variance: A measure of quantum coherence and quantum correlations for many-body systems},
   volume={94},
   ISSN={2469-9969},
   url={http://dx.doi.org/10.1103/PhysRevB.94.075121},
   DOI={10.1103/physrevb.94.075121},
   number={7},
   journal={Physical Review B},
   publisher={American Physical Society (APS)},
   author={Fr{\'e}rot, Ir{\'e}n{\'e}e and Roscilde, Tommaso},
   year={2016},
   month=aug }

@article{scheie2021witnessing,
   title={Witnessing entanglement in quantum magnets using neutron scattering},
   volume={103},
   ISSN={2469-9969},
   url={http://dx.doi.org/10.1103/PhysRevB.103.224434},
   DOI={10.1103/physrevb.103.224434},
   number={22},
   journal={Physical Review B},
   publisher={American Physical Society (APS)},
   author={Scheie, A. and Laurell, Pontus and Samarakoon, A. M. and Lake, B. and Nagler, S. E. and Granroth, G. E. and Okamoto, S. and Alvarez, G. and Tennant, D. A.},
   year={2021},
   month=jun }

@misc{gao2026spectroscopicdemarcationemergentphotons,
      title={Spectroscopic Demarcation of Emergent Photons and Spinons in a Dipolar-Octupolar Quantum Spin Liquid}, 
      author={Bin Gao and Zhengbang Zhou and Tingjun Zhang and Andrey Podlesnyak and Sang-Wook Cheong and Yong Baek Kim and Pengcheng Dai},
      year={2026},
      eprint={2601.03202},
      archivePrefix={arXiv},
      primaryClass={cond-mat.str-el},
      url={https://arxiv.org/abs/2601.03202}, 
}

\newpage

\newpage
\clearpage
\begin{center}
	\textbf{\large Supplemental Material for \\``Quantum Fisher Information as a Thermal Probe in Frustrated Magnets through Insights from Quantum Spin Ice''}
\end{center}
\setcounter{equation}{0}
\setcounter{figure}{0}
\setcounter{table}{0}
\setcounter{page}{1}
\setcounter{section}{0}

\makeatletter
\renewcommand{\theequation}{S\arabic{equation}}
\renewcommand{\thefigure}{S\arabic{figure}}
\setcounter{secnumdepth}{3}

The Supplemental Material provides details on quantum Monte Carlo simulations using a multi-directed loop (MDL) update scheme, specifically designed for quantum spin ice systems, as well as on the measurement of quantum Fisher information and other physical observables therein. It also includes detailed benchmarking between QMC, exact diagonalization, and gauge mean field theory calculations. We also discuss the conversion of local frame QFI data to that obtained from the neutron scattering cross section of dipolar-octupolar QSI material candidates in the global frame.

\section{Multi-directed loop QMC scheme for quantum spin ice model}
In this section, we briefly introduce the stochastic series expansion quantum Monte Carlo (SSE-QMC) simulation on the pyrochlore lattice~\cite{QMC2002Sandvik,Generalized2005Wessel,Computational2010Sandvik}. The basic idea is to construct the configuration space by expanding the partition function, $Z=\text{Tr}(e^{-\beta H})$, as a Taylor series into an operator string. We then sample this configuration space by proposing updates to the operator string, which fulfill the balance condition\cite{Markov2010Suwa}, $\sum_i W(A_i)P(A_i\to B)=W(B)$, 
or even the detailed balance condition, 
$W(A)P(A\to B)=W(B)P(B\to A)$.
Here, $A_i$ are all possible configurations that can be updated to configuration $B$. $W(A)$ is the weight of configuration $A$, and $P(A\to B)$ is the transition probability from configuration $A$ to $B$. To do so, one of the widely used methods is the directed loop (DL) update algorithm.


Within the SSE-QMC framework, the DL update algorithm is a highly efficient method for simulating quantum spin systems by inserting a pair of operators (e.g., $S_i^+$ and $S_i^-$), the configuration temporarily moves into an extended configuration space with these two operators evolving, and returns to the Hilbert space when the two operators meet and annihilate. However, on the pyrochlore lattice within the spin-ice regime ($0<J_{\pm}<0.052$), the DL update fails to drive the system into the quantum spin ice (QSI) regime. These are plotted as orange points in Fig.~\ref{fig:eng}(a-c). This failure can be understood from the perspective that first-order perturbations vanish, while third-order processes persist in the spin-ice regime~\cite{Pyrochlore2004Hermele}. Consequently, the simplest quantum fluctuation that connects different classical spin-ice states involves six spin flips around a hexagon, which is challenging to realize by inserting only a single operator pair. To address this problem, we develop a multi-directed loop (MDL) update that allows the insertion of multiple operator pairs, enabling efficient sampling in the QSI regime.

\begin{figure}
	\centering
	\includegraphics[width=0.7\columnwidth]{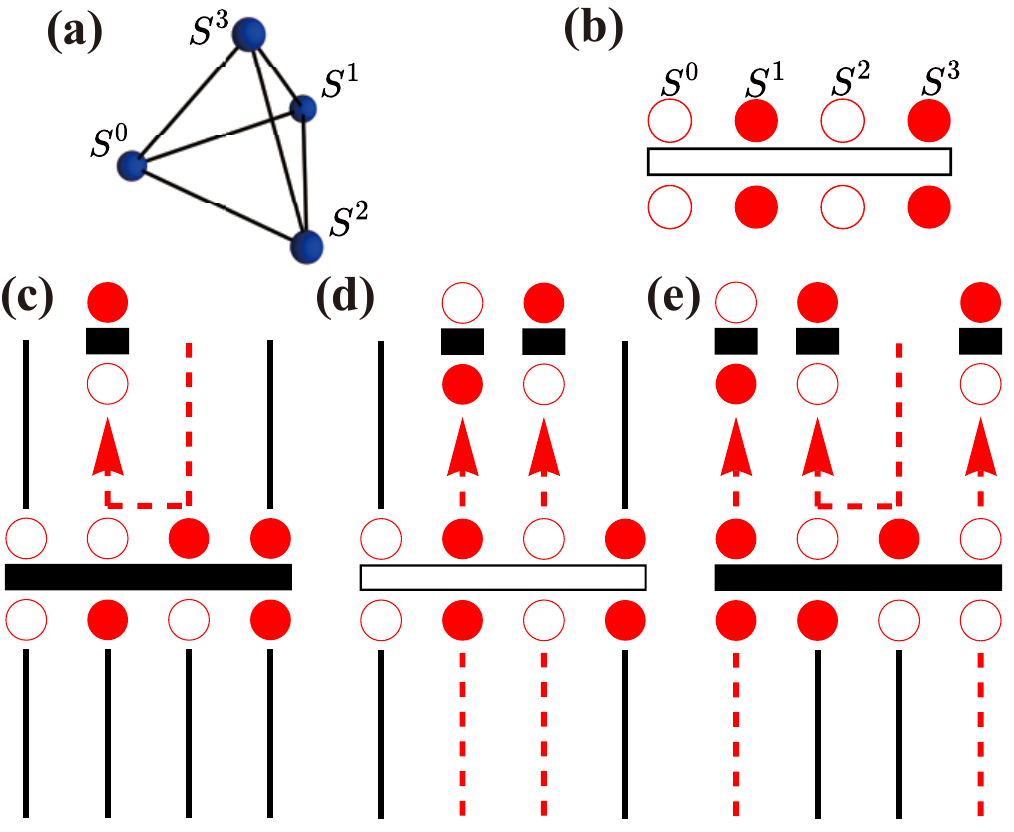}
	\caption{Schematic diagram of (a) a tetrahedron in the pyrochlore lattice and (b) the corresponding four leg operator vertex in the sampling space. (c) Illustration of the single loop update, (d) bi-loop update, and (e) tri-loop update in the multi-directed loop algorithm.}
	\label{fig:update}
\end{figure}

In our simulation, we first decompose the Hamiltonian into a sum of tetrahedron operators  (see Fig.\ref{fig:update}(a)), which is given by
\begin{equation}
	\begin{aligned}
		H=&\sum_{t}H_{t,0}+\sum_{t}H_{t,1},\\
		H_{t,0}=&J_\mathrm{z}\sum_{\langle i,j \rangle \in \hspace{-0.5mm} \tetrahedron}S_i^\mathrm{z} S_j^\mathrm{z}+H_{c,0},\\
		H_{t,1}=&-J_{\pm}\sum_{\langle i,j \rangle \in \hspace{-0.5mm} \tetrahedron}(S_i^+ S_j^- + S_i^- S_j^+).
	\end{aligned}
	\label{eq:t_ham}
\end{equation}
Here, $t$ denotes a tetrahedron ($\tetrahedron$) in the pyrochlore lattice and $\langle i,j \rangle \in \hspace{-1mm} \tetrahedron$ represents the nearest-neighbor pairs within the tetrahedron $t$. To ensure that all matrix elements are non-negative, we introduce a constant shift $H_{c,0}$ to the diagonal operator, which is $H_{c,0}=3J_\mathrm{z}/2$. 
Therefore, for a given tetrahedron state $|\alpha_i\rangle=|S_1^\mathrm{z},S_2^\mathrm{z},S_3^\mathrm{z},S_4^\mathrm{z}\rangle$, the matrix elements of the operators are $w_{i,j,x}=\langle\alpha_j|H_{t,x}|\alpha_i\rangle$ with $x=0,1$. And, the weight of a given operator string is 
\begin{equation}
	\begin{aligned}
		W=&\frac{\beta^n(L-n)!}{L!}\prod_{p=1}^n w_{i_p,j_p,x},\\
	\label{fig:weight}
	\end{aligned}
\end{equation}
where $L$ refers to the cut-off of the expansion and $n$ is the number of operators in the string. In practice, each $H_{t,x}$ is represented by a 8-leg vertex, as illustrated in Fig.~\ref{fig:update}(b). We employ the diagonal update, DL update, and MDL update to sample the configuration space efficiently. Also, we apply a thermal annealing process from inverse temperature $\beta=1$ to $\beta=1000$ with step $\delta\beta=1$. At each $\beta=\frac{1}{T}$, we perform 10,000 steps of warming and 10,000 steps of measurement. The details of these updates are as follows.

\subsubsection{Diagonal update}
In the diagonal update, we follow the detailed balance condition and apply the Metropolis algorithm. We sweep through the operator string and attempt to insert or remove diagonal operators $H_{t,0}$ at each time slice. In the insertion process, we randomly select a tetrahedron $t$ and propose to insert the diagonal operator $H_{t,0}$.
For the removal process, each diagonal operator are proposed to be removed.
Therefore, the acceptance probabilities for insertion is given by
\begin{equation}
	P^{\mathrm{insert}}_{\mathrm{accept}}=\min\left[1,\frac{N_t\beta w_{i,i,0}}{L-n}\right],
\end{equation} 
and for removal is
\begin{equation}
	P^{\mathrm{remove}}_{\mathrm{accept}}=\min\left[1,\frac{L-n+1}{N_t\beta w_{i,i,0}}\right],
\end{equation} 
with $w_{i,i,0}$ denotes the weight of $H_{t,0}$ at this time slice, and $N_t$ represents the total number of tetrahedron ($\tetrahedron$).

\subsubsection{Directed loop update}
In the DL update, we follow the balance condition and utilize the update scheme proposed in Ref.~\cite{Markov2010Suwa} to achieve the highest efficiency sampling as we know~\cite{QSL2018Wang,Dynamical2018Sun}. In practice, we first randomly select a vertex and one of its eight legs to insert a pair of operators, $S_i^+$ and $S_i^-$. The selection probability is $P_{\mathrm{select}}=1/(N_v\times 8)$, where $N_v$ is the total number of vertices in the given operator string. Then, we choose one of these pairs of operators to be the head, while the other operator becomes the tail. The head then propagates through the operator string by entering and exiting vertices until it meets the tail again, at which point the loop is closed. The propagation process is illustrated in Fig.\ref{fig:update}(c). When the head enters a vertex through one leg ($l_a$), the exiting leg ($l_b$) is chosen based on a set of exit probabilities $P_{\mathrm{exit}}$, which update the operator from $w_a$ to $w_b$. The exiting leg is selected according to the probability so that the balance condition (or the detailed balance condition) is satisfied. 

In our simulation, the exit probabilities $P_{\mathrm{exit}}$ are determined by following the balance condition and utilizing the algorithm proposed in Ref.~\cite{Markov2010Suwa} to minimize the average rejection rate, which means the head exits from the entrance leg without changing the vertex. The exit probabilities are given by
\begin{equation}
	\begin{aligned}
		P_{\mathrm{exit}}(l_a\to l_b)&=\frac{1}{w_a}\max[0,\min[\Delta_{ab},w_a+w_b-\Delta_{ab},w_a,w_b]],\\
		\Delta_{ab}&=S_a-S_{a-1}+w_1, 1\leq a,b\leq n,\\
		S_i&=\sum_{k=1}^{i}w_k, 1\leq i\leq n,\\
		S_0&=S_n.
		\label{eq:exit_prob}
	\end{aligned}
\end{equation}
Here, $n=8$ is the total number of candidate exit legs. In practice, we enumerate all nonzero vertices and iterate over all possible entrance legs to calculate and store the corresponding exit probabilities. For each vertex and entrance leg, we identify all candidate exiting legs ($n$ in total) and their associated weights $w_i$, which are sorted in descending order. The sums $\{S_i\}$ are then computed for these candidates and used to determine the exit probabilities $P_{\mathrm{exit}}(l_a\to l_i)$. All results are precomputed and stored before the simulation. During the simulation, the exiting leg $l_b$ is selected according to these stored probabilities. Such an update method achieves highly efficient sampling in various quantum spin systems, including the Balents-Fisher-Girvin model~\cite{QSL2018Wang,Dynamical2018Sun}. However, in the spin-ice regime, the DL update still fails to drive the system into the QSI phase, as shown by the orange points in Fig.~\ref{fig:eng}(a-c).

\subsubsection{Multi-Directed loop update}

Inspired by the nature that the simplest quantum fluctuation that connects different classical spin-ice states involves six spin flips around a hexagon, we develop a MDL update that allows the insertion of multiple operator pairs, enabling efficient access to the QSI regime. In this update method, we first select a hexagon in the operator string and insert three pairs of operators, $S_i^+$ and $S_i^-$, on three alternate sites of the hexagon. Then, for each pair of operators, we randomly choose one operator to be the head, while the other operator becomes the tail. The three heads then propagate through the operator string by entering and exiting vertices until all of these heads meet the tail again at the same propagation step, at which point all the loops are closed. Different from the single loop update, the propagation process of the three heads is more complicated, including one head entrance and exit a vertex (single loop update in Fig.\ref{fig:update}(c)), two heads enter and exit a vertex (bi-loop update in Fig.\ref{fig:update}(d)), and three heads enter and exit a vertex (tri-loop update in Fig.\ref{fig:update}(e)). 

To show that MDL update fulfills the detailed balance condition, we follow the worm-antiworm construction principle\cite{Generalized2005Wessel}. In MDL update, the corresponding transition probability from configuration $A$ to $B$ is given by
\begin{equation}
	P(A\to B)=P_{\mathrm{insert}} \prod_{i=1}^{N_{m}} \prod_{j=1}^{n_{v}}P_{\mathrm{prop}}(v_{i,j}\to v_{i+1,j}),
\end{equation} 
where $P_{\mathrm{insert}}$ is the probability of inserting multiple operator pairs, $N_m$ is the number of propagation steps for these operator pairs when all loops are closed. And $n_v$ is the number of vertices visited by the heads (for example, two heads entering one vertex as in Fig.~\ref{fig:update}(d), or three heads as in Fig.~\ref{fig:update}(e)). $P_{\mathrm{prop}}(v_{i,j}\to v_{i+1,j})$ is the propagation probability for updating the $j$-th visited vertex from $v_{i,j}$ to $v_{i+1,j}$, corresponding to the choice of exit legs for all heads visiting this vertex. This propagation probability is determined by listing all candidate vertices that can be updated from $v_{i,j}$, and solving for the probabilities that satisfy the detailed balance condition, $P_{\mathrm{prop}}(v_{i,j}\to v_{i+1,j})W(v_{i,j})=P_{\mathrm{prop}}(v_{i+1,j}\to v_{i,j})W(v_{i+1,j})$. 
And the transition probability for the antiworm process is given by
\begin{equation}
	P(B\to A)=P^\prime_{\mathrm{insert}} \prod_{i=1}^{N_{m}} \prod_{j=1}^{n_{v}}P_{\mathrm{prop}}(v_{N_{m}-i+1,j}\to v_{N_{m}-i,j}),
\end{equation} 
therefore, we have
\begin{equation}
	\begin{aligned}
		\frac{P(A\to B)}{P(B\to A)}&=\frac{P_{\mathrm{insert}}}{P^\prime_{\mathrm{insert}}} \prod_{i=1}^{N_{m}} \prod_{j=1}^{n_{v}}\frac{W(v_{i+1,j})}{W(v_{i,j})}\\
		&=\frac{P_{\mathrm{insert}}}{P^\prime_{\mathrm{insert}}} \frac{W(B)}{W(A)}.
	\end{aligned}
\end{equation} 
The MDL update is completed only when all three heads meet a different tail at one propagation step, ensuring that the inverse process is identical to the forward process and the insertion probabilities satisfy $P_{\mathrm{insert}}/P^\prime_{\mathrm{insert}}=1$. Therefore, the MDL update fulfills the detailed balance condition. Note that if one of the heads meets its tail and closes the corresponding loop while the other heads continue to propagate, the inverse process differs from the forward process, and the corresponding update violates the detailed balance condition. In practice, before the simulation, we enumerate all nonzero vertices and iterate over all possible entrance cases, including single-, bi-, and tri-loop cases (see Fig.\ref{fig:update}(c-e)), to calculate and store the propagation probability for each exit case. For each vertex, entrance head number, and entrance leg, we identify the propagation probability for each candidate exit leg case via Eq.~\eqref{eq:exit_prob}. All probabilities are precomputed and stored prior to the simulation.

It is worth mentioning that such a MDL update enlarged the middle extended configuration space compared to the DL update, which is beneficial for sampling the QSI regime. However, it also increases the computational cost per update. Assuming that the heads appear in each vertex leg with equal probability after a long propagation, then the probability of the heads meeting the tails and closing the loops is $3!/(4N_{v})^3$ with $N_{v}\sim L^3\beta$. Here, $L$ refers to the system size. In construct, it is $1/(4N_{v})$ for the DL update. Therefore, it is harder to close the loops for the MDL update, and the computational cost per update of MDL is roughly $(L^3\beta)^2$ times that of the DL update. This cost difference becomes even more significant in the FM phase. Therefore, we utilize both MDL and DL update to simulate the system in the QSI regime ($J_{\pm}=0.04,0.045,0.05$), while only DL in the FM regime ($J_{\pm}=0.06,0.07,0.08,0.09,0.10$). 

\subsubsection{Measurement}

\begin{figure}
	\centering
	\includegraphics[width=\columnwidth]{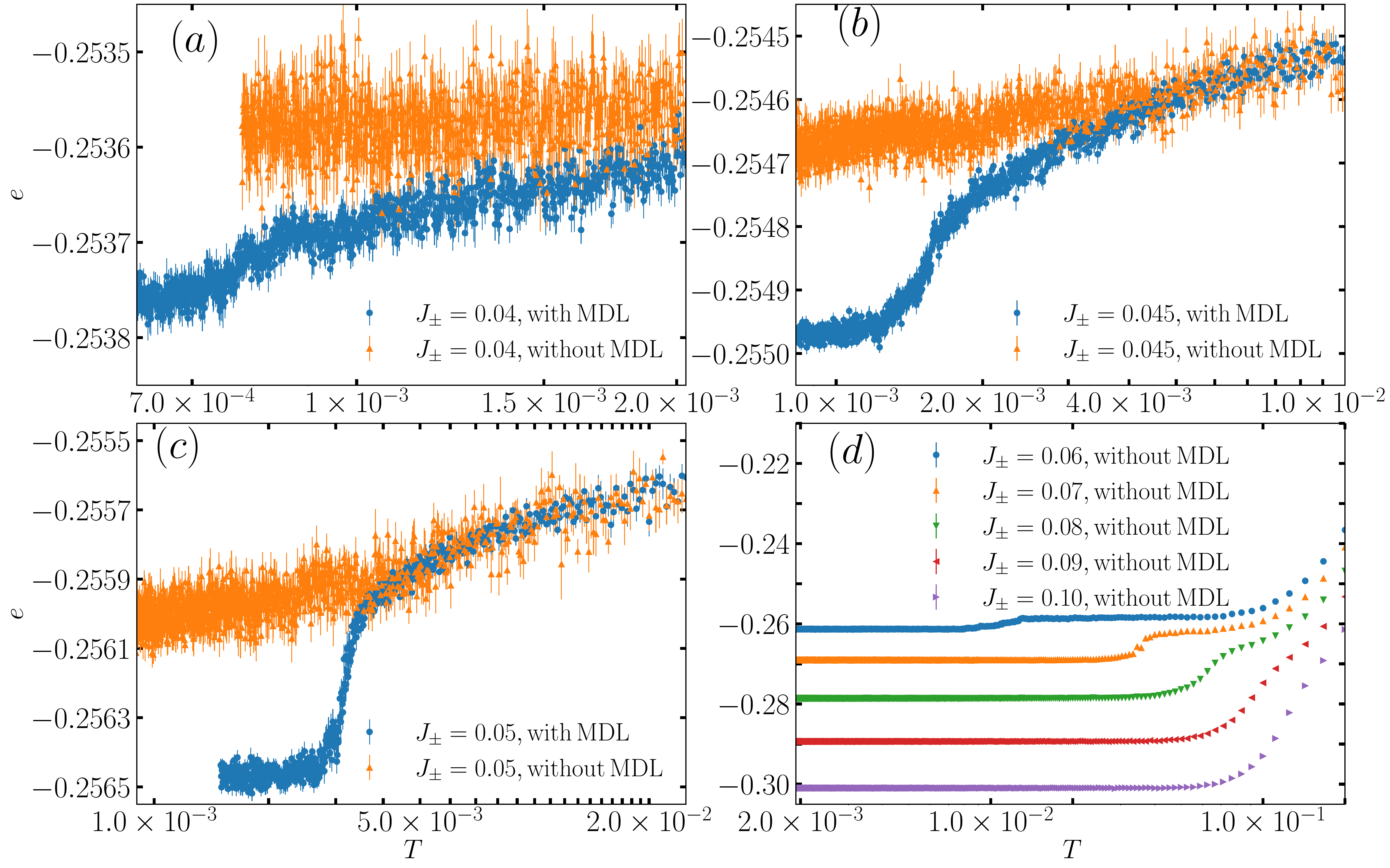}
	\caption{The energy per site $e$ as a function of temperature $T$ at (a) $J_{\pm}=0.04$, (b) $J_{\pm}=0.045$, (c) $J_{\pm}=0.05$, and (d) $J_{\pm}$ ranging from $0.06$ to $0.10$ with an increment of 0.01. In (a-c), the blue points are obtained from SSE-QMC simulation with MDL update, and the orange lines are without MDL update. In (d), all points are obtained from SSE-QMC simulation without MDL update.}
	\label{fig:eng}
\end{figure}

Here, we briefly introduce the measurement results of physical observables in our simulation. We first measure the energy per site $e$ and compare the results obtained with and without MDL updates in the QSI regime, as shown in Fig.~\ref{fig:eng}(a-c) for $J_{\pm}$ ranging from $0.04$ to $0.05$. The orange points correspond to the case without MDL update, while the blue points are with the MDL update. In all three cases, the MDL update successfully drives the system into the QSI regime at low temperature, as evidenced by a clear energy decrease, while the DL update does not. Specifically, in Fig.~\ref{fig:eng}(b), a distinct energy drop from $-0.2547$ to $-0.2550$ at $T\sim 0.0011$ is observed in the MDL update results. This small energy drop corresponds to the crossover from classical spin ice (CSI) to QSI and is consistent with the theoretical prediction $T\sim 12J_{\pm}^3 \approx 0.0011$ and previous calculations~\cite{Numerical2015Kato}. In contrast, the DL update results remain around $-0.2547$ with only a slight decreasing tendency as temperature decreases, indicating that the DL update fails to sample the QSI regime efficiently and becomes trapped in a local minimum. Also, the energy density in the FM phase is shown in Fig.~\ref{fig:eng}(d) for $J_{\pm}$ ranging from $0.06$ to $0.10$. In this regime, the DL update efficiently samples the configuration space and drives the system into the FM phase at low temperature.

\begin{figure}
	\centering
	\includegraphics[width=0.5\textwidth]{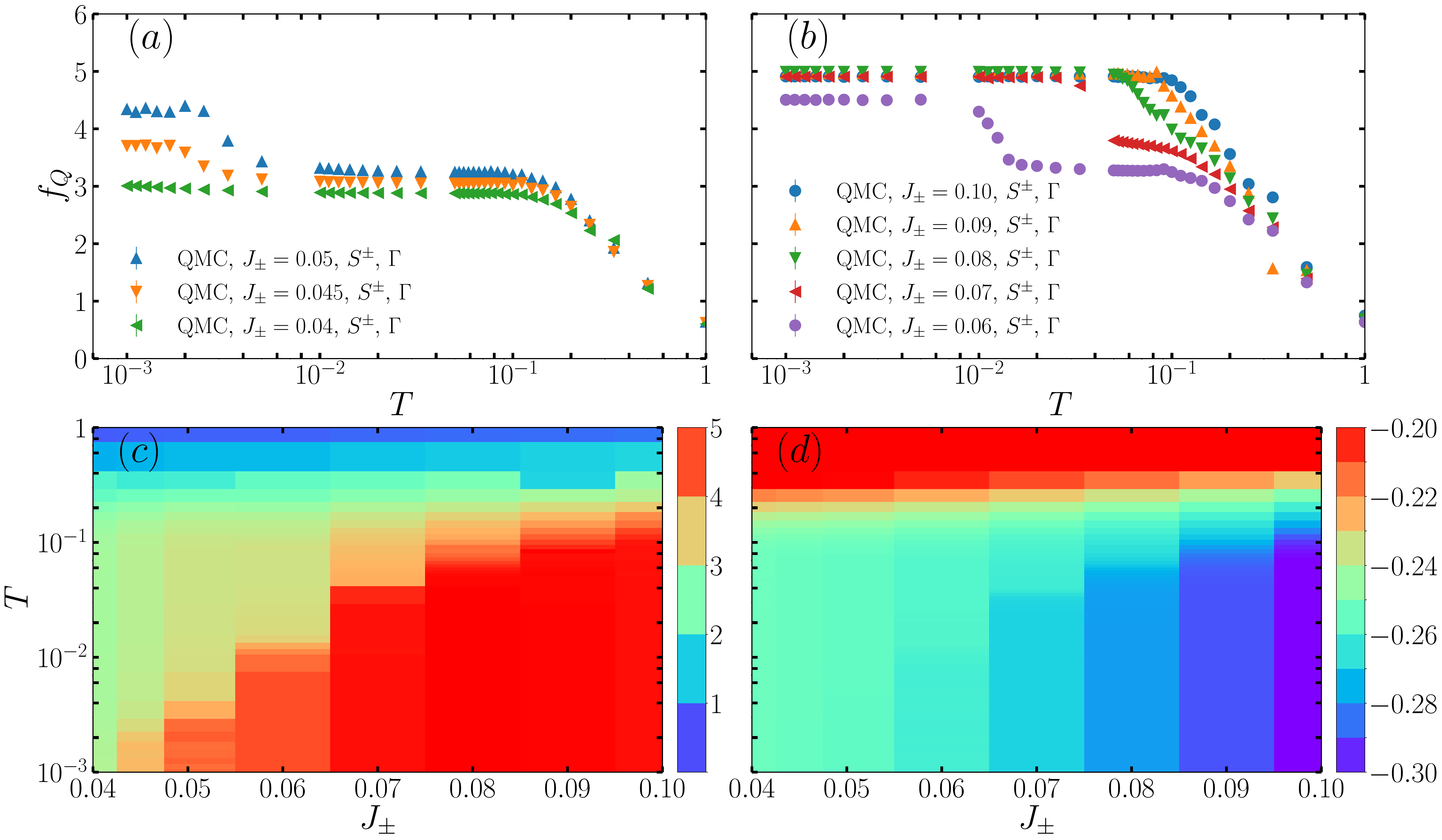}
	\caption{The QFI density $f_{Q}$ in $S^{\pm}$ channel at $\Gamma$ as a function of temperature $T$ with $J_{\pm}$ (a) ranging from $0.04$ to $0.05$ with an increment of $0.005$; (b) ranging from $0.06$ to $0.10$ with an increment of $0.01$. (c) The QFI density heat map as a function of temperature $T$ and $J_{\pm}$, ranging from $0.04$ to $0.10$, while (d) is that of the energy density $e$ heat map. }
	\label{fig:QFI_S1}
\end{figure}

Then, we measure the imaginary-time spin-spin correlation functions,
\begin{equation} 
    \begin{aligned}
        G^{\pm}&=\frac{1}{2N}\sum_{\gamma,\nu}\left\langle S^{+}_{-\mathbf{q},\gamma}(\tau) S^{-}_{\mathbf{q},\nu}(0)+S^{-}_{-\mathbf{q},\gamma}(\tau) S^{+}_{\mathbf{q},\nu}(0)\right\rangle,\\
        G^{\mathrm{zz}}&=\frac{1}{N}\sum_{\gamma,\nu}\left\langle S^{\mathrm{z}}_{-\mathbf{q},\gamma}(\tau) S^{\mathrm{z}}_{\mathbf{q},\nu}(0)\right\rangle,
    \end{aligned}
\end{equation}
where $\tau$ is the imaginary time, $\gamma$ and $\nu$ are sublattice indices, and $N$ is the total number of sites. The $G^{\pm}$ correlation is measured by tracing the head's path in the DL update process~\cite{Amplitude2021zhou}. The corresponding spectral function is then obtained via the stochastic analytic continuation (SAC) method~\cite{Progress2023shao,Amplitude2021zhou,Dynamical2018ma,Dynamics2018Huang,Dynamical2018Sun}. Finally, the QFI density $f_{Q}$ is calculated by integrating the spectral function according to Eq.~({\color{blue}4}) in the main text. Here, we focus on the QFI density at $\Gamma=(0,0,0)$, $\Gamma^\prime=(4\pi,4\pi,0)$ and $X=(0,0,2\pi)$ in both the $S^{\pm}$ and $S^{\mathrm{z}}$ channel. The results for $f_{Q}(S^{\pm}_{\Gamma},T)$ are shown in Fig.~\ref{fig:QFI_S1}: panels (a) and (b) display $f_{Q}(S^{\pm}_{\Gamma},T)$ for $J_{\pm}$ in the QSI regime ($0.04$ to $0.05$) and FM regime ($0.06$ to $0.10$), respectively. Panels (c) and (d) present heat maps of $f_{Q}(S^{\pm}_{\Gamma},T)$ and energy density $e$ as functions of temperature $T$ and $J_{\pm}$ from $0.04$ to $0.10$. These two observations agree well, but the crossover signal in the QFI density is more significant.

\begin{figure}
	\centering
	\includegraphics[width=0.5\textwidth]{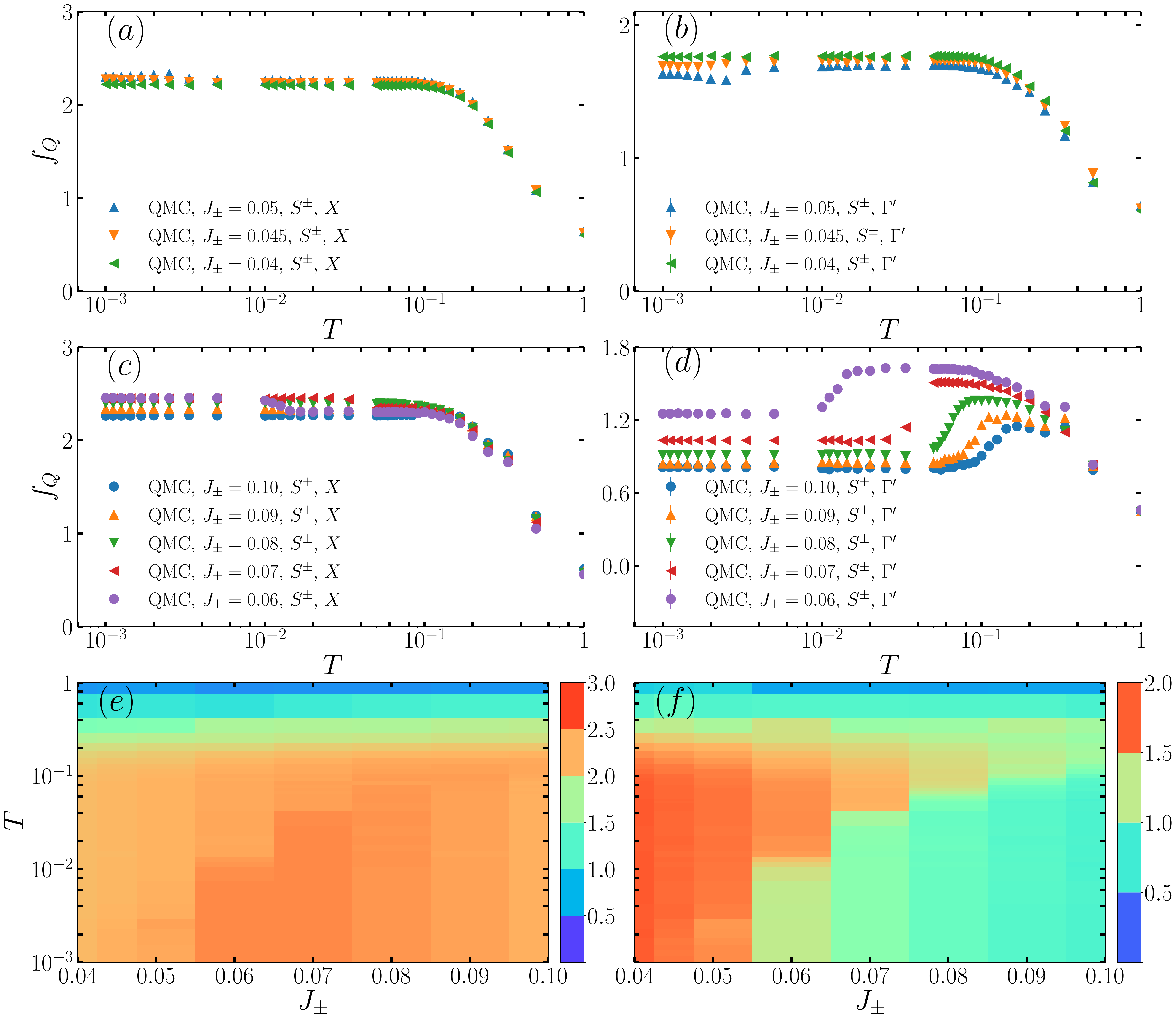}
	\caption{The QFI density $f_{Q}$ in $S^{\pm}$ channel (a, c, e) at $X$ and (b, d, f) at $\Gamma^\prime$ as a function of temperature $T$ and $J_{\pm}$. (a, b) illustrate $f_{Q}$ with $J_{\pm}$ ranging from $0.04$ to $0.05$ with an increment of $0.005$ while (c, d) is that of $J_{\pm}$ ranging from $0.06$ to $0.10$ with an increment of $0.01$. (e, f) are the QFI density heat maps as a function of temperature $T$ and $J_{\pm}$ ranging from $0.04$ to $0.10$.} 
	\label{fig:QFI_S2}
\end{figure}

Fig.~\ref{fig:QFI_S2} presents the QFI density $f_{Q}(S^{\pm}_{\mathbf{q}},T)$ at the $X$ and $\Gamma^\prime$ points. Panels (a) and (b) show $f_{Q}(S^{\pm}_{X},T)$ and $f_{Q}(S^{\pm}_{\Gamma^\prime},T)$ in the QSI regime, while panels (c) and (d) display the corresponding results in the FM phase. The heat maps in panels (e) and (f) illustrate $f_{Q}(S^{\pm}_{X},T)$ and $f_{Q}(S^{\pm}_{\Gamma^\prime},T)$ as functions of temperature $T$ and $J_{\pm}$ from $0.04$ to $0.10$. In combination with Fig.~\ref{fig:QFI_S1}, we find that $f_{Q}(S^{\pm}_{\Gamma},T)$ exhibits the strongest signal in the $S^{\pm}$ channel, while $f_{Q}(S^{\pm}_{\Gamma^\prime},T)$ captures the critical fluctuations. Therefore, we focus on these two momentum points in the main text and include the results at $X$ in the Supplemental Material for completeness.

\begin{figure}
	\centering
	\includegraphics[width=0.5\textwidth]{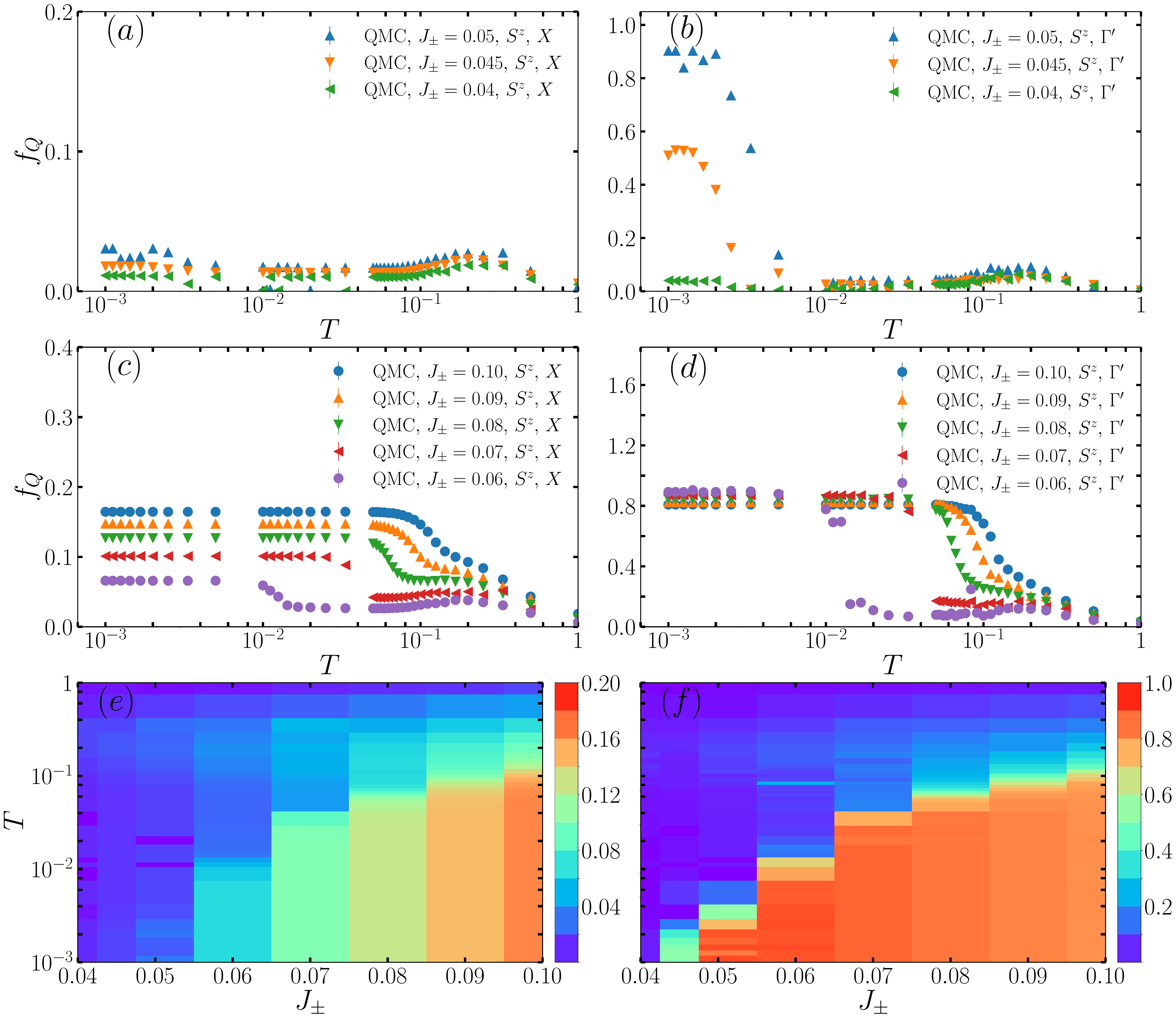}
	\caption{The QFI density $f_{Q}$ in $S^{\mathrm{z}}$ channel (a, c, e) at $X$ and (b, d, f) at $\Gamma^\prime$ as a function of temperature $T$ and $J_{\pm}$. (a, b) illustrate $f_{Q}$ with $J_{\pm}$ ranging from $0.04$ to $0.05$ with an increment of $0.005$ while (c, d) is that of $J_{\pm}$ ranging from $0.06$ to $0.10$ with an increment of $0.01$. (e, f) are the QFI density heat maps as a function of temperature $T$ and $J_{\pm}$ ranging from $0.04$ to $0.10$.}
	\label{fig:QFI_S3}
\end{figure}

Meanwhile, we also measure the QFI density $f_{Q}(S^{\mathrm{z}}_{\mathbf{q}},T)$ in the $S^{\mathrm{z}}$ channel, as shown in Fig.~\ref{fig:QFI_S3}. Panels (a) and (b) illustrate $f_{Q}(S^{\mathrm{z}}_{X},T)$ and $f_{Q}(S^{\mathrm{z}}_{\Gamma^\prime},T)$ in the QSI regime, while panels (c) and (d) show the corresponding results in the FM phase. The heat maps of $f_{Q}(S^{\mathrm{z}}_{X},T)$ and $f_{Q}(S^{\mathrm{z}}_{\Gamma^\prime},T)$ as functions of temperature $T$ and $J_{\pm}$ from $0.04$ to $0.10$ are presented in panels (e) and (f), respectively. 
In panel (f), the $f_{Q}(S^{\mathrm{z}}_{\Gamma^\prime},T)$ remains small for high temperature case, including the CSI regime. When the temperature is lowered upon entering the QSI regime, $f_{Q}(S^{\mathrm{z}}_{\Gamma^\prime},T)$ increases and reaches a plateau. This behavior is consistent with the fact that the $S^{\mathrm{z}}$ channel captures the photon sector which emerges in the QSI regime. However, compared to the $S^{\pm}$ channel (the spinon sector), the QFI density in the $S^{\mathrm{z}}$ channel captures a similar picture but with a much smaller value, which is why we only include these results in the Supplemental Material for completeness.

\begin{figure}
	\centering
	\includegraphics[width=\columnwidth]{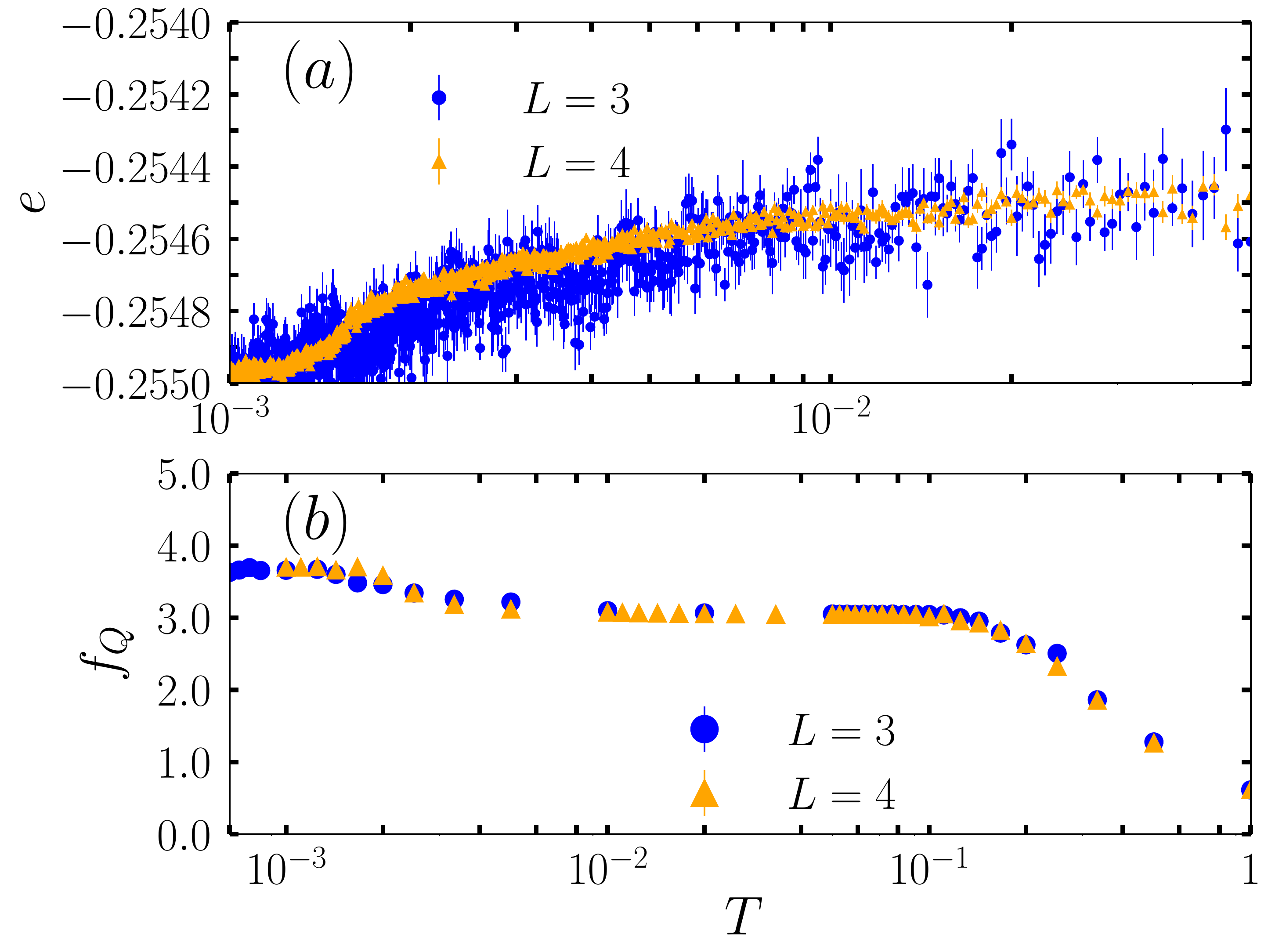}
	\caption{The comparison of the energy density ($e$) and density of QFI ($f_Q$) between the $L=3$ and $L=4$ obtained from the QMC simulations. The parameters are $J_\pm=0.045$ and $J_z=1$.}
	\label{fig:re02}
\end{figure}

Moreover, we compare the energy density $e$ and QFI density $f_Q$ between $L=3$ (without MDL) and $L=4$ (with MDL) QMC simulations, as shown in Fig.~\ref{fig:re02}. The parameters are $J_\pm=0.045$ and $J_z=1$. The results for these two system sizes are consistent, indicating that finite-size effects are not significant in our simulation.

\vspace{0.3cm}

\begin{figure*}
	\centering
	\includegraphics[width=\textwidth]{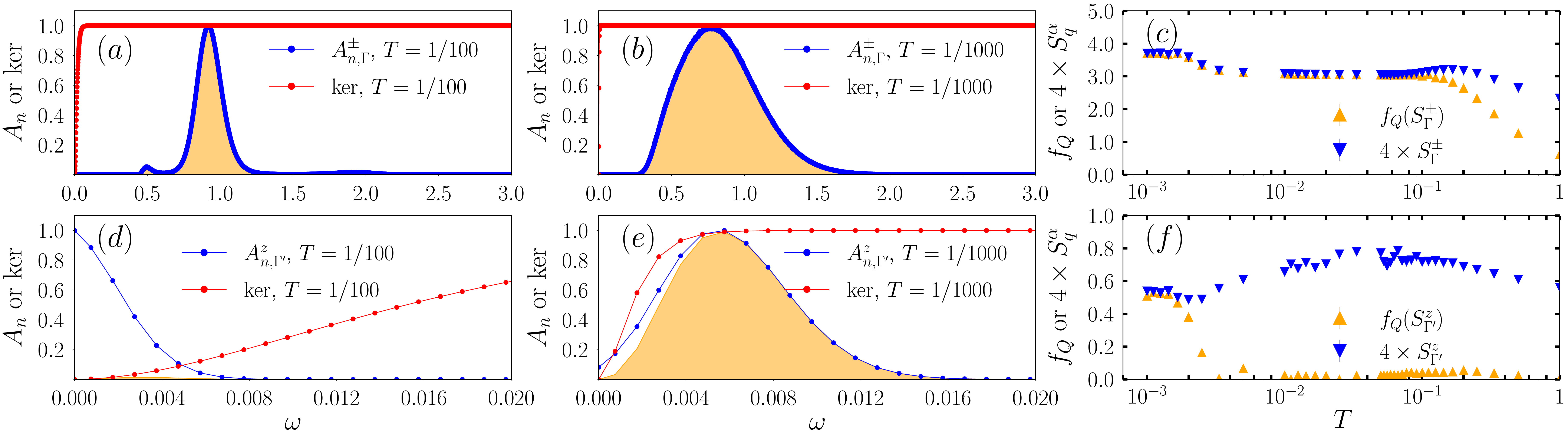}
	\caption{The comparison between QFI and static structure factor. Panel (a-c) shows the dynamical and static structure factor and the QFI for $J_\pm=0.045$ in $S^{\pm}$ channel at $\Gamma$ point. Panel (d-f) shows the same quantities for $J_\pm=0.045$ in the $S^{z}$ channel at the $\Gamma^\prime$ point. Panels (a) and (d) show the dynamical structure factor, normalized such that its highest value equals 1, at temperature $T=1/100$, while panels (b) and (e) show that at $T=1/1000$. In these four panels, the blue lines are the spectrums obtained from the QMC simulations, while the red lines are the corresponding kernel functions for the QFI calculation. And the orange shaded areas represent the contribution to the QFI. Panels (c) and (f) show the temperature dependence of the QFI (orange line), and the static structure factor (blue line) at the same momentum and channel.}
	\label{fig:re01}
\end{figure*}

Finally, we compare the QFI density with the static spin structure factor (SSSF). By defining the kernel function $\mathrm{ker}(\omega,T)$, the QFI density (Eq.~({\color{blue}4}) in the main text) can be rewritten as:
\begin{equation}
	\begin{aligned}
		f_{Q}(S^\alpha_\mathbf{q}, T) &= 4 \int_{0}^{\infty} d\omega\, A^\alpha(\mathbf{q}, \omega)\, \text{ker}(\omega,T), \\
		\text{ker}(\omega,T) &= \tanh \left(\frac{\omega}{2T}\right)\left(1-e^{- \omega/T}\right),
	\end{aligned}
	\label{eq:QFI}
\end{equation}
where the smooth thermal kernel $\mathrm{ker}(\omega,T)$ approaches unity only as $T\to 0$. This kernel function in QFI weights spectral contributions fundamentally differently from equal-time observables, such as the SSSF. The QFI kernel suppresses quasi-elastic thermal weight and highlights coherent quantum transitions, whereas the kernel for SSSF is simply unity and thus includes all spectral weight equally. This distinction is directly visible in our numerical results (see Fig.~\ref{fig:re01}), which compare QFI and SSSF for two representative cases at $J_\pm=0.045$: the $S^{\pm}$ channel at the $\Gamma$ point and the $S^{z}$ channel at the $\Gamma^\prime$ point. Panels (a-b) and (d-e) show the dynamical structure factor at two temperatures (blue curves from QMC), together with the corresponding QFI kernel (red curves). To illustrate the interplay between the dynamical structure factor and the kernel function, we have normalized the dynamical structure factor so that its maximum value equals 1, and the orange shaded regions indicate the spectral weight contributing to QFI. Panels (c) and (f) then show the temperature dependence of QFI (orange) and $4\times$SSSF (blue) for the $S^{\pm}$ channel at the $\Gamma$ point and for the $S^{z}$ channel at the $\Gamma^\prime$ point, respectively.
These results reveal the distinct temperature dependence between QFI and the $4\times$SSSF.

In the spinon-dominated case $O(\mathbf{q})=S^\pm_{\Gamma}$, panel (c) demonstrates that QFI and four times the $4\times$SSSF differ significantly in the high-temperature regime ($T>10^{-1}$), but converge in the low-temperature regime ($T<10^{-1}$). This behavior arises because the peak in the dynamical structure factor is located at $\omega\approx1$ [Fig.~\ref{fig:re01}(a-b)]. When $T$ is sufficiently low (roughly $T\lesssim 10^{-1}$), the kernel effectively encompasses the entire peak, causing QFI and $4\times$SSSF to become similar. At higher temperatures, however, the QFI kernel selectively reshapes which parts of the spectrum contribute, resulting in distinct temperature dependence between QFI and $4\times$SSSF. This illustrates that, despite the expected low-$T$ convergence, QFI and $4\times$SSSF remain fundamentally different in their sensitivity to spectral features across temperatures.

A sharper and physically most important distinction emerges in the photon sector. Choosing $O(\mathbf{q})=S^{z}_{\Gamma^\prime}$ probes emergent gauge (photon) excitations. Here, the relevant spectral peak lies at very low frequency, $\omega\approx 0.005$ [Fig.~\ref{fig:re01}(e)]. As a result, the temperature kernel ensures that QFI remains essentially zero over a broad temperature range and turns on only upon cooling into the photon energy window [Fig.~\ref{fig:re01}(f)], thereby cleanly capturing the onset of the photon mode, a defining hallmark of QSI. By contrast, the SSSF remains substantial throughout because it is dominated by thermally activated quasi-elastic low-frequency weight, and thus does not provide a comparably clean discriminator for the photon sector. This operator-resolved selectivity is precisely the added value of QFI: it isolates the onset of coherent spinonic and photonic modes and yields a direct, physically transparent diagnostic of the QSI regime that static equal-time probes can obscure.


\section{Exact Diagonalization}

The exact diagonalization scheme follows that of the microcanonical thermal pure quantum method (mTPQ)~\cite{thermal2012sugiura, canonical2013sugiura}. We first construct a state at $T\rightarrow\infty$, $|\psi_0\rangle=\sum_{i=1} c_i |i\rangle$, such that $\{c_i\}$ is a set of random complex number and normalized $\sum_i |c_i|^2=1$. Then for a given Hamiltonian $H$, the $k^{\text{th}}$ TPQ state is constructed iteratively via
\begin{equation}
    |\psi_k\rangle=(L-H)|\psi_{k-1}\rangle, \label{eq:mTPQ_iter}
\end{equation}
where $L$ is some constant value greater than the largest eigenvalue of the Hamiltonian. We can then estimate the energy and the inverse temperature via:
\begin{equation}
    E_k=\frac{\langle \psi_k |H|\psi_k\rangle}{\langle \psi_k |\psi_k\rangle}
\end{equation}
and
\begin{equation}
    \beta_k = \frac{2k}{L-E_k},
\end{equation}
for the $k^{\text{th}}$ iteration.

The microcanonical TPQ construction generates a sequence of pure states $\{|\psi_k\rangle\}$ whose energy
distribution concentrates sharply around a typical energy $E_k$ in the thermodynamic limit~\cite{thermal2012sugiura}.
In the iterative update of Eq.~\eqref{eq:mTPQ_iter}, successive applications of the filtering operator progressively
enhance the weight of lower-energy components, thereby lowering $E_k$ and scanning toward lower effective temperatures.
As shown by Sugiura and Shimizu, expectation values of few-body observables evaluated in a single TPQ state
self-average to the corresponding ensemble values, with relative fluctuations vanishing with increasing system size.
Consequently, finite-temperature observables can be obtained efficiently from
$\langle \psi_k|\mathcal{O}|\psi_k\rangle/\langle \psi_k|\psi_k\rangle$ at the $k$ corresponding to the desired
temperature, without explicitly summing over the partition function.

We compute finite-temperature spectra from an mTPQ representative state $|\psi_k\rangle$ using the broadened resolvent of a chosen operator $\mathcal{O}$~\cite{jaklic1994ftlm, prelovsek2013lanczoschapter},
\begin{equation}
    G_{\mathcal{O}}(z;\beta_k)
    =
    \frac{\langle \psi_k|\mathcal{O}^\dagger\,(z-\widetilde{H})^{-1}\mathcal{O}|\psi_k\rangle}
         {\langle \psi_k|\psi_k\rangle},
    \label{eq:resolvent_def}
\end{equation}
with $\widetilde{H}\equiv H-E$ and $z=\omega+i\eta$. Here $\eta>0$ controls the Lorentzian broadening, and the shift $E$ fixes the absolute frequency reference. We choose $E$ as the ground-state energy, obtained independently from a separate Lanczos run, so that the spectrum is reported with the correct energy offset. The associated spectral density follows as
\begin{equation}
    I_{\mathcal{O}}(\omega;\beta_k) = -\frac{1}{\pi}\,\mathrm{Im}\,G_{\mathcal{O}}(\omega+i\eta;\beta_k).
    \label{eq:spectral_def}
\end{equation}

To evaluate $G_{\mathcal{O}}$, we introduce the normalized Krylov seed and its weight,
\begin{equation}
    |f_0\rangle = \frac{\mathcal{O}|\psi_k\rangle}{\|\mathcal{O}|\psi_k\rangle\|},
    \qquad
    \mu_0 \equiv \frac{\|\mathcal{O}|\psi_k\rangle\|^2}{\langle \psi_k|\psi_k\rangle},
    \label{eq:seed_def}
\end{equation}
and perform Lanczos tridiagonalization of $\widetilde{H}$ starting from $|f_0\rangle$. This produces the usual tridiagonal representation specified by diagonal elements $\{a_n\}$ and off-diagonal elements $\{b_{n+1}\}$~\cite{gagliano1987,hallberg1995,prelovsek2013lanczoschapter}.

In the resulting Krylov basis, the matrix element of the resolvent admits the standard continued-fraction approximation~\cite{gagliano1987,hallberg1995},
\begin{equation}
    G_{\mathcal{O}}(z;\beta_k)
    \approx
    \mu_0\,
    \frac{1}{z-a_0-\cfrac{b_1^2}{z-a_1-\cfrac{b_2^2}{z-a_2-\ddots}}},
    \label{eq:continued_fraction}
\end{equation}
which we truncate after $n_{\mathrm{L}}$ Lanczos iterations. Combining Eqs.~\eqref{eq:spectral_def} and \eqref{eq:continued_fraction} then yields $I_{\mathcal{O}}(\omega;\beta_k)$, giving us the desired spectral function. We choose a Krylov dimension of 800 for careful convergence of the spectral function.

\begin{figure}
    \centering
    \includegraphics[width=0.9\linewidth]{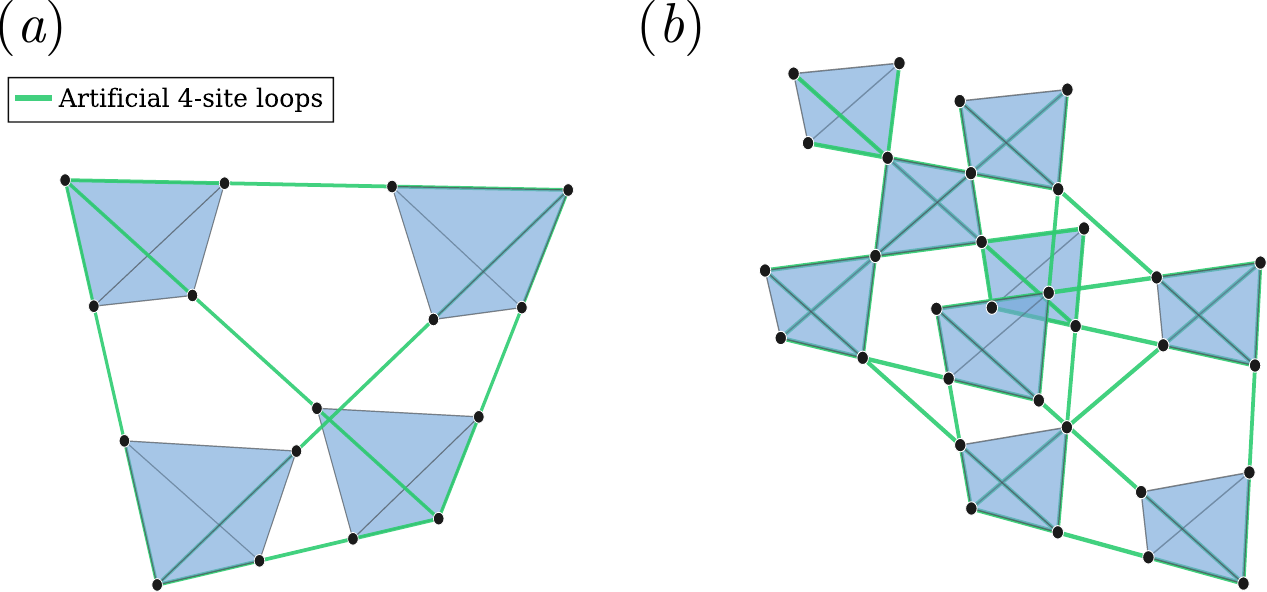}
    \caption{(a) 16 sites cubic cluster setup for exact diagonalization calculation (b) 32 sites $2\times2\times2$ cluster geometry. highlighted green edges are examples of the spurious 4-site loop enabled by finite size effects}
    \label{fig:ED_geometry}
\end{figure}

It is important to distinguish two separate sources of uncertainty in our finite-$T$ mTPQ calculations: (i) physical finite-size effects associated with working on a finite, symmetry-respecting pyrochlore cluster, and (ii) statistical (typicality) fluctuations associated with estimating thermal quantities using a finite number of random initial vectors. The mTPQ method provides an unbiased estimator of the finite-$N$ canonical (or microcanonical) ensemble, but it does not remove physical finite-size effects, which can remain significant at low temperature when the low-energy spectrum is sparse.

For the pyrochlore lattice, $N=16$ is the largest symmetry-respecting cluster that is feasible within our present finite-temperature dynamical workflow. The next symmetry-compatible cluster has $N=32$ sites and is numerically intractable here. More importantly, the dominant finite-size artifact in the quantum spin-ice regime is not a smooth $1/N$ correction, but a geometry-enabled process tied to the short linear extent of these clusters: periodic boundary conditions permit an additional four-site exchange within the spin-ice manifold along the shortest edges. This modifies the effective low-energy dynamics by enhancing the characteristic photon scale to order $9J_\pm^2/J_{zz}$, instead of the thermodynamic-limit ring-exchange scale of order $12J_\pm^3/J_{zz}^2$~\cite{gao2026spectroscopicdemarcationemergentphotons,ross2011quantum, benton2012seeing, Pyrochlore2004Hermele, savary2012coulombic}. As shown in Fig.~\ref{fig:ED_geometry}, this mechanism is a property of the cluster geometry
(short edges spanning two tetrahedra) and therefore, is not expected to be removed simply by moving to the next
symmetry-respecting cluster, where the same short-edge structure persists, and such processes remain allowed. The main consequence is a renormalization of the absolute photonic scale and hence a shift of the photon-related crossover temperature on finite clusters.

Despite this known artifact, the quantities we focus on are still well-motivated. Our interpretation of QFI is based on the hierarchy and redistribution of spectral weight in $S(\mathbf{k},\omega;T)$ across spinon- and photon-dominated frequency windows. In particular, QFI is a frequency integral of the dynamical response with a smooth, temperature-dependent kernel, so the qualitative crossover structure is controlled primarily by where spectral weight resides relative to the relevant excitation scales, rather than by precise ground-state convergence or microscopic level spacings at finite $N$. We therefore use the $N=16$ cluster as the largest symmetry-respecting finite-$T$ reference and benchmark the resulting spectral hierarchy and QFI trends against complementary thermodynamic-limit approaches discussed in the main text.

\begin{figure}
    \centering
    \includegraphics[width=\linewidth]{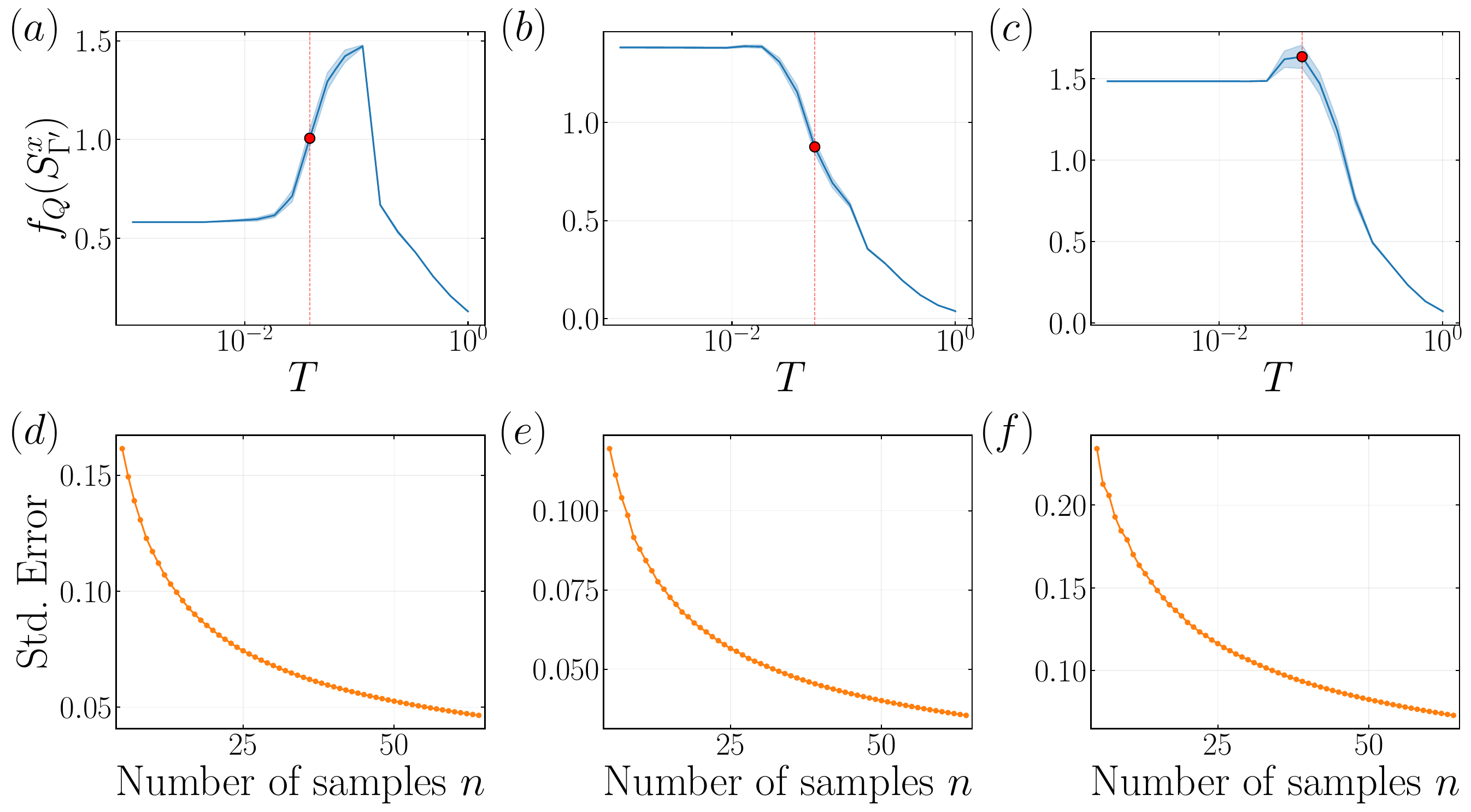}
    \caption{Quantum Fisher information density $f_Q(S^x_{\Gamma'})$ for $J_\pm=0.05$ (a), $J_\pm=-0.15$ (b), $J_\pm=-0.3$ (c) as a function of time averaged over 64 samples. The shaded blue region denotes the standard error, and the red dots highlight the temperature associated with the largest error. We then perform their respective convergence analysis at the temperature with the largest error for $J_\pm=0.05,-0.15,-0.3$ respectively in (d,e,f).}
    \label{fig:QFI_conv_supp}
\end{figure}

Separately, typicality fluctuations can increase at low temperature and must be controlled independently of physical finite-size effects. We therefore average over 64 random-vector realizations and quantify the resulting spread. To this end, we have performed rigorous convergence as shown in Fig.~\ref{fig:QFI_conv_supp}. The disorder-averaged QFI converges as $\sigma/\sqrt{n}$, and with $n = 64$ realizations the standard error of the mean remains below $1\%$ of the signal at low temperatures and below ${\sim}5\%$ even at the worst-case temperature.

\begin{figure}
	\centering
	\includegraphics[width=\columnwidth]{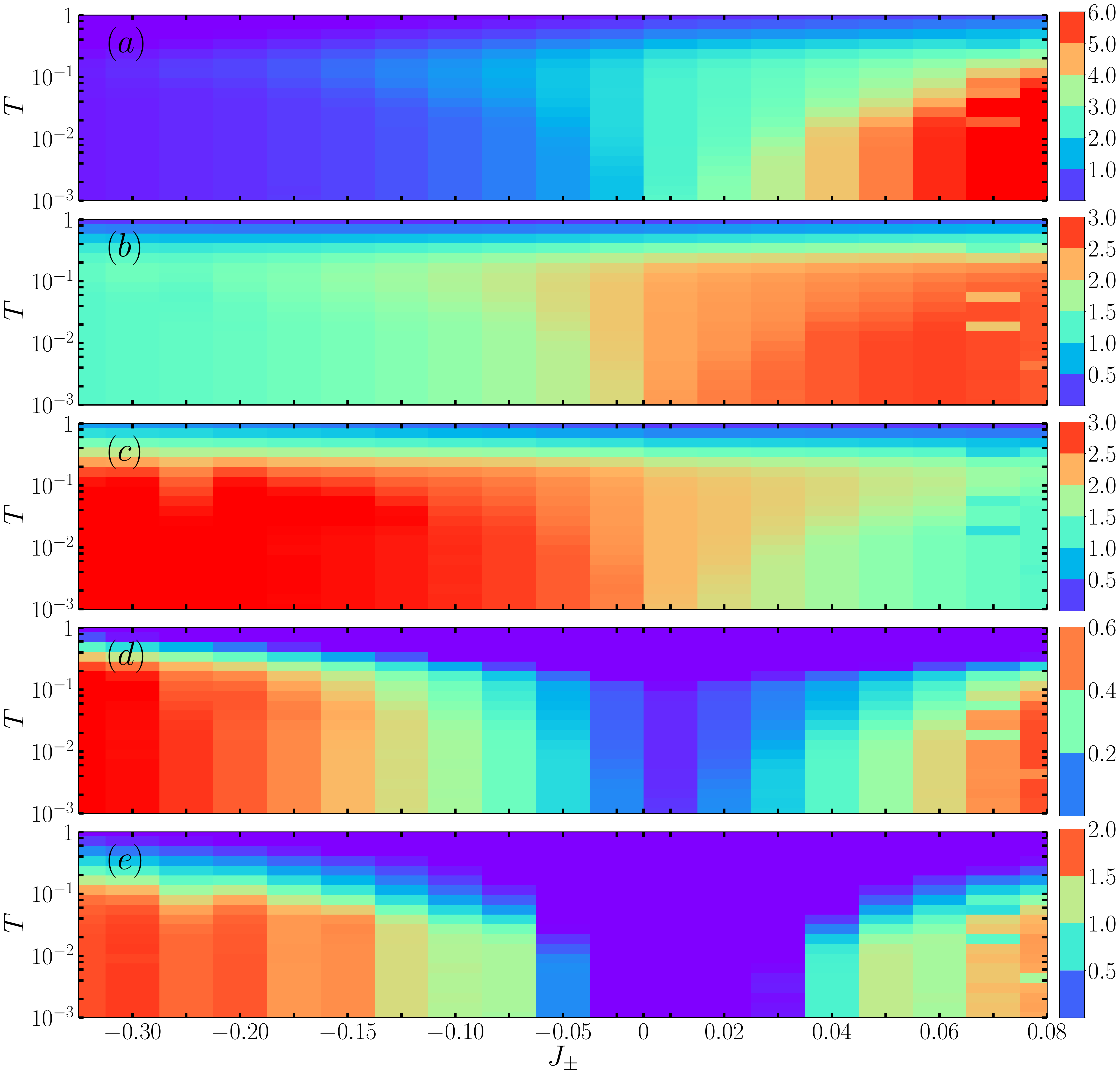}
	\caption{The ED calculation results of the QFI density $f_{Q}$ (a-c) in $S^{\pm}$ channel and (d-e) in $S^{\mathrm{z}}$ channel as a function of temperature $T$ and $J_{\pm}$. (a) illustrates $f_{Q}$ at $\Gamma=(0,0,0)$ while (b) is that at $X$ and (c) at $\Gamma^\prime=(4\pi,4\pi,0)$. (d) tells $f_{Q}$ at $X$ while (e) is that at $\Gamma^\prime=(4\pi,4\pi,0)$.}
	\label{fig:ed_res}
\end{figure}

Besides the results shown in the main text, we also include the $S^\pm$ channel at the $X$ point as well as the $S^\mathrm{z}$ channel in Fig.~\ref{fig:ed_res}. We can also clearly observe the delineation of the two temperature scales from the computed QFI values, as we discussed in the main text. Furthermore, we would like to highlight the increasing magnitude $f_Q(S^\mathrm{z}_\mathbf{q})$ as we move towards the Heisenberg point $J_\pm=-0.5$. QFI also encapsulates the total fluctuation generated by $e^{i\theta O}$ via some operator $O$. In the $U(1)$ lattice gauge theory description, $S^z\sim E$, the canonical electric field. Therefore, at the lowest temperature, $f_Q(S^\mathrm{z}_\mathbf{q})\sim\langle EE\rangle$ can be thought of as some measure of the total gauge fluctuation. 

\section{Gauge Mean Field Theory}

We now provide a detailed account of the gauge mean-field theory (GMFT) formalism, which offers a canonical mapping of the spin Hamiltonian to that of a lattice $U(1)$ gauge theory coupled to some bosonic matter field. To keep the presentation transparent, let us first focus on the regime where the coupling $J_\mathrm{z}$ is the dominant interaction, i.e. $|J_\mathrm{z}|>|J_{xx}|=|J_{yy}|$, so that we may set $J_{\mathrm{z}}=J_\mathrm{z}$.

In this formulation, one introduces a slave ``charge" degree of freedom on the sites of the parent diamond lattice, defined as
\begin{align}
Q_{\mathbf{r}_\alpha} = \sum_{\mu \in \partial t_{\mathbf{r}_\alpha}} S_{\mathbf{R}_\mu}^\text{z},
\end{align}
where $t_{\mathbf{r}_\alpha}$ denotes the tetrahedron centered at $\mathbf{r}_\alpha$, and $\partial t_{\mathbf{r}_\alpha}$ are the four pyrochlore spins forming its, $\mu$ denotes the sublattice indices. The index $\alpha\in{A,B}$ distinguishes the two diamond sublattices (up- and down-pointing tetrahedra). Notice we make an explicit distinction between the sublattice-indexed spin coordinates, denoted by $\mathbf{R}_\mu$, where $\mu\in\{0,1,2,3\}$ represents the sublattice index, and the parent diamond lattice, denoted by the lower case $\mathbf{r}_\alpha$. These two coordinate systems are related by:
\begin{align}
     \mathbf{R}_{\mu}  &= \mathbf{r}_{\alpha} + \eta_{\alpha}\mathbf{b}_{\mu}/2 \label{eq:SIPC_SIDC}
\end{align}
where $\mathbf{b}_\mu$ connects A-sublattice sites to their four nearest B-sublattice neighbors and  $\eta_A=1$ and $\eta_B=-1$.

\begin{subequations}
\begin{align}
    \mathbf{b}_0 &= -\frac{1}{4}(1,1,1)\\
    \mathbf{b}_1 &= \frac{1}{4}(-1,1,1)\\
    \mathbf{b}_2 &= \frac{1}{4}(1,-1,1)\\
    \mathbf{b}_3 &= \frac{1}{4}(1,1,-1). \label{eq:bmu}
\end{align}
\end{subequations}
By canonical construction, the conjugate variable $\varphi_{\mathbf{r}_\alpha}$ obeys $[\varphi_{\mathbf{r}_\alpha}, Q_{\mathbf{r}\alpha'}] = i\delta_{\mathbf{r}_\alpha\mathbf{r}_{\alpha'}}$, allowing one to define bosonic (spinon) raising and lowering operators
\begin{align}
\Phi^\dagger_{\mathbf{r}_\alpha} = e^{i\varphi_{\mathbf{r}_\alpha}}, \qquad \Phi_{\mathbf{r}_\alpha} = e^{-i\varphi_{\mathbf{r}_\alpha}}.
\end{align}
The original pseudospin operators can now be re-expressed in the enlarged Hilbert space $\mathcal{H}=\mathcal{H}_Q \otimes \mathcal{H}_{\text{spin}}$, where the mapping is given by
\begin{align}
S_{\mathbf{R}_\mu}^{+} &\rightarrow \Phi^\dagger_{\mathbf{r}_A} \left(\tfrac{1}{2}e^{iA_{\mathbf{r}_A, \mathbf{r}_A+\mathbf{b}_\mu}}\right)\Phi_{\mathbf{r}_A+\mathbf{b}_\mu}, \\
S_{\mathbf{R}_\mu}^{\mathrm{z}} &\rightarrow E_{\mathbf{r}_A, \mathbf{r}_A+\mathbf{b}_\mu},
\end{align}
with $A_{\mathbf{r}_A, \mathbf{r}_A+\mathbf{b}_\mu}$ and $E_{\mathbf{r}_A, \mathbf{r}_A+\mathbf{b}_\mu}$ denoting conjugate gauge and electric fields on the diamond links.

Physically, this construction makes explicit the emergent gauge structure inherent in the spin-ice manifold: spin flips map to spinon matter hopping minimally coupled to compact $U(1)$ gauge fields.

The resulting Hamiltonian contains quadratic spinon charge terms and spinon hopping terms ($J_\pm$), yielding
\begin{align}\label{eq:H_GMFT_FULL}
\mathcal{H}= & \frac{J_{\mathrm{z}}}{2} \sum_{\mathbf{r}_\alpha} Q_{\mathbf{r}_\alpha}^2-\frac{J_{ \pm}}{4} \sum_{\mathbf{r}_\alpha} \sum_{\mu, \nu \neq \mu} \Phi_{\mathbf{r}_\alpha+\eta_\alpha \mathbf{b}_\mu}^{\dagger} \Phi_{\mathbf{r}_\alpha+\eta_\alpha \mathbf{b}_\nu} \nonumber \\
&e^{i \eta_\alpha\left(A_{\mathbf{r}_\alpha, \mathbf{r}_\alpha+\eta_\alpha \mathbf{b}_\nu}-A_{\mathbf{r}_\alpha, \mathbf{r}_\alpha+\eta_\alpha \mathbf{b}_\mu}\right)},
\end{align}
where $J_\pm = -(J_{xx}+J_{yy})/4$.

At this stage, we adopt two crucial approximations: (i) the electric field $E$ is integrated out, leaving behind purely bosonic matter fields coupled to static background fluxes; and (ii) the gauge field $A$ is frozen to its mean-field value $\bar{A}$, thereby neglecting dynamical gauge fluctuations. These simplifications transform the model into a tractable quadratic bosonic theory, whose self-consistent solution captures the stability of the $U(1)$ quantum spin liquid. Evaluating the Hamiltonian at the spin ice manifold where $Q=0$, we obtained that:

\begin{align}\label{eq:H_GMFT_FULL}
\mathcal{H}= & -\frac{J_{ \pm}}{4} \sum_{\mathbf{r}_\alpha} \sum_{\mu, \nu \neq \mu} \Phi_{\mathbf{r}_\alpha+\eta_\alpha \mathbf{b}_\mu}^{\dagger} \Phi_{\mathbf{r}_\alpha+\eta_\alpha \mathbf{b}_\nu} e^{i \eta_\alpha\left(\bar{A}_{\mathbf{r}_\alpha, \mathbf{r}_\alpha+\eta_\alpha \mathbf{b}_\nu}-\bar{A}_{\mathbf{r}_\alpha, \mathbf{r}_\alpha+\eta_\alpha \mathbf{b}_\mu}\right)}. 
\end{align}

To compute QFI, we need to calculate the dynamical spin susceptibility. Under the GMFT formalism, since the photonic mode is integrated out, we only have access to the spinon dynamics. In other words, we only have access to $\langle S^+ S^-\rangle \sim \langle \Phi^\dagger \Phi \Phi^\dagger\Phi\rangle$. In particular, we can compute the spin structure factor via the following formalism:

\begin{align}
&S^{+-}_{\mu\nu}(\mathbf{q},\omega)
=\frac{1}{N}\sum_{\mathbf{R}_\mu,\mathbf{R}'_\nu}e^{i\mathbf{q}\cdot(\mathbf{R}_\mu-\mathbf{R}'_\nu)}
\int dte^{i\omega t}
\langle S^+_{\mathbf{R}_\mu}(t)S^-_{\mathbf{R}'_\nu}(0)\rangle\nonumber\\
&=\frac{1}{N} \sum_{\mathbf{r}_{A}, \mathbf{r}_A'} \sum_{\mu, \nu} e^{i\mathbf{q}\cdot\left(\mathbf{r}_{A} - \mathbf{r}_{A}' + (\mathbf{b}_{\mu} - \mathbf{b}_{\nu})/2 \right)} \nonumber \\&\int dt e^{i\omega t} \frac{1}{4} \left\langle{\Phi^{\dagger}_{\mathbf{r}_A}(t) e^{i \overline{A}_{\mathbf{r}_A,\mathbf{r}_A+\mathbf{b}_\mu}} \Phi_{\mathbf{r}_A+\mathbf{b}_\mu}(t) \Phi^{\dagger}_{\mathbf{r}_A'+\mathbf{b}_\nu}(0) e^{-i \overline{A}_{\mathbf{r}_A',\mathbf{r}_A'+\mathbf{b}_\nu}} \Phi_{\mathbf{r}_A'}(0)  }\right\rangle
\end{align}
where $N$ is the number of spins. We can then evaluate this using Wick's contraction. Within the XXZ model, it turns out that the only non-trivial term is:

\begin{align}
&S^{+-}_{\mu\nu}(\mathbf{q},\omega)
=\frac{1}{N} \sum_{\mathbf{r}_{A}, \mathbf{r}_A'} \sum_{\mu, \nu} e^{i\mathbf{q}\cdot\left(\mathbf{r}_{A} - \mathbf{r}_{A}' + (\mathbf{b}_{\mu} - \mathbf{b}_{\nu})/2 \right)} \nonumber \\&\int dt e^{i\omega t} \frac{F_{\mu\nu} }{4} \left\langle  \Phi_{\mathbf{r}_A+\mathbf{b}_\mu}(t) \Phi^{\dagger}_{\mathbf{r}_A'+\mathbf{b}_\nu}(0)\right\rangle\left\langle \Phi^{\dagger}_{\mathbf{r}_A}(t) \Phi_{\mathbf{r}_A'}(0)  \right\rangle,
\end{align}
where $F_{\mu\nu} = e^{i \left(\bar{A}_{\mathbf{r}_A, \mathbf{r}_A+\mathbf{b}_\nu}-\bar{A}_{\mathbf{r}_A, \mathbf{r}_A+\mathbf{b}_\mu}\right)}$. In other words, we can compute the dynamical spin susceptibility by evaluating the Green's function of the spinons. We can then integrate this to find the QFI of the spinon channel at zero temperature.

\begin{figure}
    \centering
    \includegraphics[width=1\linewidth]{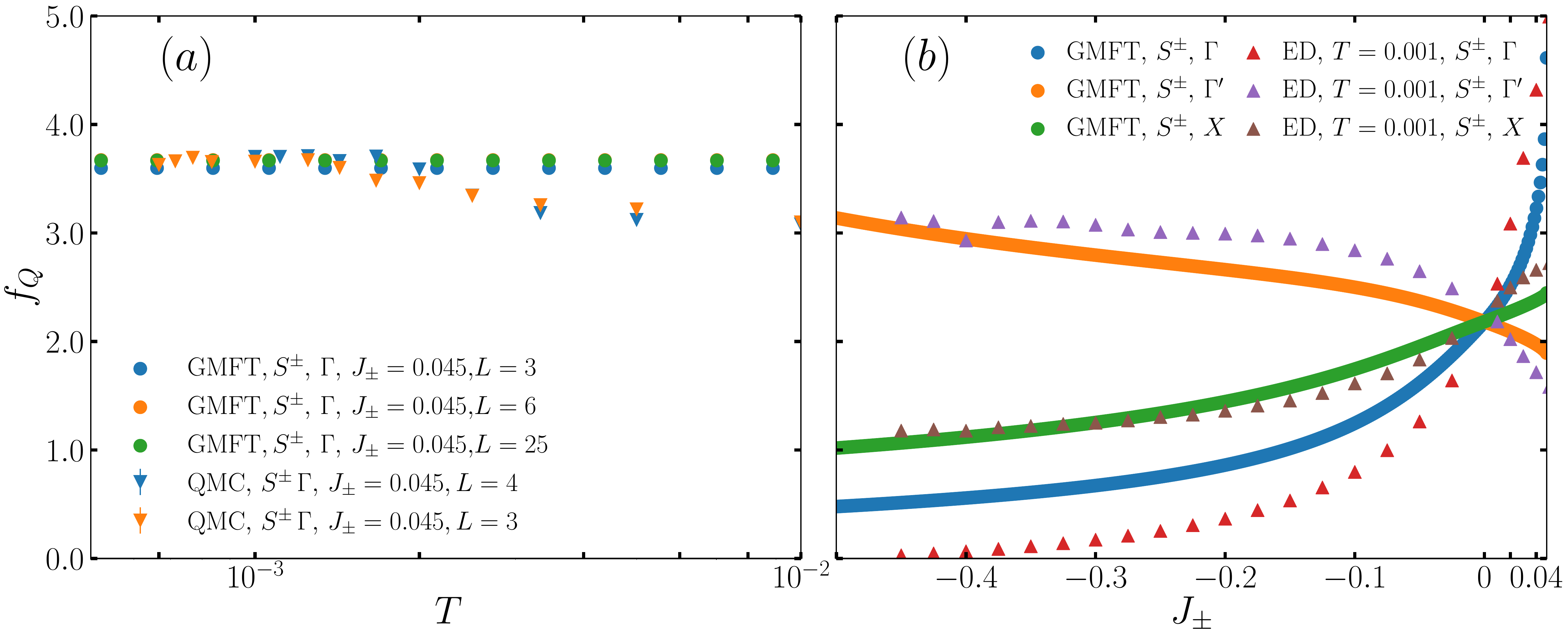}
    \caption{The GMFT calculation results of the QFI density $f_{Q}$ in the $S^{\pm}$ channel. Panel (a) shows the comparison between GMFT results (with a factor $6/7$) for various system sizes from $L=3$ to $L=25$ at $J_{\pm}=0.045$, and the QMC results for $L=3$ and $L=4$. Panel (b) shows the QFI density as a function of $J_{\pm}$ at zero temperature in the $S^{\pm}$ channel at $\mathrm{\Gamma}$, $\mathrm{\Gamma^\prime}$, and $\mathrm{X}$.}
    \label{fig:GMFT}
\end{figure}

We show the QFI computed from GMFT at zero temperature in Fig.~\ref{fig:GMFT}. Despite having made drastic assumptions in the theory, the results of GMFT actually align quite well with the QMC results (up to a factor of approximately $6/7$).

\section{Neutron Scattering Cross Section of dipolar-octupolar material candidates}

The neutron scattering cross section is obtained via Fermi's golden rule, and after tracing over all the sample spin degrees of freedom, we obtained that:
\begin{equation}
\frac{d^2\sigma}{d\Omega d\omega} \propto \frac{k_f}{k_i} \sum_{\alpha\beta}\left(\delta_{\alpha\beta}-\hat{\mathbf{q}}_\alpha\cdot \hat{\mathbf{q}}_\beta\right) M^{\alpha\beta}(\mathbf{q},\omega), \label{eq:NS_CS}
\end{equation}
where  $k_f$ ($k_i$) is the norm of the momentum for the outgoing (incoming) neutron (in what follows, we will drop the kinetic prefactor by assuming $k_f/k_i\approx 1$ for the experiment of interest), and 
\begin{equation}
    M^{\alpha\beta}(\mathbf{q},\omega) = \frac{1}{2\pi} \int dt\langle M^\alpha(\mathbf{-q}, t)M^\beta(\mathbf{q}, 0)  \rangle e^{i\omega t}. \label{eq:mab}
\end{equation}
Here $M^\alpha(\mathbf{q},\omega)$ is the Fourier transformed magnetization density operator.

In the case of DO pyrochlore compounds, the lowest lying crystal electric field doublet $\ket{\pm}$ can be modeled using pseudospins operators $\tau^z=\frac{1}{2}(|+\rangle\langle+|-|-\rangle\langle-|)$ and $\tau^{ \pm}=| \pm\rangle\langle\mp|$ that transform non-trivially under point group operation~\cite{rau2019frustrated, huang2014quantum}. In the case of Cerium-based candidates, only the pseudospin component $\tau^z$ has a dipolar magnetic charge density and linearly couples with the magnetic field (e.g., $\ket{\pm}=\ket{J=7/2,m_J=\pm 3/2}$ for Ce$_2$Zr$_2$O$_7$~\cite{gaudet2019quantum, gao2019experimental, smith2024experimental}).
The pseudospin components $\tau^z$ are defined with respect to the local $z$-axis, as determined by crystal electric field analysis. These local $ z$-axes vary according to the tetrahedron sublattice index and are given by
\begin{subequations}
\begin{align}
    \mathbf{z}_0 &= \frac{1}{\sqrt{3}}(1,1,1)\\
    \mathbf{z}_1 &= \frac{1}{\sqrt{3}}(1,-1,-1)\\
    \mathbf{z}_2 &= \frac{1}{\sqrt{3}}(-1,1,-1)\\
    \mathbf{z}_3 &= \frac{1}{\sqrt{3}}(-1,-1,1).
\end{align}
\end{subequations}
Therefore, in the dipolar approximation, the magnetization density operator is:
\begin{equation}
\begin{aligned}
    M^\alpha(\mathbf{q}, t) &\propto \frac{1}{\sqrt{N}} \sum_{\mathbf{R}_\mu} e^{i\mathbf{q}\cdot\mathbf{R}_\mu} \hat{\mathbf{z}}_\mu^\alpha \tau^z_{\mathbf{R}_\mu}(t)\\
    & =  \sum_{\mu} \hat{\mathbf{z}}_\mu^\alpha \tau^z_{\mathbf{q},\mu}(t),
\end{aligned}
\end{equation}
where we have defined $\tau^z_{\mathbf{q},\mu}(t) = \frac{1}{\sqrt{N}}\sum_{\mathbf{R}_\mu} e^{i\mathbf{q}\cdot \mathbf{R}_\mu} \tau^z_{\mathbf{R}_\mu}(t)$ via the standard Fourier transform.

Pluging this into Eq.~\eqref{eq:NS_CS} and ~\eqref{eq:mab}:
\begin{align}
\frac{d^2\sigma}{d\Omega d\omega} &\propto \frac{1}{N}\sum_{\alpha\beta \mu \nu}\left(\delta_{\alpha\beta}-\hat{\mathbf{q}}_\alpha\cdot \hat{\mathbf{q}}_\beta\right) (\hat{\mathbf{z}}_\mu)^\alpha(\hat{\mathbf{z}}_\nu)^\beta \langle\tau^{z}_{-\mathbf{q},\mu}(t)\tau^z_{\mathbf{q},\nu}(0)\rangle \nonumber \\&=\frac{1}{N}
\sum_{\mu \nu} \left(\mathbf{\hat{z}}_\mu \cdot \mathbf{\hat{z}}_\nu -\frac{\left(\mathbf{\hat{z}}_\mu \cdot \mathbf{Q}\right)\left( \mathbf{\hat{z}}_\nu \cdot \mathbf{Q}\right)}{|\mathbf{Q}|^2}\right)\langle\tau^{z}_{-\mathbf{q},\mu}(t)\tau^z_{\mathbf{q},\nu}(0)\rangle \label{eq:ins_cs_tau}  \\
&:= A^{\text{DO}} (\mathbf{q}, t). \label{eq:A_DO}
\end{align}
We can then write out explicitly the transverse projector in a DO pyrochlore system for the $\langle \tau^z \tau^z\rangle$ channel as:
\begin{equation}
    \mathcal{F}_{\mu\nu} := \mathbf{\hat{z}}_\mu \cdot \mathbf{\hat{z}}_\nu -\frac{\left(\mathbf{\hat{z}}_\mu \cdot \mathbf{Q}\right)\left(\mathbf{\hat{z}}_\nu \cdot \mathbf{Q}\right)}{|\mathbf{Q}|^2}
\end{equation}
In this work, we want to compute experimentally relevant QFI for DO compounds, which are generally described by an XYZ Hamiltonian with a finite $\tilde{J}_{xz}$:
\begin{equation}
    \mathcal{H}_{\text{DO}} = \sum_{\langle i, j\rangle} T_{\alpha\alpha} \tau_i^\alpha \tau_j^\alpha + T_{xz} (\tau^x_i \tau^z_j + \tau^z_i \tau^x_j) - g \mu_B \sum_i \mathbf{B}_i \cdot \tau_i^z \label{eq:DO_Ham},
\end{equation}
where $\alpha\in\{x,y,z\}$
We can remove the $T_{xz}$ term by applying a rotation along the y-axis: $\tau^y = \tilde{S}^y$; $\tau^x = \cos(\theta)\tilde{S}^x - \sin(\theta)\tilde{S}^z$; $\tau^z = \sin(\theta)\tilde{S}^x + \cos(\theta)\tilde{S}^z$; $\tan(2\theta)=\frac{2T_{xz}}{T_{xx}-T_{zz}}$. By doing so, we map this problem to the XYZ model:
\begin{equation}
    \mathcal{H}_{\text{XYZ}} = \sum_{\langle i, j\rangle} \tilde{J}_{\alpha\alpha} \tilde{S}_i^\alpha \tilde{S}_j^\alpha  - g \mu_B \sum_i \mathbf{B}_i \cdot (\tilde{S}_i^z\cos\theta + \tilde{S}_i^x\sin\theta).
\end{equation}
Finally, we cover the XXZ model described in Eq.~({\color{blue}1}) of the main text when under the simplifying assumption that $\tilde{J}_{\alpha\alpha}=\tilde{J}_{\beta\beta} = -2J_{\pm} < \tilde{J}_{\gamma\gamma}=J_{\mathrm{z}}$ with one dominant component, where $\alpha$, $\beta$, $\gamma$ are the 3 pseudospin components. In doing so, we map $S^z=\tilde{S}^\gamma$ and the other two components to be $x$ and $y$ in Eq.~({\color{blue}1}) of the main text. Let us define 
\begin{equation}
    C^{\alpha\beta}_{\mu\nu}(\mathbf{q},\omega) = \frac{1}{2\pi}\int dt e^{i\omega t} \langle S^\alpha_\mathbf{-q,\mu}(t) S^\beta_\mathbf{q,\nu}(0)\rangle
\end{equation}
In the case of Ce$_2$Zr$_2$O$_7$, the proposed parameter sets are $(0.062, 0.063, 0.011)\text{meV}$ and $(0.063, 0.062, 0.011)\text{meV}$ with a potentially small $\theta\lesssim0.1\pi$. These two sets have equally excellent goodness of fit~\cite{smith2023quantum}. As such, depending on whether if $J_{xx}$ or $J_{yy}$ is dominant, $\tilde{S}^z=(S^+-S^-)/2i$ or $\tilde{S}^z=(S^++S^-)/2$ as either $\tilde{S}^x$ or $\tilde{S}^y$ becomes the dominant longitudal component. Therefore, in the case of Ce$_2$Zr$_2$O$_7$:
\begin{align}
    A^{\text{DO}}(\mathbf{q},\omega) &\sim \sum_{\mu,\nu} \frac{\mathcal{F}_{\mu\nu}}{4} [ \left(C^{+-}_{\mu\nu}(\mathbf{q},\omega) + C^{-+}_{\mu\nu}(\mathbf{q},\omega)\right) \pm  \nonumber\\ 
    &\quad\left( C^{++}_{\mu\nu}(\mathbf{q},\omega) + C^{--}_{\mu\nu}(\mathbf{q},\omega)\right)],
\end{align}
where the $\pm$ sign depends on whether if $\tilde{S}^{x}$ or $\tilde{S}^y$ is dominant. Under the XXZ model, the latter two terms vanish as a consequence of the $U(1)$ symmetry of the Hamiltonian. As such, to compute $A^{\text{DO}}$:
\begin{align}
    A^{\text{DO}}(\mathbf{q},\omega) 
    &= \sum_{\mu,\nu} \mathcal{F}_{\mu\nu}\left(C^{+-}_{\mu\nu}(\mathbf{q},\omega) + C^{-+}_{\mu\nu}(\mathbf{q},\omega)\right)/4.
\end{align}

\section{Quantum Fisher Information Matrix}

\subsection{Review of the Quantum Fisher Information}

For this appendix to be as self-contained as possible, we will first review some basic facts about the QFI before discussing more specifically its relation to spin structure factors and neutron scattering cross sections.

For a differentiable family of quantum states $\rho_\theta = e^{-i\theta O}\rho e^{i\theta O}$ generated by a Hermitian operator $O$, the quantum Fisher information (QFI) is defined via the symmetric logarithmic derivative (SLD) $L$ as~\cite{liu2016quantum}
$$
F_Q[\rho,O] \equiv \mathrm{Tr}\left(\rho L^2\right), 
\qquad
\partial_\theta \rho_\theta\big|_{\theta=0}=\tfrac{1}{2}\{\rho,L\}=-i[O,\rho].
$$
In eigenbasis $\rho=\sum_n p_n |n\rangle\langle n|$, the QFI admits the Lehmann spectral representation
$$
F_Q[\rho,O]
= 2\sum_{m,n}\frac{(p_m-p_n)^2}{p_m+p_n}\big|\langle m|O|n\rangle\big|^2,
$$
where terms with $p_m+p_n=0$ are omitted. For pure states $\rho=|\psi\rangle\langle\psi|$, this reduces to
$$
F_Q[|\psi\rangle,O]=4\Big(\langle\psi|O^2|\psi\rangle-\langle\psi|O|\psi\rangle^2\Big)=4\mathrm{Var}_{\psi}(O).
$$
For thermal states $\rho \propto e^{-\beta H}$, the same expression holds with $p_n=e^{-\beta E_n}/Z$ and the $\{|n\rangle\}$ chosen as the energy eigenbasis of $H$.

A pure $N$-particle state is called $m$-producible if it factorizes into a tensor product of blocks, each containing at most $m$ parties; a mixed state is $m$-producible if it is a convex mixture of such pure states. The entanglement depth is the smallest $m$ for which the state is $m$-producible. For a generator that is a sum of local terms,
\[
\mathcal{O}=\sum_{i=1}^{N} \mathcal{O}_i,\qquad 
\Delta_i \equiv \lambda_{\max}(\mathcal{O}_i)-\lambda_{\min}(\mathcal{O}_i),
\]
where $\Delta_i$ is the spectral width of $\mathcal{O}_i$ and $\lambda_{\max(\min)}$ are the corresponding largest (smallest) eigenvalues, every $m$-producible state $\rho$ satisfies the QFI bound (Hyllus--Tóth)~\cite{hyllus2012fisher, toth2012multipartite}
\begin{equation}
F_Q[\rho,\mathcal{O}]
\le\max_{\substack{\text{partitions } \{B_\ell\}\\ |B_\ell|\le m}}
\sum_{\ell}\Big(\sum_{i\in B_\ell}\Delta_i\Big)^2,
\label{eq:kprod-general}
\end{equation}
where the maximum runs over all partitions of $\{1,\dots,N\}$ into blocks $B_\ell$ with $|B_\ell|\le m$.

If $\Delta_i=\Delta$ for all $i$, then \eqref{eq:kprod-general} becomes
\[
F_Q[\rho,\mathcal{O}] \le \Delta^2\sum_\ell |B_\ell|^2.
\]
Thus, for fixed $N=\sum_\ell |B_\ell|$ and the constraint $|B_\ell|\le m$, maximizing the bound reduces to maximizing $\sum_\ell |B_\ell|^2$. The optimal partition is therefore as many size-$m$ blocks as possible and at most one residual block. Writing $N=sm+r$ with $s=\lfloor N/m\rfloor$ and $0\le r<m$, the maximizing partition has $s$ blocks of size $m$ and, if $r>0$, one block of size $r$ (all other blocks empty). Therefore
\begin{equation}
F_Q[\rho,\mathcal{O}] \le \Delta^2\left(sm^2+r^2\right),
\label{eq:kprod-hom}
\end{equation}
with equality attained by that partition (if $r=0$, it is $s$ blocks of size $m$). 

Hence, if a measured (or computed) QFI violates \eqref{eq:kprod-hom} for some $m$, i.e.
$
F_Q[\rho,\mathcal{O}] > \Delta^2\left(sm^2+r^2\right),
$
then the state has entanglement depth at least $m{+}1$. 
\paragraph{Remark 1 (inhomogeneous local spectra; finite-$N$ bound).}
Let $\mathcal{O}=\sum_{i=1}^{N} c_i\mathcal{O}_i$ with $c_i\in\mathbb{R}$ and local spectral widths
$\Delta_i:=\lambda_{\max}(\mathcal{O}_i)-\lambda_{\min}(\mathcal{O}_i)$, so that the spectral width of
$c_i\mathcal{O}_i$ is $|c_i|\Delta_i$. Define the nonnegative weights
\[
w_i := |c_i|\Delta_i.
\]
For an $m$-producible state $\rho$, the Hyllus--T\'oth bound \eqref{eq:kprod-general} implies the tight finite-$N$
constraint
\begin{equation}
F_Q[\rho,\mathcal{O}]
\le
\max_{\substack{\text{partitions } \{B_\ell\}\\ |B_\ell|\le m}}
\sum_{\ell}\Big(\sum_{i\in B_\ell} w_i\Big)^2.
\label{eq:kprod-inhom-tight}
\end{equation}

Moreover, since $x^2$ is convex, the maximizing partition in \eqref{eq:kprod-inhom-tight} is obtained by grouping the
largest weights together. Concretely, let $w_{(1)}\ge w_{(2)}\ge\cdots\ge w_{(N)}$ be the nonincreasing rearrangement
of $\{w_i\}$, and write $N=sm+r$ with $s=\lfloor N/m\rfloor$ and $0\le r<m$. Then the tight bound can be written
explicitly as
\begin{equation}
F_Q[\rho,\mathcal{O}]
\le
\sum_{\ell=1}^{s}\Big(\sum_{j=(\ell-1)m+1}^{\ell m} w_{(j)}\Big)^2
+
\Big(\sum_{j=sm+1}^{N} w_{(j)}\Big)^2,
\label{eq:kprod-inhom-sorted}
\end{equation}
with the last term absent if $r=0$.

It is sometimes convenient to use a simpler, fully analytic (but generally non-tight) relaxation of
\eqref{eq:kprod-inhom-tight}. Applying Cauchy--Schwarz on each block $B_\ell$ gives
\begin{equation}
\Big(\sum_{i\in B_\ell} w_i\Big)^2
\le
|B_\ell|\sum_{i\in B_\ell} w_i^2.
\label{eq:block-cs}
\end{equation}
Summing \eqref{eq:block-cs} over all blocks in the partition (including the residual block of size $r$ if
$N=sm+r$ with $0\le r<m$) yields
\begin{equation}
\sum_{\ell}\Big(\sum_{i\in B_\ell} w_i\Big)^2
\le
\sum_{\ell}|B_\ell|\sum_{i\in B_\ell} w_i^2.
\label{eq:cs-sum-over-blocks}
\end{equation}
Since every admissible block satisfies $|B_\ell|\le m$ (and in particular the residual block has size $r<m$),
each term on the right-hand side obeys
\[
|B_\ell|\sum_{i\in B_\ell} w_i^2 \le m\sum_{i\in B_\ell} w_i^2,
\]
so that
\begin{equation}
\sum_{\ell}\Big(\sum_{i\in B_\ell} w_i\Big)^2
\le
m\sum_{\ell}\sum_{i\in B_\ell} w_i^2
=
m\sum_{i=1}^{N} w_i^2.
\label{eq:kprod-inhom-relaxed-derivation}
\end{equation}
Because \eqref{eq:kprod-inhom-relaxed-derivation} holds for every admissible partition, it also holds after
maximizing over partitions in \eqref{eq:kprod-inhom-tight}. We therefore obtain the relaxed, finite-$N$ bound
\begin{equation}
F_Q[\rho,\mathcal{O}]
\le
m\sum_{i=1}^{N} w_i^2
=
m\sum_{i=1}^{N} c_i^2\Delta_i^2,
\label{eq:kprod-inhom-relaxed}
\end{equation}
which is valid for arbitrary $N$ and does not require $m$ to divide $N$. We stress that the use of the factor $m$ is
a uniform upper bound that also covers the residual block (since $r<m$); keeping $|B_\ell|$ explicitly in
\eqref{eq:cs-sum-over-blocks} yields a slightly tighter but partition-dependent expression.

\paragraph{Remark 2 (sum of witnesses and spectral-width control).}
Suppose we define a composite quantity as the sum of two QFI densities,
$
f_Q^{(\Sigma)} \equiv f_Q(\mathcal{O}^{(1)}) + f_Q(\mathcal{O}^{(2)}) ,
$
with additive generators
$
\mathcal{O}^{(k)}=\sum_{i=1}^N\mathcal{O}^{(k)}_i,
\qquad
\Delta_i^{(k)}:=\lambda_{\max}\big(\mathcal{O}^{(k)}_i\big)-\lambda_{\min}\big(\mathcal{O}^{(k)}_i\big).
$
For any $m$-producible state $\rho$ one has, by the standard block-partition argument and Cauchy--Schwarz,
\begin{align}
f_Q(\mathcal{O}^{(k)}) 
&\le m\overline{(\Delta^{(k)})^2},
\qquad
\overline{(\Delta^{(k)})^2} :=\frac{1}{N}\sum_{i=1}^N \big(\Delta_i^{(k)}\big)^2,
\label{eq:single-chan-inhom}
\end{align}
hence the sum bound
\begin{equation}
f_Q^{(\Sigma)}
\le m\Big(
\overline{(\Delta^{(1)})^2}
+\overline{(\Delta^{(2)})^2}
\Big).
\label{eq:sum-bound}
\end{equation}
Therefore, the normalized sum witness
\begin{align}
\frac{f_Q^{(\Sigma)}}
{\overline{(\Delta^{(1)})^2}+\overline{(\Delta^{(2)})^2}}
&> m 
\Longrightarrow
\text{entanglement depth } \ge m+1.
\end{align}

Equation \eqref{eq:sum-bound} is likewise a convenient sufficient (generally non-tight) bound obtained from
Cauchy--Schwarz and block-size control; the corresponding tight finite-$N$ bound follows from applying
\eqref{eq:kprod-general} separately to each generator and maximizing over partitions.

\subsection{Spin Structure Factors\label{sec:QFIM_SSF}}
In the main text, we have use $S_\mathbf{q}^\alpha=\sum_i S_i^\alpha e^{i\mathbf{q}\cdot\mathbf{R}_i}$ as the operator associated with the QFI. One would note that the usual definition of QFI requires an operator $\mathcal{O}$ to be Hermitian such that the induced generator $e^{i\theta \mathcal{O}}$ is unitary and thereby generates norm-preserving dynamics. Although $S_\mathbf{q}^\alpha$ is not Hermitian, we show below that the definition of QFI in Eq.~({\color{blue}4}) of the main text still encompasses all the important properties of the conventional QFI. Namely, it is a robust measure of fluctuations, and we can still derive a meaningful lower bound of entanglement depth. 

To keep our results general, let us consider some Hermitian operator $\mathcal{O}_{\mathbf{R}_\mu}$ on each site under the coordinate system specified in Eq.~\eqref{eq:SIPC_SIDC} such that the operator we associate with QFI has the form $\mathcal{O}_\mathbf{q}=\sum_{\mathbf{R}_\mu} \mathcal{O}_{\mathbf{R}_\mu} e^{i\mathbf{q}\cdot \mathbf{R}_\mu}$, where $\mathcal{O}_{\mathbf{R}_\mu}$ is some local observable with homogeneous spectral width $\Delta_{\mathbf{R}_\mu}=\Delta$. For later discussion, let us denote such a homogeneous spectral width as $\Delta(\mathcal{O}_{\mathbf{R}_\mu})$ to be explicit about the underlying local operators. We can plug in $\mathcal{O}_{\mathbf{R}_\mu} = S^{\alpha}_{\mathbf{R}_\mu}$ where $\alpha\in\{\mathrm{x},\mathrm{y},\mathrm{z}\}$ at the end to make a direct connection with the definitions used in the main text. To begin, let us split $\mathcal{O}_\mathbf{q}=\mathcal{O}_{c;\mathbf{q}} +i\mathcal{O}_{s;\mathbf{q}}$ into an Hermitian and antihermitian parts, where $\mathcal{O}_{c;\mathbf{q}}$ and $\mathcal{O}_{s;\mathbf{q}}$ are Hermitian operators:
\begin{align}
    \mathcal{O}_{c;\mathbf{q}} &=\frac{\mathcal{O}_\mathbf{q}+\mathcal{O}_\mathbf{q}^\dagger}{2} =\sum_{\mathbf{R}_\mu} O_{\mathbf{R}_\mu} \cos(\mathbf{q}\cdot{\mathbf{R}_\mu}) \label{eq:O_c}\\
    \mathcal{O}_{s;\mathbf{q}} & =\frac{\mathcal{O}_\mathbf{q}-\mathcal{O}_\mathbf{q}^\dagger}{2i} =\sum_{\mathbf{R}_\mu} O_{\mathbf{R}_\mu} \sin(\mathbf{q}\cdot{\mathbf{R}_\mu})\label{eq:O_s}
\end{align}
Now we construct a QFI matrix (QFIM) $F$~\cite{liu2019quantum,fiderer2021general,dominik2018simple}, whose elements are specified by:
\begin{equation}
	F_{ab} (T):=4 \int d \omega \tanh \left(\frac{\omega}{2T}\right)\left(1-e^{- \omega/T}\right) \mathcal{A}_{ab}(\mathbf{q}, \omega).
	\label{eq:QFIM}
\end{equation}
where
\begin{equation}
    \mathcal{A}_{ab}(\mathbf{q}, \omega) = \frac{1}{2\pi N}\int dt e^{i\omega t}\langle \mathcal{O}_{a,\mathbf{q}}(t) \mathcal{O}_{b,\mathbf{q}}(0)\rangle.
\end{equation}
with $a,b\in\{c,s\}$. Now we wish to show that $f_Q(\mathcal{O}_\mathbf{q}, T)$ defined in the main text via Eq.~({\color{blue}4}) is related to the trace of this QFIM. Since $\mathcal{O}_\mathbf{q}=\mathcal{O}_{c;\mathbf{q}} +i\mathcal{O}_{s;\mathbf{q}}$,
\begin{align}
\mathcal{O}_\mathbf{q}^\dagger \mathcal{O}_\mathbf{q} &= \mathcal{O}_{c;\mathbf{q}} \mathcal{O}_{c;\mathbf{q}} +\mathcal{O}_{s;\mathbf{q}} \mathcal{O}_{s;\mathbf{q}} + i [\mathcal{O}_{c;\mathbf{q}},\mathcal{O}_{s;\mathbf{q}}]\notag\\
&=\mathcal{O}_{c;\mathbf{q}} \mathcal{O}_{c;\mathbf{q}} +\mathcal{O}_{s;\mathbf{q}} \mathcal{O}_{s;\mathbf{q}},
\end{align}
as $[\mathcal{O}_{c;\mathbf{q}},\mathcal{O}_{s;\mathbf{q}}]=0$. Therefore, we see that
\begin{align}
    A(\mathbf{q},\omega) &:=\frac{1}{2\pi N}\int dt\langle \mathcal{O}^{\dagger}_\mathbf{q}(t)\mathcal{O}_{\mathbf{q}}(0)\rangle e^{i\omega t} \notag\\
    &=\frac{1}{2\pi N}\int dte^{i\omega t} \langle \mathcal{O}_{c;\mathbf{q}}(t) \mathcal{O}_{c;\mathbf{q}}(0)\rangle+\langle\mathcal{O}_{s;\mathbf{q}}(t) \mathcal{O}_{s;\mathbf{q}}(0) \rangle \notag\\
    &= \mathcal{A}_{cc} + \mathcal{A}_{ss}.
\end{align} 
As a result,
\begin{align}
    f_{Q}(\mathcal{O}_\mathbf{q}, T) &= \int d\omega \tanh(\frac{\omega}{2T})(1-e^{-\omega/T})A(\mathbf{q}, \omega)\notag\\
    &= \int d\omega \tanh(\frac{\omega}{2T})(1-e^{-\omega/T})(\mathcal{A}_{cc}(\mathbf{q}, \omega)+ \mathcal{A}_{ss}(\mathbf{q}, \omega))\notag\\
    &=f_{Q}(\mathcal{O}_{c;\mathbf{q}}, T) + f_{Q}(\mathcal{O}_{s;\mathbf{q}}, T) = \operatorname{Tr}(F).
\end{align}
In other words, the QFI used in the main text is precisely the trace of the QFI matrix introduced above. Interpreting $F$ as a covariance matrix, this shows that the $f_{Q}(\mathcal{O}_\mathbf{q}, T)$ equals the total variance --- the sum of the variances carried by all fluctuation modes. 

Furthermore, $f_{Q}(\mathcal{O}_\mathbf{q}, T)$ provides a robust lower bound on the entanglement depth as $f_{Q}(\mathcal{O}_\mathbf{q}, T) =f_{Q}(\mathcal{O}_{c;\mathbf{q}}, T) + f_{Q}(\mathcal{O}_{s;\mathbf{q}}, T)$ is the well-bounded sum witness discussed in remark 2. Using the result outlined in the remark, for an $m$-producible state, we have
\begin{equation}
f_{Q}(\mathcal{O}_\mathbf{q}, T)\leq m\left((\overline{\Delta\mathcal{O}_{c,\mathbf{q}})^2} + (\overline{\Delta\mathcal{O}_{s,\mathbf{q}})^2}\right)\label{eq:best_bound}.
\end{equation}
Finally, by negation,
\begin{equation}
   f_{Q}(\mathcal{O}_\mathbf{q}, T)>m\left((\overline{\Delta\mathcal{O}_{c,\mathbf{q}})^2} + (\overline{\Delta\mathcal{O}_{s,\mathbf{q}})^2}\right)
\end{equation}
implies the state is at least $(m+1)$-partite entangled.

As such, all that remains is to obtain a bound for these two channels. Using Eq.~\eqref{eq:kprod-inhom-relaxed} with inhomogeneous spectral width for each local operator $\mathcal{O}_{\mathbf{R}_\mu}$, $\Delta_{\mathbf{R}_\mu}=\Delta(\mathcal{O}_{\mathbf{R}_\mu})$, and $c_{\mathbf{R}_\mu}=\cos(\mathbf{q}\cdot\mathbf{R}_\mu)$,
\begin{equation}
    \overline{(\Delta\mathcal{O}_{c,\mathbf{q}})^2} = \frac{\left(\Delta(\mathcal{O}_{\mathbf{R}_\mu})\right)^2}{N}\sum_{\mathbf{R_\mu}}\cos^2(\mathbf{q}\cdot \mathbf{R}_\mu).
\end{equation}
Therefore, by arguing the same for $\mathcal{O}_{s,\mathbf{q}}$, we can finally bound $f_Q(\mathcal{O}_\mathbf{q})$ for any $m$-producible state:
\begin{align}
    f_Q(\mathcal{O}_\mathbf{q})  &\leq \frac{m\left(\Delta(\mathcal{O}_{\mathbf{R}_\mu})\right)^2}{N} \sum_{\mathbf{R}_\mu} \left( \cos^2(\mathbf{q}\cdot \mathbf{R}_\mu) + \sin^2(\mathbf{q}\cdot \mathbf{R}_\mu) \right) \notag \\
    &= m\left(\Delta(\mathcal{O}_{\mathbf{R}_\mu})\right)^2 \label{eq:bound_on_S_alpha}.
\end{align}
As such, we find that $f_Q(\mathcal{O}_\mathbf{q})$ is bounded above by $m\left(\Delta(\mathcal{O}_{\mathbf{R}_\mu})\right)^2$. This is precisely the definition of the entanglement bound for conventional QFI~\cite{Witnessing2021Scheie, Tutorial2025Scheie,liu2016quantum} as if we have ignored the exponential part $e^{i\mathbf{q}\cdot \mathbf{R}_\mu}$ altogether. 

To conclude, for an $m$-producible state, any operator of the form $\mathcal{O}_\mathbf{q}=\sum_{\mathbf{R}_\mu} \mathcal{O}_{\mathbf{R}_\mu} e^{i\mathbf{q}\cdot \mathbf{R}_\mu}$, where $\mathcal{O}_{\mathbf{R}_\mu}$ is some observable with homogeneous spectrum $\pm1/2$ across all sites, has the corresponding QFI bound
\begin{equation}
f_Q(\mathcal{O}_\mathbf{q},T)\leq m \left(\Delta(\mathcal{O}_{\mathbf{R}_\mu})\right)^2. \label{eq:generalized_QFI_bound}
\end{equation}
Therefore, for $\mathcal{O}_\mathbf{q} = S^\alpha_{\mathbf{q}}$ where $\alpha\in\{\mathrm{x},\mathrm{y},\mathrm{z}\}$ in a spin-$1/2$ system:
\begin{equation}
f_Q(S^\alpha_\mathbf{q},T) > m  (\Delta(S^\alpha_{\mathbf{R}_\mu}))^2 = m  (1/2-(-1/2))^2 = m,
\end{equation}
implies that the state is at least $(m+1)$-partite entangled, regardless of momentum position $\mathbf{q}$.

Now, since we have established a well-constructed QFI for non-Hermitian operators, we can extend this even further for $\alpha=\pm$ in the main text. As discussed in the main text, $f_Q(S^\pm_\mathbf{q},T) = f_Q(S^x_\mathbf{q},T) + f_Q(S^y_\mathbf{q},T)$ is a sum witness. Therefore, by remark 2, for an $m$-producible state:
\begin{equation}
    f_Q(S^\pm_\mathbf{q},T) = f_Q(S^x_\mathbf{q},T) + f_Q(S^y_\mathbf{q},T) \leq m (1+1) = 2m.
\end{equation}
Equivalently,  this is to say that if the spinon channel has nQFI
\begin{equation}
\text{nQFT}(S^\alpha_\mathbf{q}) = f_Q(S^\pm_\mathbf{q},T) / 2 >m,
\end{equation}
then the state is at least $(m+1)$-partite entangled.

\subsection{Neutron Scattering Cross Section}
Similarly, for $A^{\text{DO}}$, to have a meaningful interpretation of the entanglement depth, we would have to determine $\lambda_{max}$ and $\lambda_{min}$ of the operator in which we are evaluating our QFI to determine the QFI spectral bound $\eta$. To do so, first, we should determine the exact operator $\mathcal{O}$ that the neutron scattering cross section contains. Let us define operators
\begin{subequations}
\begin{align}
    S_\mathbf{q}^{\text{NSF}} &= \sum_{\mathbf{R}_\mu} \left(\hat{\mathbf{p}}\cdot\hat{\mathbf{z}}_\mu \right)\tau^z_{\mathbf{R}_\mu}e^{i\mathbf{q}\cdot\mathbf{R}_\mu} \label{eq:op_NSF}\\
    S_\mathbf{q}^{\text{SF}} &= \sum_{\mathbf{R}_\mu} \left(\hat{\mathbf{v}}\cdot\hat{\mathbf{z}}_\mu \right)\tau^z_{\mathbf{R}_\mu}e^{i\mathbf{q}\cdot\mathbf{R}_\mu} \label{eq:op_SF}.
\end{align}
\end{subequations}
Here $\hat{\mathbf{p}}$ denotes the neutron polarization, which is perpendicular to the momentum transfer ($\hat{\mathbf{v}}\cdot \hat{\mathbf{p}}=0$), and $\hat{\mathbf{v}}$ is a vector perpendicular to both the polarization vector and the neutron momentum transfer ($\hat{\mathbf{v}}\cdot \hat{\mathbf{p}}=\hat{\mathbf{v}}\cdot \hat{\mathbf{q}}=0$). These channels correspond to polarized neutron scattering experiments in the spin-flip/non-spin-flip (SF/NSF) channels with respect to the neutron polarization vector $\hat{\mathbf{p}}$. By definition,
\begin{align}
    A^{\text{DO}} \sim \langle S^{\text{NSF}^\dagger}_\mathbf{q} S^{\text{NSF}}_\mathbf{q}\rangle   + \langle S^{\text{SF}^\dagger}_\mathbf{q} S^{\text{SF}}_\mathbf{q}\rangle,
\end{align}
where we sum up all transverse modes, which is equivalent to taking the transverse projector in Eq.\eqref{eq:ins_cs_tau}. Accordingly, we assemble a $2\times 2$ QFIM $F^{\mathrm{DO}}$ from the two transverse components—taken here as the NSF and SF channels—so that the cross section $A^{\mathrm{DO}}$ defined in Eq.~\eqref{eq:ins_cs_tau} is identified with $\operatorname{Tr} F^{\mathrm{DO}}$. This construction is necessary because the transverse projector has two independent components. Hence, there is no single scalar operator $S^{\mathrm{DO}}$ with $A^{\mathrm{DO}}\sim\langle S^{\mathrm{DO}\dagger} S^{\mathrm{DO}}\rangle$. Nevertheless, we still denote the QFI related to $A^{\text{DO}}$ as $f_Q(S^{\text{DO}}_\mathbf{q}, T)$ for consistency's sake. The operators entering $F^{\mathrm{DO}}$ are generally non-Hermitian, which is admissible since the bounds derived in the previous section apply directly. Applying Eq.~\eqref{eq:generalized_QFI_bound} to each channel then yields a lower bound on the entanglement depth. Finally, to bound the QFI induced by unpolarized neutron scattering, let us denote the corresponding QFI from the $A^{\text{DO}}$ channel as the sum witness in remark 2 by
\begin{equation}
    f_Q(S^{\text{DO}}_\mathbf{q}) = f_Q(S^{\text{NSF}}_\mathbf{q}) + f_Q(S^{\text{SF}}_\mathbf{q}).
\end{equation}

Therefore, let us bound the NSF and SF channels to bound the total unpolarized neutron scattering cross section. To do so, we have to determine $\lambda_{max}$ and $\lambda_{min}$ of each basis operator to get an actual interpretation of the entanglement depth. To do so, we need to evaluate at a specific momentum position $\mathbf{q}$. For clarity of later discussion, let us define vector $\Pi_{NSF}$ and $\Pi_{SF}$ with four components corresponding to the four pyrochlore sublattices $\mu$ as
\begin{subequations}
\begin{align}
\Pi^\mu_{\text{NSF}}(\mathbf{q})&=\hat{\mathbf{p}}\cdot\hat{\mathbf{z}}_\mu \\
\Pi^\mu_{\text{SF}}(\mathbf{q})&=\hat{\mathbf{v}}\cdot\hat{\mathbf{z}}_\mu.
\end{align}
\end{subequations}
Since, in principle, the operator $S_\mathbf{q}^{\text{(N)SF}}$ is non-Hermitian, we should break it down into the same cosine and sine channels again and discuss the sum of the spectral width on each channel. By Eq.~\eqref{eq:best_bound}, we need to bound the cosine and the sine channel again. Let us denote $S^{\text{(N)SF}}_{c(s),\mathbf{q}}$ as the corresponding cosine and sine channels. Then, using Eq.~\eqref{eq:kprod-inhom-relaxed} with $c_{\mu}=\Pi_{\text{(N)SF}}^\mu$:
\begin{equation}
    \overline{(\Delta S^{\text{(N)SF}}_{c,\mathbf{q}})^2}=\frac{1}{N}\sum_{\mathbf{R}_\mu} \cos^2(\mathbf{q}\cdot \mathbf{R}_\mu) |\Pi^{\mu}_{\text{(N)SF}}|^2.
\end{equation}

Therefore, using Eq.~\eqref{eq:best_bound}, we can derive a QFI bound on the NSF and SF channel separately for an $m$-producible state:
\begin{align}
    f_Q(S^{\text{(N)SF}}_\mathbf{q}, T) &\leq  \frac{m}{N}\sum_{\mathbf{R}_\mu} \left(\cos^2(\mathbf{q}\cdot \mathbf{R}_\mu) +\sin^2(\mathbf{q}\cdot \mathbf{R}_\mu) \right)|\Pi^{\mu}_{\text{(N)SF}}|^2 \notag \\
    &= \frac{m}{N}\sum_{\mathbf{R}_\mu} |\Pi^{\mu}_{\text{(N)SF}}|^2 = \frac{m}{4} \sum_{\mu} |\Pi^{\mu}_{\text{(N)SF}}|^2.\label{eq:DO_QFI_bound}
\end{align}
In the last line, we have used the fact that $\Pi^{\mu}_{\text{(N)SF}}$ depends only on the sublattice but not on the unit cell position. We can then sum over all unit cells, which gives a prefactor of $N/4$
Finally, the upper bound on $f_{Q}(S^{\text{DO}}_\mathbf{q})$ for an $m$-producible state is
\begin{equation}
 f_Q(S^{\text{DO}}_\mathbf{q}) = f_Q(S^{\text{NSF}}_\mathbf{q}) + f_Q(S^{\text{SF}}_\mathbf{q}) \leq  \frac{m}{4} \sum_{\mu} \left( |\Pi^{\mu}_{\text{NSF}}|^2 +|\Pi^{\mu}_{\text{SF}}|^2 \right)  .
\end{equation}
Now, to actually evaluate this bound, let us write
\begin{align}
\sum_{\mu} \bigl|\Pi^{\mu}_{\mathrm{(N)SF}}\bigr|^2
=\sum_{\mu}\bigl|\hat{\mathbf a}\cdot\hat{\mathbf z}_\mu\bigr|^2
=\hat{\mathbf a}^{\top}\Bigl(\sum_{\mu}\hat{\mathbf z}_\mu \hat{\mathbf z}_\mu^{\top}\Bigr)\hat{\mathbf a}
=\hat{\mathbf a}^{\top} A\hat{\mathbf a},
\end{align}
where $A:=\sum_{\mu}\hat{\mathbf z}_\mu \hat{\mathbf z}_\mu^{\top}$ and $\hat{\mathbf{a}}=\hat{\mathbf{p}}$ if NSF and $\hat{\mathbf{a}}=\hat{\mathbf{v}}$ if SF.

For a single pyrochlore tetrahedron, take the four local $\hat{\mathbf{z}}_\mu$, a direct sum gives
\begin{equation}
A=\sum_{\mu=0}^{3}\hat{\mathbf z}_\mu \hat{\mathbf z}_\mu^{\top}
=\frac{4}{3} \mathds{1}_{3\times 3}, 
\end{equation}
since all off-diagonal entries cancel by symmetry (the signs appear equally often with opposite parity) and
$\operatorname{Tr}A=\sum_{\mu}\|\hat{\mathbf z}_\mu\|^2=4$ fixes the proportionality constant.
Hence
\begin{equation}
\sum_{\mu} \bigl|\Pi^{\mu}_{\mathrm{(N)SF}}\bigr|^2
=\frac{4}{3}||\hat{\mathbf{a}}||^2=\frac{4}{3},
\end{equation}
which is independent of the direction of the unit vector $\hat{\mathbf a}$. As a result, for an $m$-producible state, QFI upper bounds on NSF, SF, and the total scattering channel are given by
\begin{subequations}
\begin{align}
    f_Q(S^{\text{(N)SF}}_\mathbf{q}, T) &\leq \frac{m}{4} \sum_{\mu} |\Pi^{\mu}_{\text{(N)SF}}|^2 = \frac{m}{3}.\\
    f_Q(S^{\text{DO}}_\mathbf{q}, T) &\leq \frac{2m}{3}.
\end{align}
\end{subequations}
To summarize, we derived a bound independent of $\mathbf{q}$ for the NSF, SF, and the unpolarized scattering channel with corresponding nQFI:
\begin{subequations}
\begin{align}
    \text{nQFI}(S^{\text{(N)SF}}_\mathbf{q}, T) &= 3f_Q(S^{\text{(N)SF}}_\mathbf{q}, T)\\
    \text{nQFI}(S^{\text{DO}}_\mathbf{q}, T) &= \frac{3}{2}f_Q(S^{\text{(N)SF}}_\mathbf{q}, T).
\end{align}
\end{subequations}
We again stress that this bound is not the optimal bound in general but an upper bound of such via Cauchy-Schwartz as shown in Eq.~\eqref{eq:DO_QFI_bound}. The true optimal bound depends strongly on the choice of $\hat{\mathbf{p}}$ and $\hat{\mathbf{v}}$ and thereby the incident neutron momentum $\hat{\mathbf{q}}$. In theory, one would need to apply Eq.~\eqref{eq:kprod-general} carefully to find the best bound.

\setcounter{page}{1}
\setcounter{equation}{0}
\setcounter{figure}{0}
\renewcommand{\theequation}{S\arabic{equation}}
\renewcommand{\thefigure}{S\arabic{figure}}

\bibliographystyle{apsrev4-1}
\newpage

\end{document}